\documentclass[aps,prb,amsmath,twocolumn,amssymb,floatfix,showpacs,superscriptaddress,nofootinbib,longbibliography]{revtex4-2}
\usepackage{mathtools}
\usepackage{braket}
\usepackage[dvipsnames]{xcolor}
\usepackage{float}
\usepackage{subfigure}
\usepackage{dsfont}
\usepackage[colorlinks=true,linktoc=page,linkcolor=blue,citecolor=blue,urlcolor=blue]{hyperref}
\newcommand{\vect}[1]{\boldsymbol{\mathrm{#1}}}
\mathchardef\mhyphen="2D % Define a "math hyphen"

\newcommand{\ie}{{i.e.,\,\,}}

\newcommand\bea{\begin{eqnarray}}
\newcommand\eea{\end{eqnarray}}
\newcommand\beq{\begin{equation}}  
\newcommand\eeq{\end{equation}}

\newcommand{\non}{\nonumber}

\newcommand{\N}{\mathcal{N}} % Nxy
\newcommand{\W}{\mathcal{W}} % W

\newcommand{\mc}{\mathcal}
\newcommand{\mbf}{\mathbf}

\usepackage[normalem]{ulem}
\definecolor{lime}{HTML}{A6CE39}
\usepackage{sidecap,tikz}
\DeclareRobustCommand{\orcidicon}{\hspace{-1.0mm}
	\begin{tikzpicture}
		\draw[lime, fill=lime] (0.0,0.0) 
		circle [radius=0.15] 
		node[white] {{\fontfamily{qag}\selectfont \tiny \,ID}};
		\draw[white, fill=white] (-0.0525,0.095) 
		circle [radius=0.007];
	\end{tikzpicture}
	\hspace{-3.0mm}
}
\foreach \x in {A, ..., Z}{\expandafter\xdef\csname orcid\x\endcsname{\noexpand\href{https://orcid.org/\csname orcidauthor\x\endcsname}
		{\noexpand\orcidicon}}
}

\AtBeginDocument{%
	\newwrite\bibnotes
	\def\bibnotesext{Notes.bib}
	\immediate\openout\bibnotes=\jobname\bibnotesext
	\immediate\write\bibnotes{@CONTROL{REVTEX41Control}}
	\immediate\write\bibnotes{@CONTROL{%
			apsrev41Control,author="08",editor="1",pages="1",title="1",year="1"}}
	\if@filesw
	\immediate\write\@auxout{\string\citation{apsrev41Control}}%
	\fi
}%

\allowdisplaybreaks

\begin{document}

%=============START of MAIN PAPER===============

%=================== Title=================%
\title{Multi higher-order Dirac and nodal line semimetals} 
%=========================================%
 
\author{Amartya Pal\orcidA{}}
\email{amartya.pal@iopb.res.in}
\affiliation{Institute of Physics, Sachivalaya Marg, Bhubaneswar-751005, India}
\affiliation{Homi Bhabha National Institute, Training School Complex, Anushakti Nagar, Mumbai 400094, India}

\author{Arnob Kumar Ghosh\orcidB{}}
\email{arnob.ghosh@physics.uu.se}
\affiliation{Department of Physics and Astronomy, Uppsala University, Box 516, 75120 Uppsala, Sweden}

%--------------------------------------------------------%
%            Abstract
%--------------------------------------------------------%
\begin{abstract}
In recent years, there has been a surge of interest in exploring higher-order topology and their semi-metallic counterparts, particularly in the context of Dirac, Weyl, and nodal line semimetals, termed as higher-order Dirac semimetal~(HODSM), higher-order Weyl semimetal, and higher-order nodal line semimetal~(HONLSM). The HODSM phase exhibits hinge Fermi arcs~(FAs) with a quantized higher-order topological invariant. Conversely, the HONLSM phase is a hybrid-order topological phase manifesting both drumhead-like surface states and hinge FAs as a signature of first- and second-order topology, and also possesses both first- and second-order topological invariants. In this work, we investigate a tight binding model for multi-HODSM~(mHODSM) hosting multiple hinge FAs having a quantized quadrupolar winding number~(QWN) greater than one. Furthermore, we obtain a multi-HONLSM~(mHONLSM) phase from the mHODSM by applying an external magnetic field, which breaks the $PT$-symmetry. The mHONLSM phase possesses both the dipolar winding number, non-vanishing only inside the nodal loops, being the representative invariant for first-order topology, and the QWN, featuring both drumhead-like surface states and multiple hinge FAs. We study the spectral properties of the mHODSM and mHONLSM in different geometries. We also investigate the hinge FA-mediated transport in HONLSM employing a two-terminal setup.
\end{abstract}
%--------------------------------------------------------
%--------------------------------------------------------

\maketitle

%======================================================
\section{Introduction} \label{Sec:I_intro}
%======================================================
%--------DSM and WSM -------%
The past decades have witnessed profound research activities in the field of gapped and gapless topological phases of matter. Among the gapless topological phases, Dirac semimetal (DSM), Weyl semimetal (WSM)~ \cite{Wan2011,Burkov2011a,Burkov2011b,Young2012,Zyuzin2012a,Zyuzin2012b,Hal2012,Liu2013,Vazifeh2013,Bzdu2015,Rao2016,Randeria2016PNAS,Yan2017,McCormick2017,Armitage2018,Khanna2014,Burkov2018}, and nodal line semimetal~(NLSM)~\cite{Chiu2014_NLSM,Bian2016_NLSM,Yamakage2016_NLSM,Schnyder2016_NLSM,Nie2019_NLSM,Yang2018NLSM,Parhizgar2020_NLSM,Zhan2023,Matsuura_2013,Zhao2013} in three spatial dimensions have been investigated extensively. At the same time, the experimental discovery of DSMs, WSMs~\cite{Xu2011,Potter2014,Lv2015a,Lv2015b,Lv2015c,Xu2015a,Xu2015b,Xu2015c,Lu2015,Sun2015,Soluyanov2015,Wang2016a,Wang2016b,Moll2016,Xu2016b}, and NLSM~\cite{Bian2016_NLSM,Schnyder2016_NLSM,Nie2019_NLSM} have prompted further interest in this area. In a DSM, the valence and conductance bands touch at isolated points in the Brillouin zone (BZ), forming Dirac nodes (DNs). Around these DNs, the energy spectrum exhibits linear dispersion in momentum hosting Dirac fermions as low-energy excitations~\cite{Armitage2018}. In WSMs, the bulk is gapless at an even number of isolated points in the BZ. The topology of the WSMs is manifested at the boundary via the gapless surface states, known as Fermi arcs (FAs)~\cite{Armitage2018,McCormick2017,Khanna2014}. However, in contrast to the DSM and WSM, the NLSM possesses a bulk with gapless nodal loops in the BZ. Interestingly, the non-trivial topology of NLSM results in flat drumhead-like surface states at the boundary~\cite{Bian2016_drumhead,Weng2015_drumhead} whereas surface states of WSMs are generally dispersive in nature. Topological protection of NLSM requires the presence of crystalline symmetries, e.g., mirror reflection, in addition to internal symmetries~\cite{Chiu2014_NLSM}. Nonetheless, recently, a chiral symmetric NLSM phase has been proposed, which does not require any crystalline symmetries except translation invariance~\cite{Abdulla2023_NLSM1,Abdulla2023_NLSM2}.

%------HOTI, HODSM, HOWSM, HONLSM Definition
In recent times, the discovery of higher-order topological insulators~(HOTIs) has further stimulated research in topological materials~\cite{benalcazar2017,benalcazarprb2017,Song2017,schindler2018,MannaPRBL2022,Ghosh_2024}. Conventional topological insulators (TIs) in $d$-dimension support $(d-1)$-dimensional topologically protected boundary states~\cite{kane2005quantum,BHZPRL2006,bernevig2006quantum}. In contrast, an $n$-th order HOTI in $d$-dimensions manifests topologically protected boundary states at their $(d-n)$-dimensional boundaries~\cite{benalcazar2017,benalcazarprb2017,Song2017,schindler2018}. For example, a second-order TI~(SOTI) harbors zero-dimensional~(0D) localized corner and one-dimensional~(1D) gapless hinge states in two and three dimensions, respectively. In comparison, a third-order TI in three dimensions manifests 0D localized corner states. The notion of higher-order topology has also been transmuted into higher-order topological semimetals, namely higher-order DSM~(HODSM)~\cite{Lin2018,Bernevig2020NatComm,Wu2020,Qui2021,Wang2022,ChenQCHODSM2023,rafiulislam2024engineering}, higher-order WSM (HOWSM)~\cite{GhorashiPRL2020,WangPRL2020,Rui2022,Tanaka2022}, and higher-order NLSM~(HONLSM)~\cite{Wang2019,Wang2020,Chen2022,Qiu2024_HONLSM,Ma2024}. The HODSM hosts 1D hinges states as a signature of second-order topology and is characterized by the bulk quadrupole moment~\cite{Lin2018}. On the other hand, the HOWSM embodies both first-order (conventional TIs) and second-order topological phases, manifesting a hybrid-order topology~\cite{GhorashiPRL2020,WangPRL2020,Rui2022,Tanaka2022,WangPRBL2023}. At the same time, the HONLSM exhibits bulk bands with gapless nodal loops in the BZ, reminiscent of the bulk of NLSM~\cite{Wang2019,Wang2020,Chen2022,Qiu2024_HONLSM,Ma2024}. However, in contrast to the first-order NLSM, the HONLSM hosts both the drumhead-like surface states and along with hinge FAs~\cite{Wang2019,Wang2020,Chen2022,Qiu2024_HONLSM,Ma2024}. While the topological characterization of the HONLSM requires computing both the winding number or Zak phase and bulk quadrupole moment as the representative of first and second-order topology, respectively~\cite{Chen2022,Abdulla2023_NLSM1,Abdulla2023_NLSM2,Ma2024,Qiu2024_HONLSM}. The HONLSM phase has also been realized in different meta-material platforms such as acoustic setup~\cite{Qiu2024_HONLSM} and phononic crystals~\cite{Ma2024}.

Furthermore, HOTIs can also be generalized to a multi-HOTI~(mHOTI) harboring multiple states per boundary~\cite{Benalcazar2022Nxy}. The mHOTIs are protected by chiral symmetry and are topologically characterized by employing a multipole chiral winding number~(MCWN)~\cite{Benalcazar2022Nxy}. However, these mHOTIs require long-range hoppings and hence have been realized only in meta-material platforms so far~\cite{Benalcazar2022Nxy,WangPRLMCN2023,LiuPRApp2023, LiPRApp2023,LiPRBMCN2023}. Very recently, MCWN has been utilized to predict multi-HODSM~(mHODSM)~\cite{Qi2024HOmDSM} where DNs are associated with MCWN and the system supports multiple FAs per hinge. However, a multi-HONLSM~(mHONLSM) phase featuring MCWN and multiple hinge FAs along with drumhead surface states is yet to be reported. Moreover, having multiple hinge FAs provides us with multiple available conducting channels for FA-mediated transport~\cite{Pal2024}. Motivated by the above fact, in this work, we seek the answers to the following questions- (1) Is it possible to obtain the mHONLSM phase with a higher MCWN starting from the mHODSM phase? (2) What are the signatures of mHONLSM in the band structure and at the boundary of the system? (3) Can we obtain FA-mediated enhanced conductance in a nanowire geometry consisting of mHONLSM? In this manuscript, we address these interesting questions.

%---- Answer----
In this work, we consider a three-dimensional~(3D) tight-binding model Hamiltonian that hosts the mHODSM phase, protected by the $PT$-symmetry. The Hamiltonian consists of both first- and second-nearest neighbor hopping, while the latter is referred to as the long-range hopping throughout our manuscript. First, we investigate the topological phase diagram of this model employing the MCWN, in particular, the quadrupolar winding number~(QWN) $\mc{N}_{xy}$. We analyze the bulk Fermi surface featuring the DNs of the bulk Hamiltonian and also for a system with semi-infinite geometries. Next, we break the $PT$-symmetry by applying an external magnetic field to the mHODSM, which in turn transmutes the system to the mHONLSM. We observe that the nodal points of the mHODSM phase get inflated into nodal loops resembling the nature of NLSM. Interestingly, the system also posses chiral symmetry, and we compute the dipolar winding number (DWN) $\mc{W}$ to characterize the first-order topology and the QWN $\mc{N}_{xy}$ to characterize the second-order topology of the mHONLSM phase. We obtain two kinds of mHONLSM phases with (i) $\N_{xy}=1,4$ and $\W=-1$ and (ii) $\N_{xy}=4$ and $\W=-1$. We analyze the band structure of these two kinds of mHONLSM phases in different geometries. We also explore the FA-mediated transport properties of mHONLSM by employing a two-terminal setup.

%--------Structure------%
The remainder of the paper is organized as follows. In Sec.~\ref{Sec:II_HODSM}, we introduce the tight-binding model for the mHODSM phase and study the corresponding phase diagram and band structures. Sec.~\ref{Sec:III_HOMNLSM} is devoted to the main results of this manuscript, where we discuss the mHONLSM phase and characterize it by computing $\W$ and $\N_{xy}$. We study the band structure of the mHONLSM and also investigate the FA-mediated conductance signature employing a two-terminal setup. Finally, we summarize our article and provide a brief outlook in Sec.~\ref{Sec:IV_conclusion}.\\

%======================================================
\section{\lowercase{m}HODSM}\label{Sec:II_HODSM}
%======================================================
In this section, we introduce the model Hamiltonian exhibiting the mHODSM phase. We topologically characterize the phase using QWN and obtain the phase diagram. Afterward, we analyze the band structure of the system in different geometries.

%---------------------------------------------------------------------
\subsection{Model Hamiltonian} \label{DSM_model}
%---------------------------------------------------------------------
We consider a model for the 3D mHODSM, which consists of stacking of 2D SOTI layers along the $z$-direction~\cite{schindler2018,Roy2019,Qi2024HOmDSM}. The system can be described by the following tight binding Hamiltonian on a cubic lattice as
\begin{widetext}
\begin{align}
    \mc{H}(\mbf{k}) = & \lambda_1 (\sin k_x \Gamma_1 + \sin k_y \Gamma_2) + (m-t_x \cos k_x - t_y \cos k_y)\Gamma_3 + \Delta (\cos k_x - \cos k_y)\Gamma_4  \non \\
    & +(\lambda_2 + t_z \cos k_z)[\sin 2k_x \Gamma_1 + \sin 2k_y \Gamma_2 - (\cos 2k_x + \cos 2k_y)\Gamma_3 + (\cos 2k_x - \cos 2k_y)\Gamma_4] \ ,
    \label{Eq.mHODSM_Hamiltonian}
\end{align}
\end{widetext}
where, $t_x$, $t_y$, and $t_z$ represent the nearest-neighbour hopping amplitudes along $x$-, $y$-, and $z$-directions, respectively. The on-site crystal field splitting, spin-orbit coupling, and Wilson-Dirac (WD) mass term are denoted by $m$, $\lambda_1$, and $\Delta$, respectively~\cite{schindler2018,Roy2019}. The terms associated with $\lambda_2$ generate the long-range hopping along $x$- and $y$-directions. The $4 \times 4$ $\mbf{\Gamma}$ matrices are chosen as $\Gamma_1=\sigma_1 s_3$, $\Gamma_2=\sigma_2 s_0$, $\Gamma_3=\sigma_3 s_0$, and $\Gamma_4=\sigma_1 s_1$; where the Pauli matrices $\vect{\sigma}$ and $\vect{s}$ act on the orbital and spin subspace, respectively. The WD mass term is important to obtain higher-order topology in the system. For $\lambda_2,t_z=0$, the Hamiltonian represent 2D SOTI hosting 0D corner modes for $m<\lvert t_x +t_y \rvert$. On the other hand, $\lambda_2\neq 0$ and $t_z=0$, also represent a 2D SOTI, however, with next-nearest-neighbor hopping~\cite{ghosh2024NHlongrange}. In this case, the SOTI model also represents a multi-SOTI that supports multiple modes per corner~\cite{ghosh2024NHlongrange}. At the same time, a non-zero $t_z$ allows coupling between two stacked SOTI layers. When $\Delta \neq 0$, the Hamiltonian $\mc{H}(\mbf{k})$ breaks TRS $\mc{T}=i\sigma_0 s_2 \mc{K}$: $\mc{T}\mc{H}(\mbf{k})\mc{T}^{-1} \neq \mc{H}(-\mbf{k})$, inversion symmetry $\mc{P}=\sigma_3 s_0$: $\mc{P}\mc{H}(\mbf{k})\mc{P}^{-1} \neq \mc{H}(-\mbf{k})$, and the four-fold rotational symmetry around the $z$-axis $C_4= e^{-\frac{i\pi}{4}\sigma_3 s_3}$: $C_4\mc{H}(\mbf{k}) C_4^{-1} \neq \mc{H}(-k_y,k_x,k_z)$, with $\mc{K}$ being the complex conjugation operator. However, $\mc{H}(\mbf{k})$ preserves the combined $PT$-symmetry $PT=\mc{P}\mc{T}=i\sigma_3 s_2 \mc{K}$: $ \left( PT \right) \mc{H}(\mbf{k}) \left( PT \right)^{-1} = \mc{H}(\mbf{k})$. Thus, the system supports four-fold degenerate $PT$-symmetric DNs~\cite{WangPRBL2023}. Additionally, $\mc{H}(\mbf{k})$ preserves the combined $C_4 \mc{T}$-symmetry: $(C_4\mc{T}) \mc{H}(\mbf{k}) (C_4 \mc{T})^{-1} = \mc{H}(k_y,-k_x,-k_z)$. Importantly, $\mc{H}(\mbf{k})$ also respects chiral symmetry $\mc{S}=\sigma_1 s_2$: $\mc{S}\mc{H}(\mbf{k})\mc{S}^{-1}=-\mc{H}(\mbf{k})$. Here, chiral symmetry plays a crucial role in defining the QWN, which we employ to topologically characterize the phase. Without loss of generality, we consider $\lambda_1=t_x=t_y=\Delta=\lambda_2=1$ and $t_z=2$ for the rest of the manuscript. Nevertheless, our analysis is not affected by this specific choice of parameters.

%---------------------------------------------------------------------
\subsection{Topological invariant and phase diagram}\label{DSM_Invariant}
%---------------------------------------------------------------------

Having introduced the model Hamiltonian, we now study the topological phase diagram of the mHODSM. The HODSM phase embodies the signature of second-order topology as it constitutes the stacking of 2D SOTI layers, possessing nontrivial quadrupole moment~\cite{WangPRBL2023}. However, for a mHODSM, the quadrupole moment vanishes due to the presence of multiple hinge states~\cite{Benalcazar2022Nxy}. Nevertheless, the Hamiltonian $\mathcal{H}(\vect{k})$ [Eq.~\eqref{Eq.mHODSM_Hamiltonian}] preserves chiral symmetry, and we utilize the chiral symmetry-based QWN to topologically characterize the mHODSM~\cite{Benalcazar2022Nxy}. Here, we briefly provide an outline of how to compute the QWN. 

In this work, we consider stacking of 2D SOTI layers in the $z$-direction [see Eq.~(\ref{Eq.mHODSM_Hamiltonian})]. To compute the QWN, we consider a 3D system that is finite but obeys periodic boundary conditions~(PBC) in $x$- and $y$-directions but infinite in the $z$-direction. We represent the Hamiltonian in the semi-infinite geometry by the symbol $\mc{H}(k_z)$. The Hamiltonian $\mc{H}(k_z)$ respects the chiral symmetry generated by $S$. Here, $S$ represents the chiral symmetry operator in the semi-infinite geometry and is constructed from $\mc{S}$ employing the same basis for $\mc{H}(k_z)$. We can represent $\mc{H}(k_z)$ in an anti-diagonal form by considering the basis $U_S$, that diagonalizes the operator $S$ as $U_S S U_S^\dagger={\rm diag}\left(1,1,\cdots,1,-1,-1,\cdots, -1\right)$. In this basis $\mc{H}$ reads
\begin{align}
    \tilde{\mc{H}}(k_z)= U_S \mc{H}(k_z) U_S^\dagger=
    \begin{pmatrix}
        0 &  h(k_z) \\
        h^\dagger(k_z) & 0 \\
    \end{pmatrix} \ .
\end{align}
The chiral symmetry divides the system into two sublattice degrees of freedom $A$ and $B$ corresponding to $+1$ and $-1$ eigenvalues of $S$, respectively. Here, $S=U^S_A-U^S_B$; with $U_A^S=\sum \ket{A}\bra{A}$ and $U_B^S=\sum \ket{B}\bra{B}$. Now, we perform a singular value decomposition of $h(k_z)$ such that $h(k_z)=U_A \Sigma U_B^\dagger$, where the matrix $\Sigma$ contains the singular values and $U_{A,B}$ represent a matrix containing singular vectors. We utilize $U_{A,B}$ and $U^S_{A,B}$ to define the QWN as~\cite{Benalcazar2022Nxy}
\begin{align}
    \mc{N}_{xy} = \frac{1}{2\pi i}\mathrm{Tr} \ln \left(\bar{Q}_A \bar{Q}_{B}^\dagger\right)  \ ,
    \label{Eq:Nxy}
\end{align}
where $\bar{Q}_{A,B}$ are sublattice quadrupole operator, defined as $\bar{Q}_{A,B}=U_{A,B}^\dagger U^S_{A,B} Q U^S_{A,B} U_{A,B}^\dagger$ with the quadrupole operator reads as $Q=\exp \left(-2\pi i xy/L_xL_y\right)$. Here, $L_x (L_y)$ represent system's dimension in $x(y)$-direction. We employ $\mc{N}_{xy}$ in Eq.~(\ref{Eq:Nxy}) to topologically characterize the mHODSM phase and later the mHONLSM phase as well.

%~~~~~~~~~~~~~~~~~~~~~~~~~~~~~~~~~~~~~~~~~~~~~~~~~~~~~~
%~~~~~~~~~~~~~~~~~~~~~~~~~~~~~~~~~~~~~~~~~~~~~~~~~~~~~~
\begin{figure}
    \centering
    \subfigure{\includegraphics[width=0.42\textwidth]{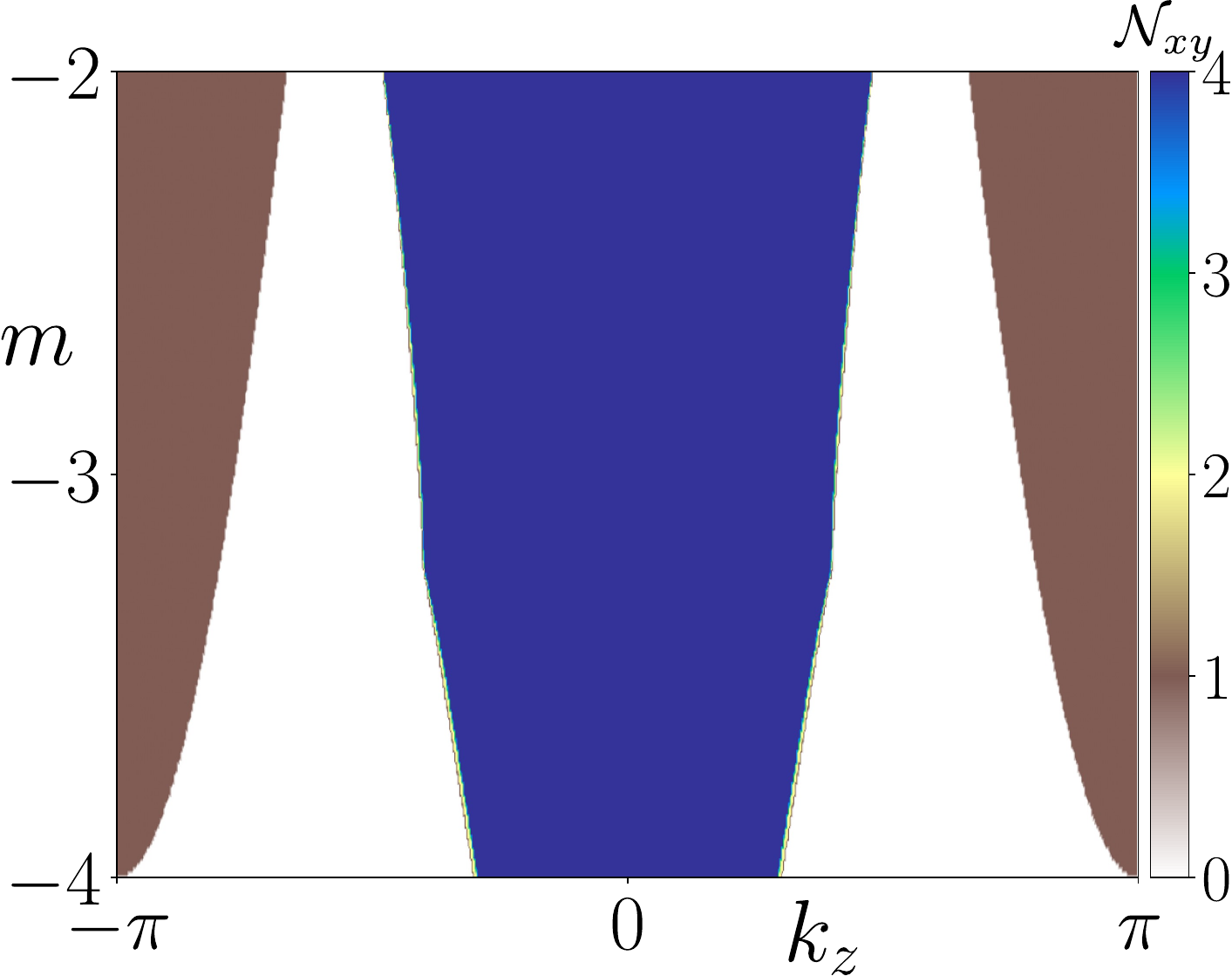}}
    \caption{Phase diagram of mHODSM in terms of the QWN $\N_{xy}$, is depicted in the $m-k_z$ plane. The colorbar represents the value of $\N_{xy}$. We consider $L_x=L_y=30$.}
    \label{Fig.Nxy_HODSM}
\end{figure}
%~~~~~~~~~~~~~~~~~~~~~~~~~~~~~~~~~~~~~~~~~~~~~~~~~~~~~~
%~~~~~~~~~~~~~~~~~~~~~~~~~~~~~~~~~~~~~~~~~~~~~~~~~~~~~~

Upon introducing $\mc{N}_{xy}$, we now focus on the phase diagram of mHODSM. In particular, we compute $\mc{N}_{xy}$ in the $k_z \mhyphen m$ plane and show the same in Fig.~\ref{Fig.Nxy_HODSM}. We observe that $\mc{N}_{xy}$ take values $1$ and $4$. Here, $\mc{N}_{xy}=4$ indicates the mHODSM with multiple hinge FAs~\cite{Qi2024HOmDSM}. Upon investigating the phase diagram Fig.~\ref{Fig.Nxy_HODSM}, we identify two types of mHODSM phases: at $m=-4$, we observe a phase with only $\mc{N}_{xy}=4$. At the same time, for any other values of $m$, $\mc{N}_{xy}$ takes values $1$ and $4$, which vary as a function of $k_z$. Nevertheless, to gain more insights into the different mHODSM phases, we focus on two different values of $m$: $m=-2$ and $-4$, and study the band structure and the features of the mHODSM in different geometries. We emphasize that the presence of chiral symmetry is necessary to characterize the topological phase with $\mc{N}_{xy}$ greater than one \cite{Benalcazar2022Nxy}.

%---------------------------------------------------------------------
\subsection{Band structures and signature of mHODSM}\label{DSM_BandStructure}
%---------------------------------------------------------------------

%~~~~~~~~~~~~~~~~~~~~~~~~~~~~~~~~~~~~~~~~~~~~~~~~~~~~~~~~
\begin{figure}
    \includegraphics[scale=0.6]{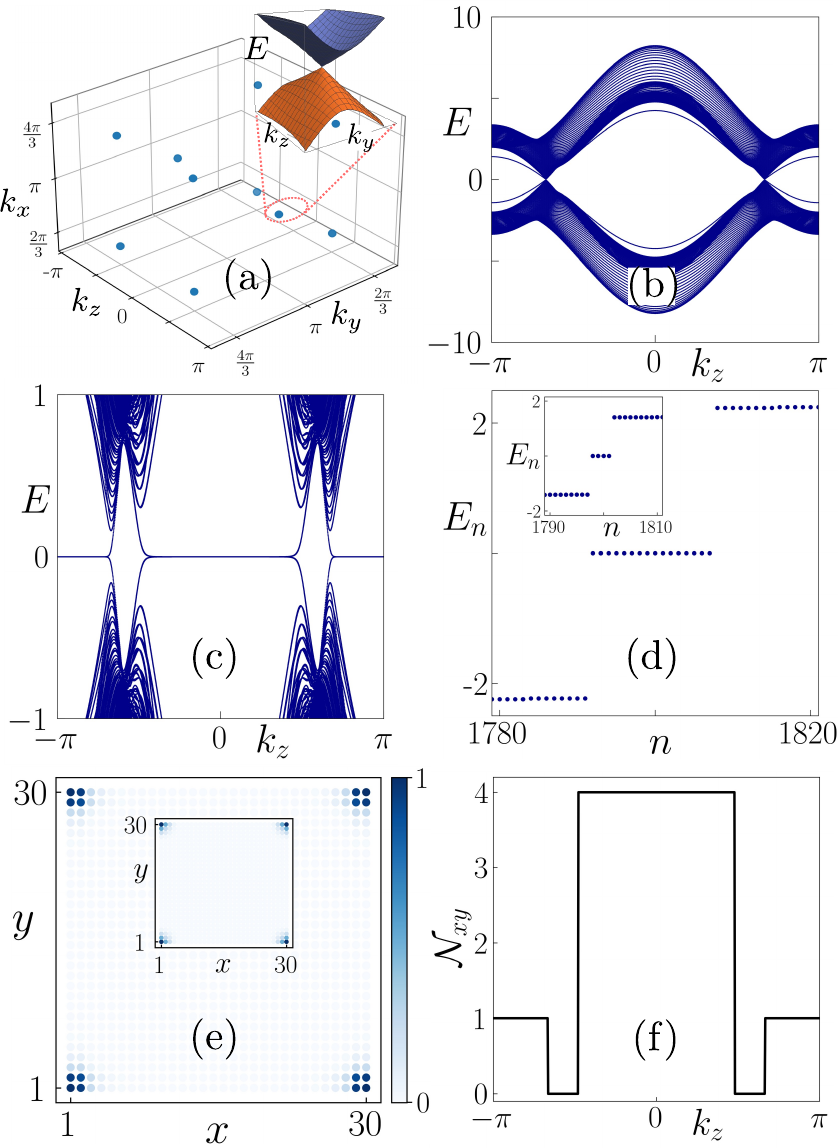}
    \caption{(a) Bulk Fermi surface of HODSM consisting of ten DNs at ($k_x,k_y,k_z) = \left(\pi \pm \frac{\pi}{3},\pi \pm \frac{\pi}{3},~\pm \frac{\pi}{2}\right),~\left(\pi,\pi,\pm \frac{2\pi}{3}\right)$ is shown in the $(k_x \mhyphen k_y \mhyphen k_z)$ space. In the inset, we illustrate the linear band dispersion in the $k_y\mhyphen k_z$ plane around the DN located at $(\pi,\pi,\frac{2\pi}{3})$.  (b) Band structure of the system in slab geometry as a function of $k_z$ with OBC along $x$-direction ($L_x=60$) and fixing $k_y=\pi$. (c) Band structure in rod geometry as a function of $k_z$ with OBCs along $x$- and $y$-direction. (d) Eigenvalue spectrum, $E_n$, is shown as a function of state index, $n$, corresponding to $k_z=0$ in (c). Inset represents $E_n$ vs $n$ for $k_z=\pi$ in (c). (e) LDOS for the zero-energy states in (d). (f) $\N_{xy}$ is plotted as a function of $k_z$. For (c)-(f), we choose $L_x=L_y=30$ and $m=-2$ for all the panels.}
    \label{Fig.HODSM_m_m2}
\end{figure}
%~~~~~~~~~~~~~~~~~~~~~~~~~~~~~~~~~~~~~~~~~~~~~~~~~~~~~~~~

%~~~~~~~~~~~~~~~~~~~~~~~~~~~~~~~~~~~~~~~~~~~~~~~~~~~~~~~~
\begin{figure}
    \includegraphics[scale=0.6]{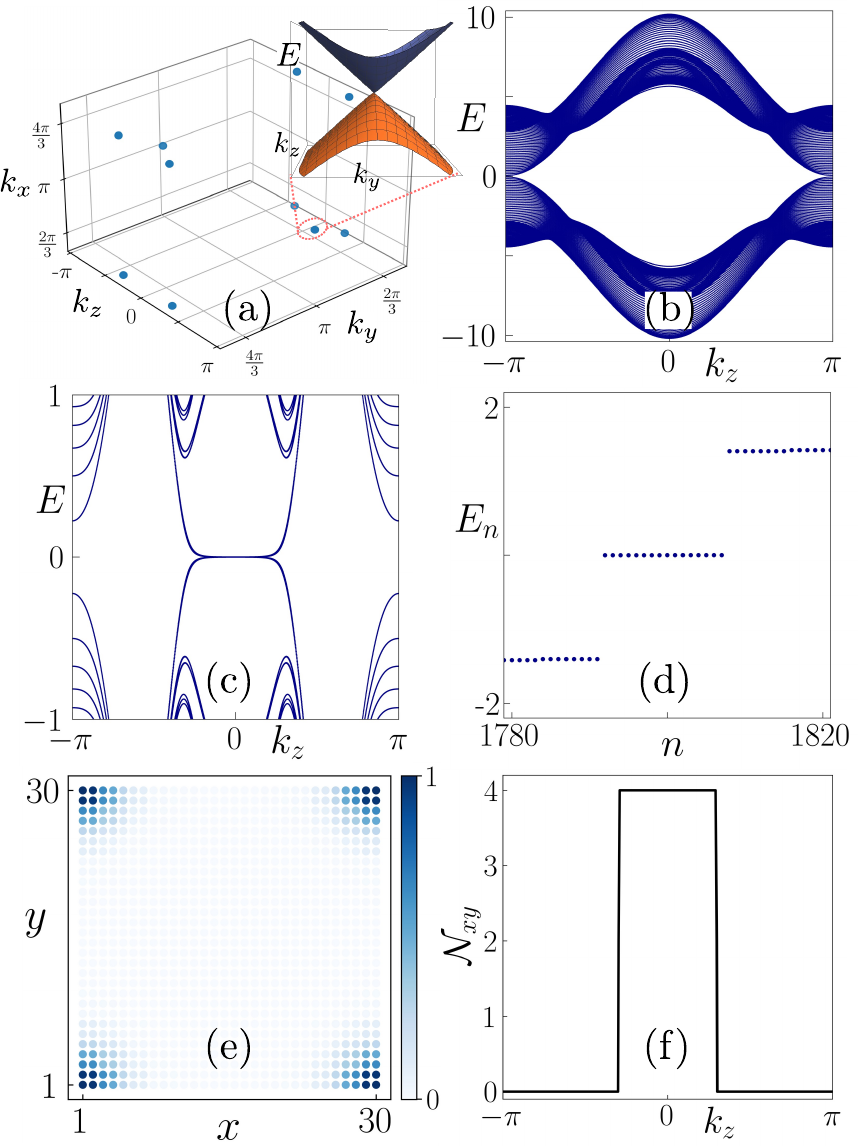}
    \caption{(a) Bulk Fermi surface of HODSM consisting of nine DNs at $(k_x,k_y,k_z)= (\pi, \pi, \pm \pi ),~(\pi \pm k_{x0}, \pi \pm k_{y0}, \pm k_{z0} )$ is shown in the $(k_x \mhyphen k_y \mhyphen k_z)$ space with $k_{x0}=k_{y0}\simeq1.318,~k_{z0}\simeq 1.047$ (The DNs located at $(\pi, \pi, \pi )$ and $(\pi, \pi, -\pi )$ are equivalent). In the inset, we show the Dirac cone around the DN $(\pi,\pi,\pi)$ in the $k_y \mhyphen k_z$ plane. (b) Band structure of the system in slab geometry as a function of $k_z$ with OBC along $x$ direction ($L_x=60$) and at $k_y=\pi$. (c) Band structure in rod geometry as a function of $k_z$ with OBC along $x$- and $y$-direction. (d) Eigenvalue spectrum, $E_n$ as a function of state index $n$ for $k_z=0$ in (c). (e) LDOS for the zero-energy states in (d). (f) $\N_{xy}$ is plotted as a function of $k_z$. For (c)-(f), we choose $L_x=L_y=30$ and $m=-4$ for all the panels.}
    \label{Fig.HODSM_m_m4}
\end{figure}
%~~~~~~~~~~~~~~~~~~~~~~~~~~~~~~~~~~~~~~~~~~~~~~~~~~~~~~~~

%.......................................
%\subsubsection{Case I: $m=-2$}
%.......................................
\textit{Case I- $m=-2$}: Here, we concentrate on a single line at $m=-2$ in the phase diagram Fig.~\ref{Fig.Nxy_HODSM} and demonstrate the results in Fig.~\ref{Fig.HODSM_m_m2}. In particular, in Fig.~\ref{Fig.HODSM_m_m2}(a), we illustrate the bulk Fermi surface i.e. the zero energy excitations of the Hamiltonian, $\mc{H}(\mbf{k})$, in Eq.~(\ref{Eq.mHODSM_Hamiltonian}). We observe that the valence and the conduction bands touch each other at ten isolated DNs in the BZ with the location of DNs being $(k_x,k_y,k_z) = \left(\pi \pm \frac{\pi}{3},\pi \pm \frac{\pi}{3}, \pm \frac{\pi}{2}\right),~\left(\pi,\pi,\pm \frac{2\pi}{3}\right)$. We also explicitly show the linear dispersion of the bulk around one of the DNs at $(\pi,\pi, \frac{2\pi}{3})$ in the inset of Fig.~\ref{Fig.HODSM_m_m2}(a) resembling the feature of Dirac cones. Next, we compute the band structure of the system in slab geometry, \ie we consider open boundary condition (OBC) along the $x$-direction and PBC along $y$- and $z$-directions to understand the surface states. Fixing $k_y=\pi$, we show the band structure in slab geometry in Fig.~\ref{Fig.HODSM_m_m2}(b). We see the existence of surface states connecting the DNs. We can see such surface states in a first-order DSM also~\cite{Randeria2016PNAS}. At the same time, a 3D HODSM possesses the features of SOTI. To probe the same, we consider rod geometry such that the system obeys OBC along $x$- and $y$-directions and PBC along $z$-direction such that the system is infinite in the stacking direction. We demonstrate the band structure as a function of $k_z$ employing rod geometry in Fig.~\ref{Fig.HODSM_m_m2}(c). From Fig.~\ref{Fig.HODSM_m_m2}(c), we see the existence of hinge FAs. However, we observe hinge FAs that are divided into two segments and separated by gaps. To understand the number of states in these two segments, we choose a slice from Fig.~\ref{Fig.HODSM_m_m2}(c) fixing $k_z=0$ and plot the eigenvalue spectrum $E_n$ as a function of the eigenvalue index $m$ in Fig.~\ref{Fig.HODSM_m_m2}(d). We observe sixteen states close to zero-energy, implying four FAs per hinge. In the inset of Fig.~\ref{Fig.HODSM_m_m2}(d), we show eigenvalue $E_n$ vs $n$ for $k_z=\pi$, and we notice four states at zero-energy corresponding to single FA per hinge. Afterward, we investigate the normalized local density of states~(LDOS) of these hinge FA eigenvalues at zero-energy. On this account, we show the LDOS of the zero-energy states as a function of the system's dimension in $x$- and $y$-directions, for $k_z=0$ in Fig.~\ref{Fig.HODSM_m_m2}(e). We can identify that the FAs are localized at the corner of the 2D cross-section plane perpendicular to the stacking direction ($z$-direction) of the system, implying the FAs are the hinge states. In the inset of Fig.~\ref{Fig.HODSM_m_m2}(e), we show the LDOS of the hinge FAs at $k_z=\pi$, which again suggests localization of the FAs at the hinges of the system. Having discussed the band structure of the mHODSM and boundary signatures of the FAs, we associate a topological invariant with them. In particular, we compute QWN $\mc{N}_{xy}$ defined in Eq.~(\ref{Eq:Nxy}) as a function of $k_z$ and depict the same in Fig.~\ref{Fig.HODSM_m_m2}(f). The QWN $\mc{N}_{xy}$ take the values $0$, $1$, and $4$ as a function of $k_z$, which apparently matches with the flat FAs that appear at zero-energy in Fig.~\ref{Fig.HODSM_m_m2}(c). Here, a non-zero $\mc{N}_{xy}$ signifies the existence of hinge FAs, with $\mc{N}_{xy}=1$ signifies one FA per hinge and $\mc{N}_{xy}=4$ implies the presence of four FAs per hinge. The existence of more than one FAs per hinge with $\N_{xy}\ge 1$ is a signature of the mHODSM~\cite{Qi2024HOmDSM}.

%.......................................
%\subsubsection{Case II: $m=-4$}
%.......................................

\textit{Case II- $m=-4$:} Here, we discuss another mHODSM phase that is accessible from the phase diagram Fig.~\ref{Fig.Nxy_HODSM}. We show the bulk Fermi surface consisting of nine DNs as shown in  Fig.~\ref{Fig.HODSM_m_m4}(a). Location of the DNs are $(k_x,k_y,k_z)= (\pi, \pi, \pi ),~(\pi \pm k_{x0}, \pi \pm k_{y0}, \pm k_{z0} )$ with $k_{x0}=k_{y0}\simeq1.318,~k_{z0}\simeq 1.047$ (The DNs located at $(\pi, \pi, \pi )$ and $(\pi, \pi, -\pi )$ are equivalent). We depict the linearly dispersing Dirac cone around the DN $(\pi, \pi, \pi )$ in the inset of Fig.~\ref{Fig.HODSM_m_m4}(a). The surface states are shown in Fig.~\ref{Fig.HODSM_m_m4}(b) as a function of $k_z$ with $k_y=\pi$. Moreover, we plot the rod geometry band structure in Fig.~\ref{Fig.HODSM_m_m4}(c) as a function of $k_z$ for a system obeying OBCs along $x$- and $y$-directions. In contrast to the previous case, we only observe a single set of degenerate hinge FAs here. To know the degeneracy of the FAs, we also plot the eigenvalue spectrum at $k_z=0$ in Fig.~\ref{Fig.HODSM_m_m4}(d). We notice sixteen modes at zero-energy, signifying four FAs per hinge. To verify the hinge localization of the FAs, we show the LDOS associated with the zero-energy states in Fig.~\ref{Fig.HODSM_m_m4}(e). We see that the FAs are localized at the hinges of the system. Moreover, we compute the $\mathcal{N}_{xy}$ as a function of $k_z$ in Fig.~\ref{Fig.HODSM_m_m4}(f). Here, $\mathcal{N}_{xy}$ takes a value of $4$ for the $k_z$ points, which supports the hinge FAs and thereby corroborates the second-order topological nature of the system.

In summary, in this section, we establish, by considering a stacking of SOTI Hamiltonians with long-range hopping in the $z$-direction that the system supports a mHODSM with multiple hinge FAs. We show two kinds of mHODSM in Figs.~\ref{Fig.HODSM_m_m2} (one and four hinge FAs) and \ref{Fig.HODSM_m_m4} (only four hinge FAs), with the multiple hinge FAs being topologically characterized employing the QWN.

%======================================================
\section{\lowercase{m}HONLSM}\label{Sec:III_HOMNLSM}
%======================================================
This section is devoted to the main results of this work. Here, we discuss the mHONLSM with multiple hinge FAs along with drumhead-like surface states. To this end, we first introduce the Hamiltonian that embodies the mHONLSM. The mHONLSM phase being a hybrid-order topological phase, we topologically characterize the phase employing both the DWN and the QWN. Next, we discuss the band structure in bulk, slab, and rod geometry. We also study the transport properties of the hinge FAs of the system using two-terminal conductance.

%---------------------------------------------------------------------
\subsection{Model Hamiltonian} \label{WSM_model}
%---------------------------------------------------------------------
In literature, there exist various proposals for the realization of the first-order NLSMs. Conventionally, the NLSMs are protected by some crystalline symmetries~\cite{Chiu2014_NLSM}, internal symmetries~\cite{Zhao2013,Matsuura_2013}, or even only chiral symmetry~\cite{Abdulla2023_NLSM1,Abdulla2023_NLSM2}. In our case, the model Hamiltonian that we employ in Eq.~\eqref{Eq.mHODSM_Hamiltonian} breaks the explicit TRS but rather respects the $PT$-symmetry and chiral symmetry. We obtain the mHONLSM phase by breaking the $PT$-symmetry explicitly while preserving the chiral symmetry. To break the $PT$-symmetry, we apply an external magnetic field along the negative $x$-direction, such that the total Hamiltonian reads
%~~~~~~~~~~~~~~~%
\begin{equation}
	\mc{H}'(\vect{k})=\mathcal{H}(\vect{k})-B_x \Lambda \ ,\label{Eq.HONLSM_Hamiltonian}
\end{equation}
%~~~~~~~~~~~~~~~%
 where, $\Lambda=\sigma_0 s_1$. Here, the Hamiltonian $\mc{H}'(\vect{k})$ breaks the $PT$-symmetry such that $ \left( PT \right) \mc{H}'(\mbf{k}) \left( PT \right)^{-1} \neq \mc{H}'(\mbf{k})$. Due to the broken $PT$-symmetry, the DNs of HODSM inflates into nodal loops in mHONLSM. Note that we only consider the coupling of the external magnetic field with spin degrees of freedom of the electrons~\cite{Vazifeh2013, Khanna2014,McCormick2017}. Furthermore, the external magnetic field also breaks the combined $C_4\mc{T}$ symmetry. However, the Hamiltonian $\mc{H}'(\vect{k})$ continues to satisfy the chiral symmetry $S$, which allows us to topologically characterize the system based on chiral symmetry. We consider the Hamiltonian $\mc{H}'(\vect{k})$ for the topological characterization and also to obtain the signature of the mHONLSM in different geometries.

%---------------------------------------------------------------------
\subsection{Topological invariant and phase diagram}\label{WSM_Invariant}
%---------------------------------------------------------------------
%~~~~~~~~~~~~~~~~~~~~~~~~~~~~~~~~~~~~~~~~~~~~~~~~~~~~~~~~
\begin{figure}
    \centering
    \includegraphics[scale=0.37]{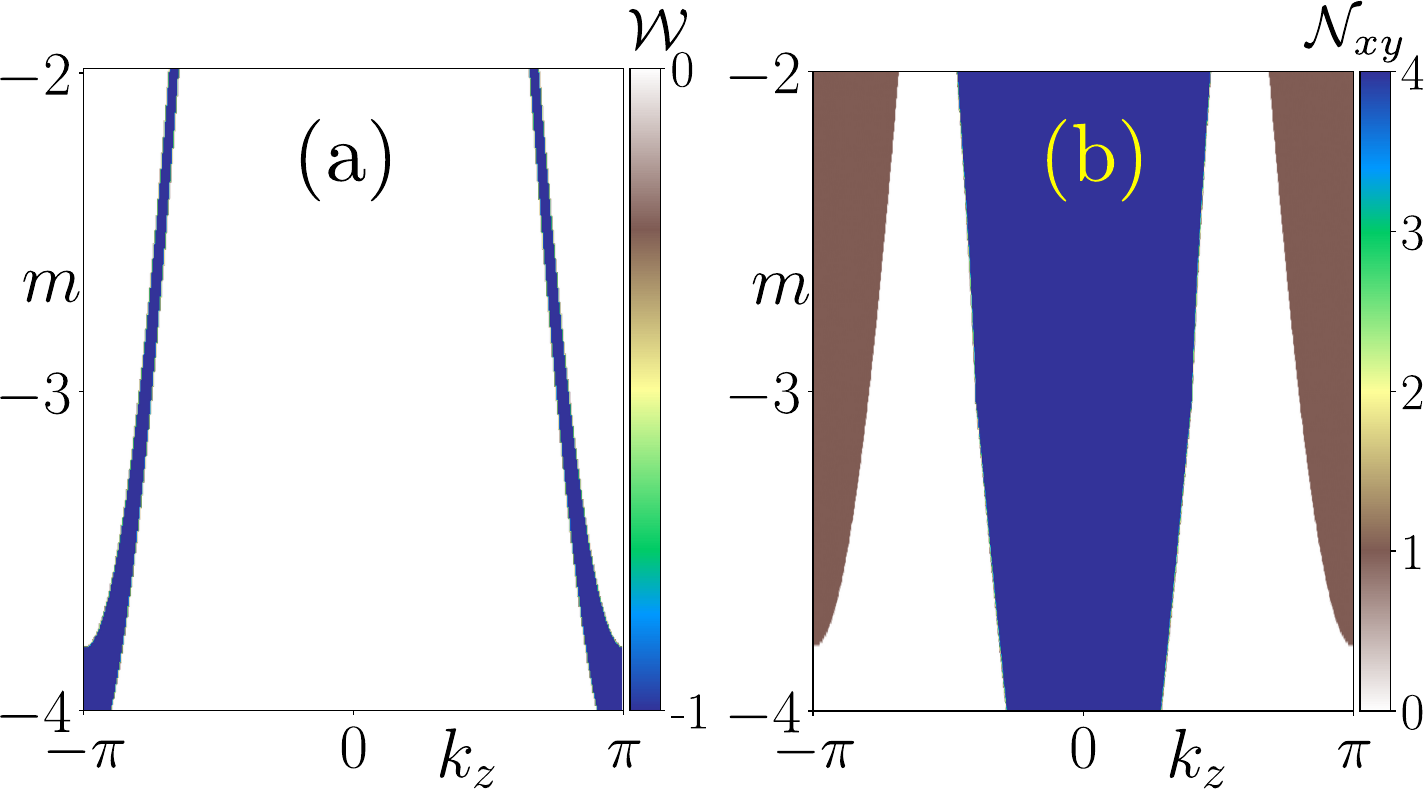}
    \caption{(a) The DWN $\mc{W}$ is shown in the $m \mhyphen k_z$ plane. We fix $k_y=\pi$ and consider $L_x=60$. (b) The QWN $\N_{xy}$ is depicted in the $m \mhyphen k_z$ plane. We consider $L_x=L_y=30$. Here, $B_x=0.2$ in both the panels.}
    \label{Fig.Nxy_HONLSM}
\end{figure}
%~~~~~~~~~~~~~~~~~~~~~~~~~~~~~~~~~~~~~~~~~~~~~~~~~~~~~~~~

%~~~~~~~~~~~~~~~~~~~~~~~~~~~~~~~~~~~~~~~~~~~~~~~~~~~~~~~~
\begin{figure*}
	\includegraphics[scale=0.6]{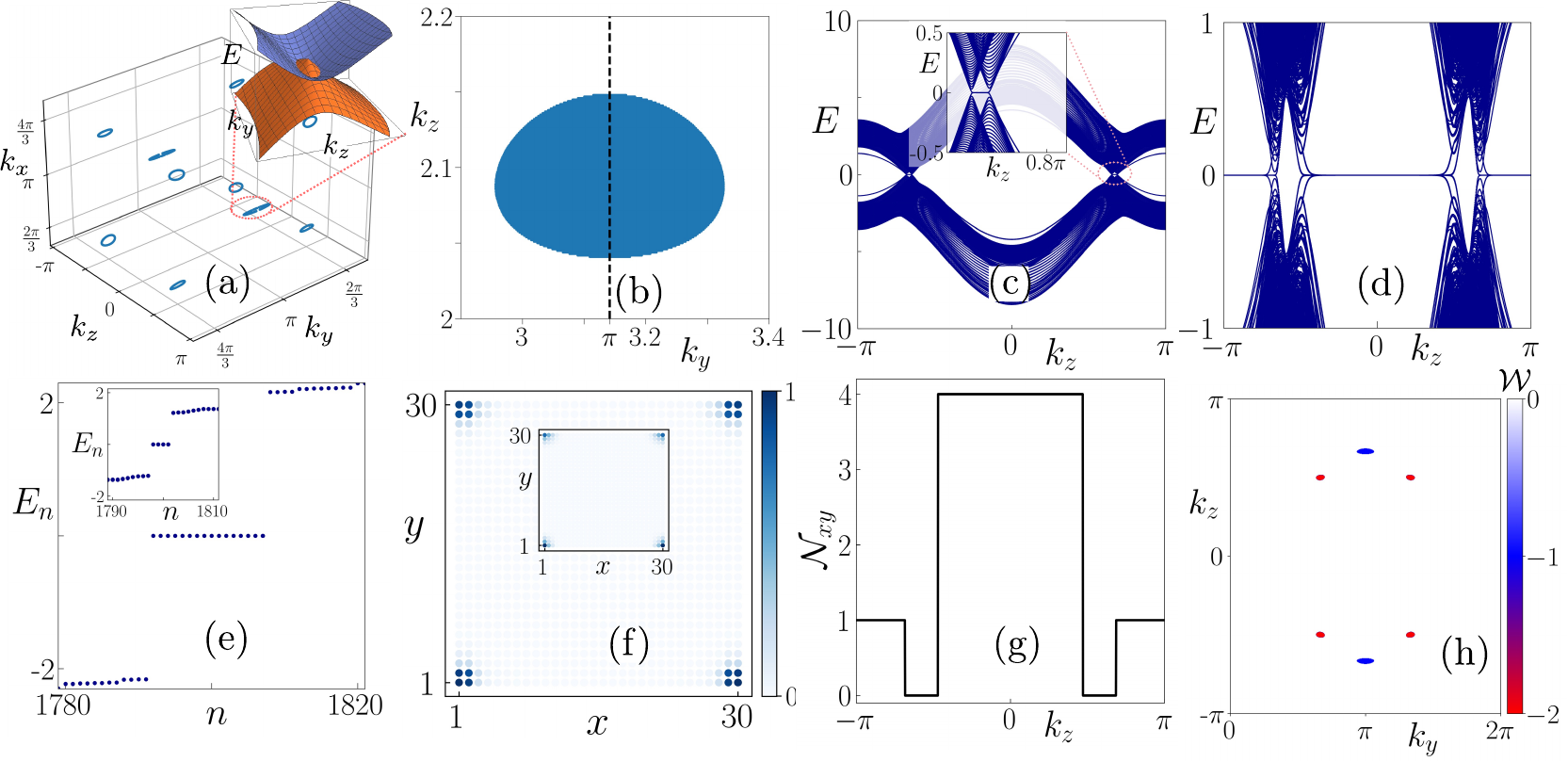}
	\caption{(a) Bulk Fermi surface of mHONLSM comprised of ten nodal loops is shown in the $(k_x \mhyphen k_y \mhyphen k_z)$ space. In the inset, we show the band structure for the nodal loop centered at $(\pi,\pi,2\pi/3)$ in the $k_y \mhyphen k_z$ plane fixing $k_x=\pi$. (b) We illustrate the drumhead-like surface states, i.e., the zero energy states corresponding to the nodal loop highlighted in (a) in the $k_y \mhyphen k_z$ plane employing slab geometry with finite system size along $x$-direction taking $L_x=60$. (c) Band structure in the slab geometry as a function of $k_z$ with OBC along $x$-direction ($L_x=60$) and fixing $k_y=\pi$, corresponding to the black dashed line in (b). (d) Band structure in the rod geometry as a function of $k_z$ with OBC along $x$- and $y$-direction. (e) Eigenvalue spectrum $E_n$ as a function of state index $n$ for $k_z=0$ in (d), while the inset corresponds to $k_z=\pi$. (f) The LDOS associated with the zero-energy states in (e) is shown as a function of the system's dimension. (g) The QWN $\N_{xy}$ is plotted as a function of $k_z$. (h) DWN $\W$, is shown in the ($k_y \mhyphen k_z$) plane assuming $x$-direction to be finite with $L_x=60$. For (d-g), we choose $L_x=L_y=30$. We fix $m=-2$ and $B_x=0.2$ for all the panels.}
	\label{Fig.HONLSM_m_m2}
\end{figure*}
%~~~~~~~~~~~~~~~~~~~~~~~~~~~~~~~~~~~~~~~~~~~~~~~~~~~~~~~~

Conventionally, topological protection of first-order NLSMs is demonstrated by associating a winding number~\cite{Abdulla2023_NLSM1,Abdulla2023_NLSM2} with each of the nodal loops. To point out the first-order topological nature of our system, we indulge in characterizing the phase using the DWN $\mc{W}$, which is used to characterize the chiral symmetric first-order topological phase~\cite{LinChiralWinding2021,MondalPRB2023}, which we outline below. Note that the DWN is utilized to characterize the first-order topology in a 1D system. In our case, we obtain an effective 1D system along the $x$-direction while treating values of $k_y$ and $k_z$ as the parameters. This allows us to compute the DWN as a function of $k_y, k_z$ and other parameters of the model Hamiltonian. We compute the DWN using the similar prescription mentioned for QWN (see Sec.~\ref{DSM_Invariant})~\cite{LinChiralWinding2021}. To this end, we consider the system to be finite in $x$-direction but obeys PBC. Now, employing the basis in which the chiral symmetry operator is diagonal, we can recast $\mc{H}'(k_y,k_z)$ as
\begin{align}
	\tilde{\mc{H}}'(k_y,k_z)=
	\begin{pmatrix}
		0 &  h'(k_y,k_z) \\
		h'^\dagger(k_y,k_z) & 0 \\
	\end{pmatrix} \ .
\end{align}
As discussed in the previous section, we perform a singular value decomposition of $h'(k_y,k_z)$: $h'(k_y,k_z)=U_A'\Sigma'U_B'^\dagger$ with $\Sigma'$ containing the singular values and $U'_{A,B}$ being the singular vectors. Now, we can define the DWN as~\cite{LinChiralWinding2021} 
\begin{equation}
	\mc{W} = \frac{1}{2\pi i} \mathrm{Tr\, ln} (\bar{P}_A\bar{P}_B^\dagger) \ ,
\end{equation}
where $\bar{P}_{A}$ and $\bar{P}_{B}$ are the sublattice dipole moment operator restricted to the sublattice $A$ and $B$, respectively, and defined as $\bar{P}_{A,B} = U'^\dagger_{A,B}U'^{S\dagger}_{A,B} P U'^S_{A,B} U'_{A,B}$ with the dipole moment operator   $P= \mathrm{exp}(-2\pi i\, x/L_x)$. Here, $L_x$ represents the system size along the $x$-direction. The matrices $U'^S_{A,B}$ are defined in the same way as done in Sec.~\ref{DSM_Invariant}. However, the dimensions of these matrices are now different (less) compared to the previous case (Sec.~\ref{DSM_Invariant}) as we consider an effective 1D system. In addition to the first-order topology, the mHONLSM also exhibits the second-order topology~\cite{WangPRBL2023}. Thus, to fully characterize the mHONLSM, we need to compute both the $\mc{W}$ and $\N_{xy}$.

%~~~~~~~~~~~~~~~~~~~~~~~~~~~~~~~~~~~~~~~~~~~~~~~~~~~~~~~~
\begin{figure*}
	\includegraphics[scale=0.6]{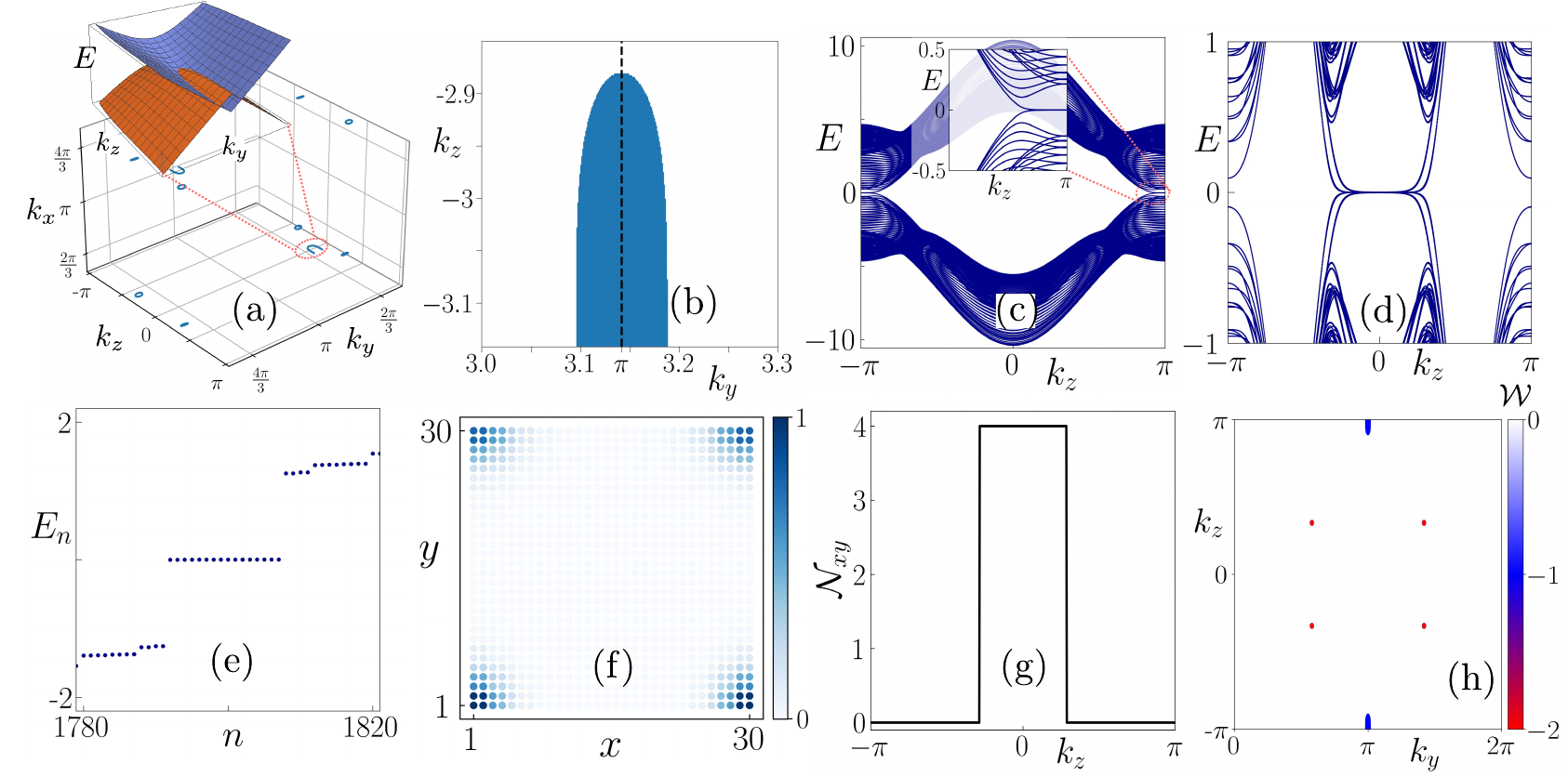}
	\caption{(a) Bulk Fermi surface of the mHONLSM comprising of nine nodal loops is shown in the $(k_x \mhyphen k_y \mhyphen k_z)$ space. In the inset, we demonstrate the band structure for nodal loops centered at $(\pi,\pi,\pi)$ in the $k_y-k_z$ plane fixing $k_x=\pi$. (b) Illustration of drumhead-like surface states in the $k_y \mhyphen k_z$ plane with finite system size along $x$-direction ($L_x=60$), corresponding to the nodal loop highlighted in (a). (c) Band structure in the slab geometry as a function of $k_z$ with OBC along $x$-direction ($L_x=60$) and fixing $k_y=\pi$, corresponding to the black dashed line in (b). (d) Band structure in the rod geometry as a function of $k_z$ with OBC along $x$- and $y$-direction. (e) Eigenvalue spectrum, $E_n$, is shown as a function of state index $n$, corresponding to $k_z=0$ in (d), while the inset corresponds to $k_z=\pi$. (f) The LDOS associated with the zero-energy states in (e) is shown as a function of the system's dimension. (g) The QWN $\N_{xy}$ is plotted as a function of $k_z$. (h) DWN $\W$, is shown in the $k_y \mhyphen k_z$ plane assuming $x$-direction to be finite with $L_x=60$. For (d-g), we choose $L_x=L_y=30$. We fix $m=-4$ and $B_x=0.2$ for all the panels.}
	\label{Fig.HONLSM_m_m4}
\end{figure*}
%~~~~~~~~~~~~~~~~~~~~~~~~~~~~~~~~~~~~~~~~~~~~~~~~~~~~~~~~

We compute $\W$ and $\N_{xy}$ and illustrate them in the $m$-$k_z$ plane in Fig.~\ref{Fig.Nxy_HONLSM}(a) and (b), respectively. In Fig.~\ref{Fig.Nxy_HONLSM}(a), we notice the appearance of a regime with $\W=-1$, which suggests the existence of first-order topology. Note that, in Fig.~\ref{Fig.Nxy_HONLSM}(a), we fix $k_y=\pi$, to show the behavior of $\W$ in the $m \mhyphen k_z$ plane. In the next subsection, choosing particular strengths of $m$, we fully explore the first-order topology of mHONLSM in the $k_y \mhyphen k_z$ plane. In addition to $\W$, $N_{xy}$ also acquires nonzero values of $1$ and $4$, as shown in Fig.~\ref{Fig.Nxy_HONLSM}(b). Interestingly, for each value of $m$, the non-zero values of $\W$ and $N_{xy}$ are separated in $k_z$. At this point, we also mention that if we apply the external magnetic field in the positive $x$-direction, we observe overlapping regions of $\W$ and $N_{xy}$ in the $m$-$k_z$ plane. Nonetheless, the coexistence of $\W$ and $N_{xy}$ suggest a mHONLSM phase with a hybrid-order topology. Furthermore, in Fig.~\ref{Fig.Nxy_HONLSM}(b), the $N_{xy}$ also takes values which is greater than one, implying a mHONLSM phase. Similar to the mHODSM, the mHONLSM phase is also divided into two parts- mHONLSM with $N_{xy}=1,4$ and $N_{xy}=4$ only. To shine further light on these two different kinds of mHONLSM, we choose two values of $m=-2$ and $-4$, and discuss the band structures in different geometries in the next subsection.

%---------------------------------------------------------------------
\subsection{Band structures and signature of mHONLSM}\label{NLSM_BandStructure}
%---------------------------------------------------------------------
\textit{Case I- $m=-2$}: Considering a fixed value of $m=-2$ from the phase diagram Fig.~\ref{Fig.Nxy_HONLSM}, we first investigate the bulk band structure of mHONLSM in Fig.~\ref{Fig.HONLSM_m_m2}(a). We observe each of the DNs in Fig.~\ref{Fig.HODSM_m_m2}(a) gets inflated into nodal loops totaling ten nodal loops in Fig.~\ref{Fig.HONLSM_m_m2}(a). Concentrating on the nodal loop centered around $\left(\pi,\pi,\frac{2\pi}{3}\right)$, we explicitly illustrate the nature of bulk bands in the $k_y \mhyphen k_z$ plane fixing $k_x=\pi$ in the inset of Fig.~\ref{Fig.HONLSM_m_m2}(a). This affirms that the bulk of the Hamiltonian (Eq.~\eqref{Eq.HONLSM_Hamiltonian}) indeed represent the signature of NLSM. Now, to see the boundary manifestation of the mHONLSM, we first investigate the drumhead-like surface states being the signature of NLSMs. On this account, we consider slab geometry by employing PBC along $y$- and $z$-directions and OBC along $x$-direction (with system size $L_x=60$). In Fig.~\ref{Fig.HONLSM_m_m2}(b), we plot surface states at the zero energy in the $k_y-k_z$ plane associated with the highlighted nodal loop in Fig.~\ref{Fig.HONLSM_m_m2}(a) exhibiting `Fermi drumhead'. Thus, from Fig.~\ref{Fig.HONLSM_m_m2}(b), we can infer that the mHONLSM phase manifests drumhead-like surface states. Furthermore, we focus on a particular value of $k_y=\pi$ (shown by a dashed line in Fig.~\ref{Fig.HONLSM_m_m2}(b)), and show the band structure of the system in slab geometry in Fig.~\ref{Fig.HONLSM_m_m2}(c) and observe the flat dispersionless surface states at zero energy. Moreover, the mHONLSM phase also possesses the characteristics of second-order topology, manifesting hinge FAs. To resolve this, we consider rod geometry by employing OBC along $x$- and $y$-directions (with system size $L_x=L_y=30$) and PBC along $z$-direction. We demonstrate the band structure in rod geometry in Fig.~\ref{Fig.HONLSM_m_m2}(d). We observe the hinge FAs as the states at $E=0$, and the FAs are divided into two segments. We note that the surface FAs are gapped out in rod geometry due to finite size effect as already established in literature~\cite{Kaladzhyan2019,Igarashi2017,Lin2017,Wang2018a,Pal2024}. Nevertheless, to understand the number of hinge FAs in each segment, we show the eigenvalue spectrum $E_n$ as a function of the eigenvalue index $n$ in rod geometry for fixed values of $k_z=0$ in Fig.~\ref{Fig.HONLSM_m_m2}(e) and $k_z=\pi$ in the inset. We notice sixteen eigenvalues at $E_n=0$ \ie four FAs per hinge and four eigenvalues at $E_n=0$ \ie one FA per hinge in Fig.~\ref{Fig.HONLSM_m_m2}(e) and its inset, respectively. To further corroborate the location of the FAs, we also compute the LDOS associated with these FAs appearing at zero-energy and show them in Fig.~\ref{Fig.HONLSM_m_m2}(f). From Fig.~\ref{Fig.HONLSM_m_m2}(f) and its inset, we can identify that the FAs are localized at the hinges of the system. Having investigated the emergence of the mHONLSM phase from the spectral analysis, we now study the corresponding topological invariants as discussed before. In particular, we show the $\N_{xy}$ as a function of $k_z$ in Fig.~\ref{Fig.HONLSM_m_m2}(g). Here, $\N_{xy}$ shows a jump between $0$, $1$, and $4$ and follows a similar structure as the hinge FAs shown in Fig.~\ref{Fig.HONLSM_m_m2}(d). Nevertheless, the $\N_{xy}$ values greater than one signify a mHONLSM phase. Furthermore, to fully explore the first-order topology of mHONLSM, we also plot the DWN $\W$ in the $k_y \mhyphen k_z$ plane in Fig.~\ref{Fig.HONLSM_m_m2}(h). We observe the DWN $\W$ takes values $-1$ and $-2$. Interestingly, the regions with nonzero values of $\W$ indicate the presence of topological nodal loops, and $\W$ is nonzero only inside the nodal loops. Here, $\W=-1$ and $\W=-2$ signify the presence of single and double nodal loops, respectively, when projected onto the $k_y \mhyphen k_z$ plane.

\textit{Case II- $m=-4$}:
Similar to the previous case, we show another mHONLSM phase by fixing another value of $m=-4$. We illustrate the bulk Fermi surface structure in Fig.~\ref{Fig.HONLSM_m_m4}(a) featuring nine nodal loops at zero located at zero energy. We also plot the bulk band structure corresponding to the nodal loop centered at $(\pi,\pi,\pi)$. The drum head surface states at $E=0$ is shown in Fig.~\ref{Fig.HONLSM_m_m4}(b) in the $k_y \mhyphen k_z$ plane employing the slab geometry. In Fig.~\ref{Fig.HONLSM_m_m4}(c), we illustrate the surface states as a function of $k_z$ fixing $k_y=\pi$ (corresponding to the dashed black line in Fig.~\ref{Fig.HONLSM_m_m4}(b)). In Fig.~\ref{Fig.HONLSM_m_m4}(d), we demonstrate the rod geometry band structure featuring the hinge FAs. Here, we observe only one type of hinge FAs in contrast to the previous case (see Fig.~\ref{Fig.HONLSM_m_m2}(d)). For a fixed value of $k_z=0$, we show the eigenvalue spectrum $E_n$ as a function of the state index $n$ in Fig.~\ref{Fig.HONLSM_m_m4}(e). We see sixteen zero-energy modes corresponding to four states per hinge. The LDOS associated with the zero-energy states are shown in Fig.~\ref{Fig.HONLSM_m_m4}(f), corroborating the hinge localization of the zero-energy states. In Fig.~\ref{Fig.HONLSM_m_m4}(g), we exhibit the QWN $\N_{xy}$ as a function of $k_z$ where $\N_{xy}$ takes the values $0$ and $4$. The higher values of $\N_{xy}(\geq 1)$ signifies a mHONLSM phase. In Fig.~\ref{Fig.HONLSM_m_m4}(h), $\W$ is shown in the $k_y \mhyphen k_z$ plane. Similar to the $m=-2$ case, here $\W$ takes two non-zero values \ie $-1$ and $-2$, signifying the presence of single and double nodal loops, respectively, projected onto the $k_y \mhyphen k_z$ plane. The non-zero DWN $\W$ is the manifestation of first-order topology in the system. 

Note that, in the rod geometry band structure [Figs.~\ref{Fig.HONLSM_m_m2}(d) and \ref{Fig.HONLSM_m_m4}(d)], the first-order surface FA states are gapped out due to the quantum confinement effect, whereas the second-order hinge localized states are gapless and present at the zero energy states. Therefore, utilizing the band structure in slab and rod geometry, we can distinguish between the first- and second-order FAs. On the other hand, the hinge FAs that we obtain are dispersionless \,  i.e., flat in nature, and thus do not carry any chirality~\cite{GhorashiPRL2020}.

%~

%======================================================
\subsection{Two terminal transport}\label{SubSec:Transport}
\begin{figure}
	\includegraphics[scale=0.32]{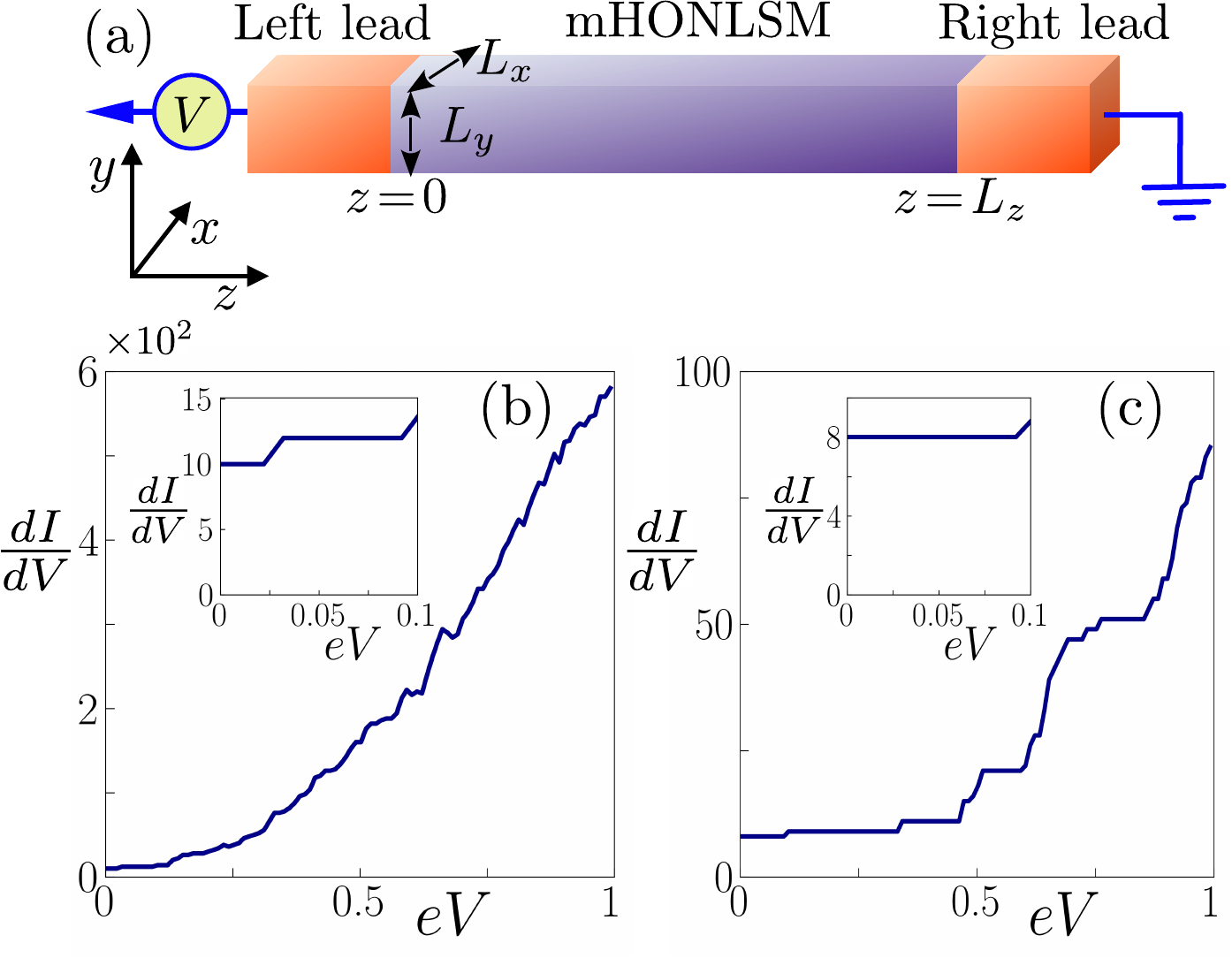}
	\caption{(a) The two terminal transport setup consisting of the mHONLSM as the device (middle blue region) and two mHONLSM leads. (b,c) Differential conductance $\frac{dI}{dV}$ (in units of $e^2/h$) as a function of the bias voltage $eV$ for $m=-2$ (corresponding to Fig.~\ref{Fig.HONLSM_m_m2}) and $m=-4$ (corresponding to Fig.~\ref{Fig.HONLSM_m_m4}), respectively. The device region has the dimensions $L_x=L_y=30$, and $L_z=80$. The inset shows $\frac{dI}{dV}$ close to $eV=0$. The rest of the parameters take the same value as Fig.~\ref{Fig.HONLSM_m_m2}.}
	\label{Fig.Transport}
\end{figure}

%======================================================
The significance of mHONLSM over NLSM and WSM becomes most evident when studying the transport properties in a nanowire (NW) geometry. The FA surface states of a NLSM and WSM are gapped out in an NW due to the quantum confinement in a finite system~\cite{Kaladzhyan2019,Gorbar2016,Igarashi2017,Pal2024,Wang2018a,Chen2020,Lin2017}. As a result, no states are available to conduct current if Fermi energy lies in the middle of the band. In contrast, in a mHONLSM, the existence of gapless hinge modes in the NW geometry leads to a nonzero density of states at Fermi energy, thereby contributing to the transport signature. The differential conductance near the Fermi energy is solely given by the number of hinge states that are present in the system. To this end, we consider a two-terminal transport setup as shown in Fig.~\ref{Fig.Transport}(a), comprising of the mHONLSM as the device (blue; from $z=0$ to $z=L_z$) and two leads (orange), which are also considered to be mHONLSM, so that there is no mismatch of bands of the leads and device. We employ the Landauer formalism to compute the differential conductance $\frac{dI}{dV}$ as~\cite{datta1997electronic}
\begin{align}
    \frac{dI}{dV}=\frac{e^2}{h} \mathcal{T}_{\rm LR} (V_{\rm L}-V_{\rm R}) \ ,
\end{align}
where $\frac{e^2}{h}$ defines the unit of quantum conductance and $\mathcal{T}_{\rm LR} (V_{\rm L}-V_{\rm R})$ represents the transmission probability from left to the right lead at a voltage bias $(V_{\rm L}-V_{\rm R})$. The right lead in our setup is grounded ($V_{\rm R}=0$), and we apply a voltage bias in the left lead ($V_{\rm L}=V$). We compute the differential conductance as a function of the voltage difference between the left and the right lead $eV=(V_{\rm L}-V_{\rm R})$. We employ the python package KWANT to calculate the transmission probability $\mathcal{T}_{\rm LR}$~\cite{Groth_2014}.

We show the differential conductance $\frac{dI}{dV}$ (in units of $e^2/h$) as a function of the bias voltage difference $eV$ for the two kinds of mHONLSM shown in Figs.~\ref{Fig.HONLSM_m_m2} and \ref{Fig.HONLSM_m_m4} in Fig.~\ref{Fig.Transport}(b) and (c), respectively. In particular, we first focus on Fig.~\ref{Fig.Transport}(b) (corresponding to Fig.~\ref{Fig.HONLSM_m_m2}). In the inset, we show the $\frac{dI}{dV}$ close to Fermi energy. At zero bias $eV=0$, we observe that the differential conductance takes a quantized value of $10$. To understand this, we investigate Fig.~\ref{Fig.HONLSM_m_m2}(d,e), which exhibits two kinds of hinge states: four and one states per hinge, totaling twenty conducting channels available from all the hinges. However, in Fig.~\ref{Fig.Transport}(b), we only show the conductance for bias voltage $eV \geq 0$. Thus, we only obtain the contribution from half of the number of available channels, which justifies the quantized differential conductance at $eV=0$ to be $10$. Moreover, in Fig.~\ref{Fig.Transport}(c), we study the transport properties of the hinge FAs of the mHONLSM phase shown in Fig.~\ref{Fig.HONLSM_m_m4}. From the inset, we observe a quantized differential conductance of $8$ at zero bias ($eV=0$), which corresponds to a total of sixteen hinge FA channels available in this phase (see Fig.~\ref{Fig.HONLSM_m_m4}(d,e)). Nevertheless, from Fig.~\ref{Fig.Transport}(b) and (c), we establish that, in the NW geometry, we obtain the contribution from the hinge FAs to the zero bias differential conductance. Moreover, the differential conductance of the mHODSM in the NW geometry will behave similarly to the mHONLSM phase as both phases host an equal number of hinge localized modes. Since the first-order Fermi surfaces are gapped out, they will not contribute to the zero-bias conductance in the transport measurement.

In summary, in this section, we show the emergence of the mHONLSM phase by breaking the $PT$-symmetry in mHODSM. We show two kinds of mHONLSM phase in Figs.~\ref{Fig.HONLSM_m_m2} (one and four FAs per hinge) and \ref{Fig.HONLSM_m_m4} (only four FAs per hinge). We compute the QWN to characterize the multiple hinge FAs and the DWN to characterize the drumhead-like Fermi surface. By computing the two terminal conductance, we also show the hinge FAs mediated differential conductance. On the other hand, the presence of multiple hinge FAs is beneficial for obtaining a high value of conductance at Fermi energy.

%======================================================
\section{Summary and Outlook}\label{Sec:IV_conclusion}
%======================================================

To summarize, in this work, we consider a tight binding model as the stacking of 2D SOTIs with long-range hopping to obtain an mHODSM phase, which exhibits multiple FAs per hinge. We employ the QWN $\N_{xy}$ to obtain the phase diagram. Furthermore, we focus on two specific parameters exhibiting two different kinds of mHODSM and plot the DNs in the $k_x \mhyphen k_y \mhyphen k_z$ space and also illustrate corresponding band structures in different geometries. Afterward, we break the $PT$-symmetry by applying a magnetic field to obtain the mHONLSM. To probe the hybrid order topology in the mHONLSM phase, we compute both the QWN $\N_{xy}$ and the DWN $\W$ and draw the phase diagram. Akin to the mHODSM phase, we also find two kinds of mHONLSM phase: one with $\N_{xy}=1,4$ and the other with $\N_{xy}=4$ only. We show the nodal loops in the $k_x \mhyphen k_y \mhyphen k_z$ space and study drumhead-like surface states and hinge FAs employing slab and rod geometry, respectively. We also demonstrate that the DWN is finite inside the nodal loops and zero otherwise. We further employ a two-terminal transport setup to obtain the hinge FA-mediated signature as a quantized conductance at Fermi energy.

In this work, long-range hopping plays a crucial role in obtaining the mHODSM and mHONLSM phases. Thus, we can also ask if we can obtain the mHODSM and mHONLSM phases without needing long-range hopping. Driven systems can be a suitable platform to engineer this. Since we can generate long-range hopping in a driven system~\cite{MikamiBW2016}, it creates a possibility to obtain the mHODSM and mHONLSM phases in a driven system without starting with a system with long-range hopping. Moreover, in a driven system, we can also have the possibility to break the $PT$-symmetry without needing the magnetic field~\cite{WangPRBL2023}. 

%=====================================================
\subsection*{Data Availability}
%======================================================
The data that support the findings of this article are openly available at Ref.~\cite{pal2024multi}.

%=====================================================
\subsection*{Acknowledgments}
%======================================================
We are grateful for the fruitful discussions with Arijit Saha and Tanay Nag. A.P. also acknowledges Faruk Abdullah, Krishanu Roychowdhury, and Soujanya Dutta for useful discussions. A.P. acknowledges SAMKHYA: High-Performance Computing facility provided by the Institute of Physics, Bhubaneswar, for numerical computations.

\bibliographystyle{apsrev4-2mod}
\bibliography{bibfile.bib}

%apsrev4-2.bst 2015-08-30 from 4.21a (PWD, AO, DPC/HNN) hacked
\begin{thebibliography}{93}%
\makeatletter
\providecommand \@ifxundefined [1]{%
 \@ifx{#1\undefined}
}%
\providecommand \@ifnum [1]{%
 \ifnum #1\expandafter \@firstoftwo
 \else \expandafter \@secondoftwo
 \fi
}%
\providecommand \@ifx [1]{%
 \ifx #1\expandafter \@firstoftwo
 \else \expandafter \@secondoftwo
 \fi
}%
\providecommand \natexlab [1]{#1}%
\providecommand \enquote  [1]{``#1''}%
\providecommand \bibnamefont  [1]{#1}%
\providecommand \bibfnamefont [1]{#1}%
\providecommand \citenamefont [1]{#1}%
\providecommand \href@noop [0]{\@secondoftwo}%
\providecommand \href [0]{\begingroup \@sanitize@url \@href}%
\providecommand \@href[1]{\@@startlink{#1}\@@href}%
\providecommand \@@href[1]{\endgroup#1\@@endlink}%
\providecommand \@sanitize@url [0]{\catcode `\\12\catcode `\$12\catcode
  `\&12\catcode `\#12\catcode `\^12\catcode `\_12\catcode `\%12\relax}%
\providecommand \@@startlink[1]{}%
\providecommand \@@endlink[0]{}%
\providecommand \url  [0]{\begingroup\@sanitize@url \@url }%
\providecommand \@url [1]{\endgroup\@href {#1}{\urlprefix }}%
\providecommand \urlprefix  [0]{URL }%
\providecommand \Eprint [0]{\href }%
\providecommand \doibase [0]{http://dx.doi.org/}%
\providecommand \selectlanguage [0]{\@gobble}%
\providecommand \bibinfo  [0]{\@secondoftwo}%
\providecommand \bibfield  [0]{\@secondoftwo}%
\providecommand \translation [1]{[#1]}%
\providecommand \BibitemOpen [0]{}%
\providecommand \bibitemStop [0]{}%
\providecommand \bibitemNoStop [0]{.\EOS\space}%
\providecommand \EOS [0]{\spacefactor3000\relax}%
\providecommand \BibitemShut  [1]{\csname bibitem#1\endcsname}%
\let\auto@bib@innerbib\@empty
%</preamble>
\bibitem [{\citenamefont {Wan}\ \emph {et~al.}(2011)\citenamefont {Wan},
  \citenamefont {Turner}, \citenamefont {Vishwanath},\ and\ \citenamefont
  {Savrasov}}]{Wan2011}%
  \BibitemOpen
  \bibfield  {author} {\bibinfo {author} {\bibfnamefont {X.}~\bibnamefont
  {Wan}}, \bibinfo {author} {\bibfnamefont {A.~M.}\ \bibnamefont {Turner}},
  \bibinfo {author} {\bibfnamefont {A.}~\bibnamefont {Vishwanath}}, \ and\
  \bibinfo {author} {\bibfnamefont {S.~Y.}\ \bibnamefont {Savrasov}},\
  }\bibfield  {title} {\emph {\enquote {\bibinfo {title} {Topological semimetal
  and Fermi-arc surface states in the electronic structure of pyrochlore
  iridates},}\ }}\href {\doibase 10.1103/PhysRevB.83.205101} {\bibfield
  {journal} {\bibinfo  {journal} {Phys. Rev. B}\ }\textbf {\bibinfo {volume}
  {83}},\ \bibinfo {pages} {205101} (\bibinfo {year} {2011})}\BibitemShut
  {NoStop}%
\bibitem [{\citenamefont {Burkov}\ and\ \citenamefont
  {Balents}(2011)}]{Burkov2011a}%
  \BibitemOpen
  \bibfield  {author} {\bibinfo {author} {\bibfnamefont {A.~A.}\ \bibnamefont
  {Burkov}}\ and\ \bibinfo {author} {\bibfnamefont {L.}~\bibnamefont
  {Balents}},\ }\bibfield  {title} {\emph {\enquote {\bibinfo {title} {Weyl
  Semimetal in a Topological Insulator Multilayer},}\ }}\href {\doibase
  10.1103/PhysRevLett.107.127205} {\bibfield  {journal} {\bibinfo  {journal}
  {Phys. Rev. Lett.}\ }\textbf {\bibinfo {volume} {107}},\ \bibinfo {pages}
  {127205} (\bibinfo {year} {2011})}\BibitemShut {NoStop}%
\bibitem [{\citenamefont {Burkov}\ \emph {et~al.}(2011)\citenamefont {Burkov},
  \citenamefont {Hook},\ and\ \citenamefont {Balents}}]{Burkov2011b}%
  \BibitemOpen
  \bibfield  {author} {\bibinfo {author} {\bibfnamefont {A.~A.}\ \bibnamefont
  {Burkov}}, \bibinfo {author} {\bibfnamefont {M.~D.}\ \bibnamefont {Hook}}, \
  and\ \bibinfo {author} {\bibfnamefont {L.}~\bibnamefont {Balents}},\
  }\bibfield  {title} {\emph {\enquote {\bibinfo {title} {Topological nodal
  semimetals},}\ }}\href {\doibase 10.1103/PhysRevB.84.235126} {\bibfield
  {journal} {\bibinfo  {journal} {Phys. Rev. B}\ }\textbf {\bibinfo {volume}
  {84}},\ \bibinfo {pages} {235126} (\bibinfo {year} {2011})}\BibitemShut
  {NoStop}%
\bibitem [{\citenamefont {Young}\ \emph {et~al.}(2012)\citenamefont {Young},
  \citenamefont {Zaheer}, \citenamefont {Teo}, \citenamefont {Kane},
  \citenamefont {Mele},\ and\ \citenamefont {Rappe}}]{Young2012}%
  \BibitemOpen
  \bibfield  {author} {\bibinfo {author} {\bibfnamefont {S.~M.}\ \bibnamefont
  {Young}}, \bibinfo {author} {\bibfnamefont {S.}~\bibnamefont {Zaheer}},
  \bibinfo {author} {\bibfnamefont {J.~C.~Y.}\ \bibnamefont {Teo}}, \bibinfo
  {author} {\bibfnamefont {C.~L.}\ \bibnamefont {Kane}}, \bibinfo {author}
  {\bibfnamefont {E.~J.}\ \bibnamefont {Mele}}, \ and\ \bibinfo {author}
  {\bibfnamefont {A.~M.}\ \bibnamefont {Rappe}},\ }\bibfield  {title} {\emph
  {\enquote {\bibinfo {title} {Dirac Semimetal in Three Dimensions},}\ }}\href
  {\doibase 10.1103/PhysRevLett.108.140405} {\bibfield  {journal} {\bibinfo
  {journal} {Phys. Rev. Lett.}\ }\textbf {\bibinfo {volume} {108}},\ \bibinfo
  {pages} {140405} (\bibinfo {year} {2012})}\BibitemShut {NoStop}%
\bibitem [{\citenamefont {Zyuzin}\ \emph {et~al.}(2012)\citenamefont {Zyuzin},
  \citenamefont {Wu},\ and\ \citenamefont {Burkov}}]{Zyuzin2012a}%
  \BibitemOpen
  \bibfield  {author} {\bibinfo {author} {\bibfnamefont {A.~A.}\ \bibnamefont
  {Zyuzin}}, \bibinfo {author} {\bibfnamefont {S.}~\bibnamefont {Wu}}, \ and\
  \bibinfo {author} {\bibfnamefont {A.~A.}\ \bibnamefont {Burkov}},\ }\bibfield
   {title} {\emph {\enquote {\bibinfo {title} {Weyl semimetal with broken time
  reversal and inversion symmetries},}\ }}\href {\doibase
  10.1103/PhysRevB.85.165110} {\bibfield  {journal} {\bibinfo  {journal} {Phys.
  Rev. B}\ }\textbf {\bibinfo {volume} {85}},\ \bibinfo {pages} {165110}
  (\bibinfo {year} {2012})}\BibitemShut {NoStop}%
\bibitem [{\citenamefont {Zyuzin}\ and\ \citenamefont
  {Burkov}(2012)}]{Zyuzin2012b}%
  \BibitemOpen
  \bibfield  {author} {\bibinfo {author} {\bibfnamefont {A.~A.}\ \bibnamefont
  {Zyuzin}}\ and\ \bibinfo {author} {\bibfnamefont {A.~A.}\ \bibnamefont
  {Burkov}},\ }\bibfield  {title} {\emph {\enquote {\bibinfo {title}
  {Topological response in Weyl semimetals and the chiral anomaly},}\ }}\href
  {\doibase 10.1103/PhysRevB.86.115133} {\bibfield  {journal} {\bibinfo
  {journal} {Phys. Rev. B}\ }\textbf {\bibinfo {volume} {86}},\ \bibinfo
  {pages} {115133} (\bibinfo {year} {2012})}\BibitemShut {NoStop}%
\bibitem [{\citenamefont {Hal\'asz}\ and\ \citenamefont
  {Balents}(2012)}]{Hal2012}%
  \BibitemOpen
  \bibfield  {author} {\bibinfo {author} {\bibfnamefont {G.~B.}\ \bibnamefont
  {Hal\'asz}}\ and\ \bibinfo {author} {\bibfnamefont {L.}~\bibnamefont
  {Balents}},\ }\bibfield  {title} {\emph {\enquote {\bibinfo {title}
  {Time-reversal invariant realization of the Weyl semimetal phase},}\ }}\href
  {\doibase 10.1103/PhysRevB.85.035103} {\bibfield  {journal} {\bibinfo
  {journal} {Phys. Rev. B}\ }\textbf {\bibinfo {volume} {85}},\ \bibinfo
  {pages} {035103} (\bibinfo {year} {2012})}\BibitemShut {NoStop}%
\bibitem [{\citenamefont {Liu}\ \emph {et~al.}(2013)\citenamefont {Liu},
  \citenamefont {Ye},\ and\ \citenamefont {Qi}}]{Liu2013}%
  \BibitemOpen
  \bibfield  {author} {\bibinfo {author} {\bibfnamefont {C.-X.}\ \bibnamefont
  {Liu}}, \bibinfo {author} {\bibfnamefont {P.}~\bibnamefont {Ye}}, \ and\
  \bibinfo {author} {\bibfnamefont {X.-L.}\ \bibnamefont {Qi}},\ }\bibfield
  {title} {\emph {\enquote {\bibinfo {title} {Chiral gauge field and axial
  anomaly in a Weyl semimetal},}\ }}\href {\doibase 10.1103/PhysRevB.87.235306}
  {\bibfield  {journal} {\bibinfo  {journal} {Phys. Rev. B}\ }\textbf {\bibinfo
  {volume} {87}},\ \bibinfo {pages} {235306} (\bibinfo {year}
  {2013})}\BibitemShut {NoStop}%
\bibitem [{\citenamefont {Vazifeh}\ and\ \citenamefont
  {Franz}(2013)}]{Vazifeh2013}%
  \BibitemOpen
  \bibfield  {author} {\bibinfo {author} {\bibfnamefont {M.~M.}\ \bibnamefont
  {Vazifeh}}\ and\ \bibinfo {author} {\bibfnamefont {M.}~\bibnamefont
  {Franz}},\ }\bibfield  {title} {\emph {\enquote {\bibinfo {title}
  {Electromagnetic Response of Weyl Semimetals},}\ }}\href {\doibase
  10.1103/PhysRevLett.111.027201} {\bibfield  {journal} {\bibinfo  {journal}
  {Phys. Rev. Lett.}\ }\textbf {\bibinfo {volume} {111}},\ \bibinfo {pages}
  {027201} (\bibinfo {year} {2013})}\BibitemShut {NoStop}%
\bibitem [{\citenamefont {Bzdu\ifmmode~\check{s}\else \v{s}\fi{}ek}\ \emph
  {et~al.}(2015)\citenamefont {Bzdu\ifmmode~\check{s}\else \v{s}\fi{}ek},
  \citenamefont {R\"uegg},\ and\ \citenamefont {Sigrist}}]{Bzdu2015}%
  \BibitemOpen
  \bibfield  {author} {\bibinfo {author} {\bibfnamefont {T.~c.~v.}\
  \bibnamefont {Bzdu\ifmmode~\check{s}\else \v{s}\fi{}ek}}, \bibinfo {author}
  {\bibfnamefont {A.}~\bibnamefont {R\"uegg}}, \ and\ \bibinfo {author}
  {\bibfnamefont {M.}~\bibnamefont {Sigrist}},\ }\bibfield  {title} {\emph
  {\enquote {\bibinfo {title} {Weyl semimetal from spontaneous inversion
  symmetry breaking in pyrochlore oxides},}\ }}\href {\doibase
  10.1103/PhysRevB.91.165105} {\bibfield  {journal} {\bibinfo  {journal} {Phys.
  Rev. B}\ }\textbf {\bibinfo {volume} {91}},\ \bibinfo {pages} {165105}
  (\bibinfo {year} {2015})}\BibitemShut {NoStop}%
\bibitem [{\citenamefont {Rao}(2016)}]{Rao2016}%
  \BibitemOpen
  \bibfield  {author} {\bibinfo {author} {\bibfnamefont {S.}~\bibnamefont
  {Rao}},\ }\bibfield  {title} {\emph {\enquote {\bibinfo {title} {Weyl
  Semi-Metals: A Short Review},}\ }}\href
  {http://journal.iisc.ernet.in/index.php/iisc/article/view/4611} {\bibfield
  {journal} {\bibinfo  {journal} {Journal of the Indian Institute of Science}\
  }\textbf {\bibinfo {volume} {96}},\ \bibinfo {pages} {145} (\bibinfo {year}
  {2016})}\BibitemShut {NoStop}%
\bibitem [{\citenamefont {Kargarian}\ \emph {et~al.}(2016)\citenamefont
  {Kargarian}, \citenamefont {Randeria},\ and\ \citenamefont
  {Lu}}]{Randeria2016PNAS}%
  \BibitemOpen
  \bibfield  {author} {\bibinfo {author} {\bibfnamefont {M.}~\bibnamefont
  {Kargarian}}, \bibinfo {author} {\bibfnamefont {M.}~\bibnamefont {Randeria}},
  \ and\ \bibinfo {author} {\bibfnamefont {Y.-M.}\ \bibnamefont {Lu}},\
  }\bibfield  {title} {\emph {\enquote {\bibinfo {title} {Are the surface Fermi
  arcs in Dirac semimetals topologically protected?}}\ }}\href {\doibase
  10.1073/pnas.1524787113} {\bibfield  {journal} {\bibinfo  {journal} {Proc.
  Natl. Acad. Sci.}\ }\textbf {\bibinfo {volume} {113}},\ \bibinfo {pages}
  {8648} (\bibinfo {year} {2016})}\BibitemShut {NoStop}%
\bibitem [{\citenamefont {Yan}\ and\ \citenamefont {Felser}(2017)}]{Yan2017}%
  \BibitemOpen
  \bibfield  {author} {\bibinfo {author} {\bibfnamefont {B.}~\bibnamefont
  {Yan}}\ and\ \bibinfo {author} {\bibfnamefont {C.}~\bibnamefont {Felser}},\
  }\bibfield  {title} {\emph {\enquote {\bibinfo {title} {Topological
  Materials: Weyl Semimetals},}\ }}\href {\doibase
  10.1146/annurev-conmatphys-031016-025458} {\bibfield  {journal} {\bibinfo
  {journal} {Annu. Rev. Condens. Matter Phys.}\ }\textbf {\bibinfo {volume}
  {8}},\ \bibinfo {pages} {337} (\bibinfo {year} {2017})}\BibitemShut {NoStop}%
\bibitem [{\citenamefont {McCormick}\ \emph {et~al.}(2017)\citenamefont
  {McCormick}, \citenamefont {Kimchi},\ and\ \citenamefont
  {Trivedi}}]{McCormick2017}%
  \BibitemOpen
  \bibfield  {author} {\bibinfo {author} {\bibfnamefont {T.~M.}\ \bibnamefont
  {McCormick}}, \bibinfo {author} {\bibfnamefont {I.}~\bibnamefont {Kimchi}}, \
  and\ \bibinfo {author} {\bibfnamefont {N.}~\bibnamefont {Trivedi}},\
  }\bibfield  {title} {\emph {\enquote {\bibinfo {title} {Minimal models for
  topological Weyl semimetals},}\ }}\href {\doibase 10.1103/PhysRevB.95.075133}
  {\bibfield  {journal} {\bibinfo  {journal} {Phys. Rev. B}\ }\textbf {\bibinfo
  {volume} {95}},\ \bibinfo {pages} {075133} (\bibinfo {year}
  {2017})}\BibitemShut {NoStop}%
\bibitem [{\citenamefont {Armitage}\ \emph {et~al.}(2018)\citenamefont
  {Armitage}, \citenamefont {Mele},\ and\ \citenamefont
  {Vishwanath}}]{Armitage2018}%
  \BibitemOpen
  \bibfield  {author} {\bibinfo {author} {\bibfnamefont {N.~P.}\ \bibnamefont
  {Armitage}}, \bibinfo {author} {\bibfnamefont {E.~J.}\ \bibnamefont {Mele}},
  \ and\ \bibinfo {author} {\bibfnamefont {A.}~\bibnamefont {Vishwanath}},\
  }\bibfield  {title} {\emph {\enquote {\bibinfo {title} {Weyl and Dirac
  semimetals in three-dimensional solids},}\ }}\href {\doibase
  10.1103/RevModPhys.90.015001} {\bibfield  {journal} {\bibinfo  {journal}
  {Rev. Mod. Phys.}\ }\textbf {\bibinfo {volume} {90}},\ \bibinfo {pages}
  {015001} (\bibinfo {year} {2018})}\BibitemShut {NoStop}%
\bibitem [{\citenamefont {Khanna}\ \emph {et~al.}(2014)\citenamefont {Khanna},
  \citenamefont {Kundu}, \citenamefont {Pradhan},\ and\ \citenamefont
  {Rao}}]{Khanna2014}%
  \BibitemOpen
  \bibfield  {author} {\bibinfo {author} {\bibfnamefont {U.}~\bibnamefont
  {Khanna}}, \bibinfo {author} {\bibfnamefont {A.}~\bibnamefont {Kundu}},
  \bibinfo {author} {\bibfnamefont {S.}~\bibnamefont {Pradhan}}, \ and\
  \bibinfo {author} {\bibfnamefont {S.}~\bibnamefont {Rao}},\ }\bibfield
  {title} {\emph {\enquote {\bibinfo {title} {Proximity-induced
  superconductivity in Weyl semimetals},}\ }}\href {\doibase
  10.1103/PhysRevB.90.195430} {\bibfield  {journal} {\bibinfo  {journal} {Phys.
  Rev. B}\ }\textbf {\bibinfo {volume} {90}},\ \bibinfo {pages} {195430}
  (\bibinfo {year} {2014})}\BibitemShut {NoStop}%
\bibitem [{\citenamefont {Burkov}(2018)}]{Burkov2018}%
  \BibitemOpen
  \bibfield  {author} {\bibinfo {author} {\bibfnamefont {A.}~\bibnamefont
  {Burkov}},\ }\bibfield  {title} {\emph {\enquote {\bibinfo {title} {Weyl
  Metals},}\ }}\href {\doibase 10.1146/annurev-conmatphys-033117-054129}
  {\bibfield  {journal} {\bibinfo  {journal} {Annu. Rev. Condens. Matter
  Phys.}\ }\textbf {\bibinfo {volume} {9}},\ \bibinfo {pages} {359} (\bibinfo
  {year} {2018})}\BibitemShut {NoStop}%
\bibitem [{\citenamefont {Chiu}\ and\ \citenamefont
  {Schnyder}(2014)}]{Chiu2014_NLSM}%
  \BibitemOpen
  \bibfield  {author} {\bibinfo {author} {\bibfnamefont {C.-K.}\ \bibnamefont
  {Chiu}}\ and\ \bibinfo {author} {\bibfnamefont {A.~P.}\ \bibnamefont
  {Schnyder}},\ }\bibfield  {title} {\emph {\enquote {\bibinfo {title}
  {Classification of reflection-symmetry-protected topological semimetals and
  nodal superconductors},}\ }}\href {\doibase 10.1103/PhysRevB.90.205136}
  {\bibfield  {journal} {\bibinfo  {journal} {Phys. Rev. B}\ }\textbf {\bibinfo
  {volume} {90}},\ \bibinfo {pages} {205136} (\bibinfo {year}
  {2014})}\BibitemShut {NoStop}%
\bibitem [{\citenamefont {Bian}\ \emph
  {et~al.}(2016{\natexlab{a}})\citenamefont {Bian}, \citenamefont {Chang},
  \citenamefont {Sankar}, \citenamefont {Xu}, \citenamefont {Zheng},
  \citenamefont {Neupert}, \citenamefont {Chiu}, \citenamefont {Huang},
  \citenamefont {Chang}, \citenamefont {Belopolski}, \citenamefont {Sanchez},
  \citenamefont {Neupane}, \citenamefont {Alidoust}, \citenamefont {Liu},
  \citenamefont {Wang}, \citenamefont {Lee}, \citenamefont {Jeng},
  \citenamefont {Zhang}, \citenamefont {Yuan}, \citenamefont {Jia},
  \citenamefont {Bansil}, \citenamefont {Chou}, \citenamefont {Lin},\ and\
  \citenamefont {Hasan}}]{Bian2016_NLSM}%
  \BibitemOpen
  \bibfield  {author} {\bibinfo {author} {\bibfnamefont {G.}~\bibnamefont
  {Bian}}, \bibinfo {author} {\bibfnamefont {T.-R.}\ \bibnamefont {Chang}},
  \bibinfo {author} {\bibfnamefont {R.}~\bibnamefont {Sankar}}, \bibinfo
  {author} {\bibfnamefont {S.-Y.}\ \bibnamefont {Xu}}, \bibinfo {author}
  {\bibfnamefont {H.}~\bibnamefont {Zheng}}, \bibinfo {author} {\bibfnamefont
  {T.}~\bibnamefont {Neupert}}, \bibinfo {author} {\bibfnamefont {C.-K.}\
  \bibnamefont {Chiu}}, \bibinfo {author} {\bibfnamefont {S.-M.}\ \bibnamefont
  {Huang}}, \bibinfo {author} {\bibfnamefont {G.}~\bibnamefont {Chang}},
  \bibinfo {author} {\bibfnamefont {I.}~\bibnamefont {Belopolski}}, \bibinfo
  {author} {\bibfnamefont {D.~S.}\ \bibnamefont {Sanchez}}, \bibinfo {author}
  {\bibfnamefont {M.}~\bibnamefont {Neupane}}, \bibinfo {author} {\bibfnamefont
  {N.}~\bibnamefont {Alidoust}}, \bibinfo {author} {\bibfnamefont
  {C.}~\bibnamefont {Liu}}, \bibinfo {author} {\bibfnamefont {B.}~\bibnamefont
  {Wang}}, \bibinfo {author} {\bibfnamefont {C.-C.}\ \bibnamefont {Lee}},
  \bibinfo {author} {\bibfnamefont {H.-T.}\ \bibnamefont {Jeng}}, \bibinfo
  {author} {\bibfnamefont {C.}~\bibnamefont {Zhang}}, \bibinfo {author}
  {\bibfnamefont {Z.}~\bibnamefont {Yuan}}, \bibinfo {author} {\bibfnamefont
  {S.}~\bibnamefont {Jia}}, \bibinfo {author} {\bibfnamefont {A.}~\bibnamefont
  {Bansil}}, \bibinfo {author} {\bibfnamefont {F.}~\bibnamefont {Chou}},
  \bibinfo {author} {\bibfnamefont {H.}~\bibnamefont {Lin}}, \ and\ \bibinfo
  {author} {\bibfnamefont {M.~Z.}\ \bibnamefont {Hasan}},\ }\bibfield  {title}
  {\emph {\enquote {\bibinfo {title} {Topological nodal-line fermions in
  spin-orbit metal PbTaSe2},}\ }}\href {\doibase 10.1038/ncomms10556}
  {\bibfield  {journal} {\bibinfo  {journal} {Nature Communications}\ }\textbf
  {\bibinfo {volume} {7}},\ \bibinfo {pages} {10556} (\bibinfo {year}
  {2016}{\natexlab{a}})}\BibitemShut {NoStop}%
\bibitem [{\citenamefont {Yamakage}\ \emph {et~al.}(2016)\citenamefont
  {Yamakage}, \citenamefont {Yamakawa}, \citenamefont {Tanaka},\ and\
  \citenamefont {Okamoto}}]{Yamakage2016_NLSM}%
  \BibitemOpen
  \bibfield  {author} {\bibinfo {author} {\bibfnamefont {A.}~\bibnamefont
  {Yamakage}}, \bibinfo {author} {\bibfnamefont {Y.}~\bibnamefont {Yamakawa}},
  \bibinfo {author} {\bibfnamefont {Y.}~\bibnamefont {Tanaka}}, \ and\ \bibinfo
  {author} {\bibfnamefont {Y.}~\bibnamefont {Okamoto}},\ }\bibfield  {title}
  {\emph {\enquote {\bibinfo {title} {Line-Node Dirac Semimetal and Topological
  Insulating Phase in Noncentrosymmetric Pnictides CaAgX (X = P, As)},}\
  }}\href {\doibase 10.7566/JPSJ.85.013708} {\bibfield  {journal} {\bibinfo
  {journal} {Journal of the Physical Society of Japan}\ }\textbf {\bibinfo
  {volume} {85}},\ \bibinfo {pages} {013708} (\bibinfo {year}
  {2016})}\BibitemShut {NoStop}%
\bibitem [{\citenamefont {Chan}\ \emph {et~al.}(2016)\citenamefont {Chan},
  \citenamefont {Chiu}, \citenamefont {Chou},\ and\ \citenamefont
  {Schnyder}}]{Schnyder2016_NLSM}%
  \BibitemOpen
  \bibfield  {author} {\bibinfo {author} {\bibfnamefont {Y.-H.}\ \bibnamefont
  {Chan}}, \bibinfo {author} {\bibfnamefont {C.-K.}\ \bibnamefont {Chiu}},
  \bibinfo {author} {\bibfnamefont {M.~Y.}\ \bibnamefont {Chou}}, \ and\
  \bibinfo {author} {\bibfnamefont {A.~P.}\ \bibnamefont {Schnyder}},\
  }\bibfield  {title} {\emph {\enquote {\bibinfo {title}
  {${\mathrm{Ca}}_{3}{\mathrm{P}}_{2}$ and other topological semimetals with
  line nodes and drumhead surface states},}\ }}\href {\doibase
  10.1103/PhysRevB.93.205132} {\bibfield  {journal} {\bibinfo  {journal} {Phys.
  Rev. B}\ }\textbf {\bibinfo {volume} {93}},\ \bibinfo {pages} {205132}
  (\bibinfo {year} {2016})}\BibitemShut {NoStop}%
\bibitem [{\citenamefont {Nie}\ \emph {et~al.}(2019)\citenamefont {Nie},
  \citenamefont {Weng},\ and\ \citenamefont {Prinz}}]{Nie2019_NLSM}%
  \BibitemOpen
  \bibfield  {author} {\bibinfo {author} {\bibfnamefont {S.}~\bibnamefont
  {Nie}}, \bibinfo {author} {\bibfnamefont {H.}~\bibnamefont {Weng}}, \ and\
  \bibinfo {author} {\bibfnamefont {F.~B.}\ \bibnamefont {Prinz}},\ }\bibfield
  {title} {\emph {\enquote {\bibinfo {title} {Topological nodal-line semimetals
  in ferromagnetic rare-earth-metal monohalides},}\ }}\href {\doibase
  10.1103/PhysRevB.99.035125} {\bibfield  {journal} {\bibinfo  {journal} {Phys.
  Rev. B}\ }\textbf {\bibinfo {volume} {99}},\ \bibinfo {pages} {035125}
  (\bibinfo {year} {2019})}\BibitemShut {NoStop}%
\bibitem [{\citenamefont {Yang}\ \emph {et~al.}(2018)\citenamefont {Yang},
  \citenamefont {Yang}, \citenamefont {Derunova}, \citenamefont {Parkin},
  \citenamefont {Yan},\ and\ \citenamefont {Ali}}]{Yang2018NLSM}%
  \BibitemOpen
  \bibfield  {author} {\bibinfo {author} {\bibfnamefont {S.-Y.}\ \bibnamefont
  {Yang}}, \bibinfo {author} {\bibfnamefont {H.}~\bibnamefont {Yang}}, \bibinfo
  {author} {\bibfnamefont {E.}~\bibnamefont {Derunova}}, \bibinfo {author}
  {\bibfnamefont {S.~S.~P.}\ \bibnamefont {Parkin}}, \bibinfo {author}
  {\bibfnamefont {B.}~\bibnamefont {Yan}}, \ and\ \bibinfo {author}
  {\bibfnamefont {M.~N.}\ \bibnamefont {Ali}},\ }\bibfield  {title} {\emph
  {\enquote {\bibinfo {title} {Symmetry demanded topological nodal-line
  materials},}\ }}\href {\doibase 10.1080/23746149.2017.1414631} {\bibfield
  {journal} {\bibinfo  {journal} {Advances in Physics: X}\ }\textbf {\bibinfo
  {volume} {3}},\ \bibinfo {pages} {1414631} (\bibinfo {year}
  {2018})}\BibitemShut {NoStop}%
\bibitem [{\citenamefont {Parhizgar}\ and\ \citenamefont
  {Black-Schaffer}(2020)}]{Parhizgar2020_NLSM}%
  \BibitemOpen
  \bibfield  {author} {\bibinfo {author} {\bibfnamefont {F.}~\bibnamefont
  {Parhizgar}}\ and\ \bibinfo {author} {\bibfnamefont {A.~M.}\ \bibnamefont
  {Black-Schaffer}},\ }\bibfield  {title} {\emph {\enquote {\bibinfo {title}
  {Large Josephson current in Weyl nodal loop semimetals due to odd-frequency
  superconductivity},}\ }}\href {\doibase 10.1038/s41535-020-0244-2} {\bibfield
   {journal} {\bibinfo  {journal} {npj Quantum Materials}\ }\textbf {\bibinfo
  {volume} {5}},\ \bibinfo {pages} {42} (\bibinfo {year} {2020})}\BibitemShut
  {NoStop}%
\bibitem [{\citenamefont {Zhan}\ \emph {et~al.}(2023)\citenamefont {Zhan},
  \citenamefont {Li}, \citenamefont {Shi}, \citenamefont {Chen},\ and\
  \citenamefont {Sun}}]{Zhan2023}%
  \BibitemOpen
  \bibfield  {author} {\bibinfo {author} {\bibfnamefont {J.}~\bibnamefont
  {Zhan}}, \bibinfo {author} {\bibfnamefont {J.}~\bibnamefont {Li}}, \bibinfo
  {author} {\bibfnamefont {W.}~\bibnamefont {Shi}}, \bibinfo {author}
  {\bibfnamefont {X.-Q.}\ \bibnamefont {Chen}}, \ and\ \bibinfo {author}
  {\bibfnamefont {Y.}~\bibnamefont {Sun}},\ }\bibfield  {title} {\emph
  {\enquote {\bibinfo {title} {Coexistence of Weyl semimetal and Weyl nodal
  loop semimetal phases in a collinear antiferromagnet},}\ }}\href {\doibase
  10.1103/PhysRevB.107.224402} {\bibfield  {journal} {\bibinfo  {journal}
  {Phys. Rev. B}\ }\textbf {\bibinfo {volume} {107}},\ \bibinfo {pages}
  {224402} (\bibinfo {year} {2023})}\BibitemShut {NoStop}%
\bibitem [{\citenamefont {Matsuura}\ \emph {et~al.}(2013)\citenamefont
  {Matsuura}, \citenamefont {Chang}, \citenamefont {Schnyder},\ and\
  \citenamefont {Ryu}}]{Matsuura_2013}%
  \BibitemOpen
  \bibfield  {author} {\bibinfo {author} {\bibfnamefont {S.}~\bibnamefont
  {Matsuura}}, \bibinfo {author} {\bibfnamefont {P.-Y.}\ \bibnamefont {Chang}},
  \bibinfo {author} {\bibfnamefont {A.~P.}\ \bibnamefont {Schnyder}}, \ and\
  \bibinfo {author} {\bibfnamefont {S.}~\bibnamefont {Ryu}},\ }\bibfield
  {title} {\emph {\enquote {\bibinfo {title} {Protected boundary states in
  gapless topological phases},}\ }}\href {\doibase
  10.1088/1367-2630/15/6/065001} {\bibfield  {journal} {\bibinfo  {journal}
  {New Journal of Physics}\ }\textbf {\bibinfo {volume} {15}},\ \bibinfo
  {pages} {065001} (\bibinfo {year} {2013})}\BibitemShut {NoStop}%
\bibitem [{\citenamefont {Zhao}\ and\ \citenamefont {Wang}(2013)}]{Zhao2013}%
  \BibitemOpen
  \bibfield  {author} {\bibinfo {author} {\bibfnamefont {Y.~X.}\ \bibnamefont
  {Zhao}}\ and\ \bibinfo {author} {\bibfnamefont {Z.~D.}\ \bibnamefont
  {Wang}},\ }\bibfield  {title} {\emph {\enquote {\bibinfo {title} {Topological
  Classification and Stability of Fermi Surfaces},}\ }}\href {\doibase
  10.1103/PhysRevLett.110.240404} {\bibfield  {journal} {\bibinfo  {journal}
  {Phys. Rev. Lett.}\ }\textbf {\bibinfo {volume} {110}},\ \bibinfo {pages}
  {240404} (\bibinfo {year} {2013})}\BibitemShut {NoStop}%
\bibitem [{\citenamefont {Xu}\ \emph {et~al.}(2011)\citenamefont {Xu},
  \citenamefont {Weng}, \citenamefont {Wang}, \citenamefont {Dai},\ and\
  \citenamefont {Fang}}]{Xu2011}%
  \BibitemOpen
  \bibfield  {author} {\bibinfo {author} {\bibfnamefont {G.}~\bibnamefont
  {Xu}}, \bibinfo {author} {\bibfnamefont {H.}~\bibnamefont {Weng}}, \bibinfo
  {author} {\bibfnamefont {Z.}~\bibnamefont {Wang}}, \bibinfo {author}
  {\bibfnamefont {X.}~\bibnamefont {Dai}}, \ and\ \bibinfo {author}
  {\bibfnamefont {Z.}~\bibnamefont {Fang}},\ }\bibfield  {title} {\emph
  {\enquote {\bibinfo {title} {Chern Semimetal and the Quantized Anomalous Hall
  Effect in ${\mathrm{HgCr}}_{2}{\mathrm{Se}}_{4}$},}\ }}\href {\doibase
  10.1103/PhysRevLett.107.186806} {\bibfield  {journal} {\bibinfo  {journal}
  {Phys. Rev. Lett.}\ }\textbf {\bibinfo {volume} {107}},\ \bibinfo {pages}
  {186806} (\bibinfo {year} {2011})}\BibitemShut {NoStop}%
\bibitem [{\citenamefont {Potter}\ \emph {et~al.}(2014)\citenamefont {Potter},
  \citenamefont {Kimchi},\ and\ \citenamefont {Vishwanath}}]{Potter2014}%
  \BibitemOpen
  \bibfield  {author} {\bibinfo {author} {\bibfnamefont {A.~C.}\ \bibnamefont
  {Potter}}, \bibinfo {author} {\bibfnamefont {I.}~\bibnamefont {Kimchi}}, \
  and\ \bibinfo {author} {\bibfnamefont {A.}~\bibnamefont {Vishwanath}},\
  }\bibfield  {title} {\emph {\enquote {\bibinfo {title} {Quantum oscillations
  from surface Fermi arcs in Weyl and Dirac semimetals},}\ }}\href {\doibase
  10.1038/ncomms6161} {\bibfield  {journal} {\bibinfo  {journal} {Nat Commun}\
  }\textbf {\bibinfo {volume} {5}},\ \bibinfo {pages} {5161} (\bibinfo {year}
  {2014})}\BibitemShut {NoStop}%
\bibitem [{\citenamefont {Lv}\ \emph {et~al.}(2015{\natexlab{a}})\citenamefont
  {Lv}, \citenamefont {Xu}, \citenamefont {Weng}, \citenamefont {Ma},
  \citenamefont {Richard}, \citenamefont {Huang}, \citenamefont {Zhao},
  \citenamefont {Chen}, \citenamefont {Matt}, \citenamefont {Bisti},
  \citenamefont {Strocov}, \citenamefont {Mesot}, \citenamefont {Fang},
  \citenamefont {Dai}, \citenamefont {Qian}, \citenamefont {Shi},\ and\
  \citenamefont {Ding}}]{Lv2015a}%
  \BibitemOpen
  \bibfield  {author} {\bibinfo {author} {\bibfnamefont {B.~Q.}\ \bibnamefont
  {Lv}}, \bibinfo {author} {\bibfnamefont {N.}~\bibnamefont {Xu}}, \bibinfo
  {author} {\bibfnamefont {H.~M.}\ \bibnamefont {Weng}}, \bibinfo {author}
  {\bibfnamefont {J.~Z.}\ \bibnamefont {Ma}}, \bibinfo {author} {\bibfnamefont
  {P.}~\bibnamefont {Richard}}, \bibinfo {author} {\bibfnamefont {X.~C.}\
  \bibnamefont {Huang}}, \bibinfo {author} {\bibfnamefont {L.~X.}\ \bibnamefont
  {Zhao}}, \bibinfo {author} {\bibfnamefont {G.~F.}\ \bibnamefont {Chen}},
  \bibinfo {author} {\bibfnamefont {C.~E.}\ \bibnamefont {Matt}}, \bibinfo
  {author} {\bibfnamefont {F.}~\bibnamefont {Bisti}}, \bibinfo {author}
  {\bibfnamefont {V.~N.}\ \bibnamefont {Strocov}}, \bibinfo {author}
  {\bibfnamefont {J.}~\bibnamefont {Mesot}}, \bibinfo {author} {\bibfnamefont
  {Z.}~\bibnamefont {Fang}}, \bibinfo {author} {\bibfnamefont {X.}~\bibnamefont
  {Dai}}, \bibinfo {author} {\bibfnamefont {T.}~\bibnamefont {Qian}}, \bibinfo
  {author} {\bibfnamefont {M.}~\bibnamefont {Shi}}, \ and\ \bibinfo {author}
  {\bibfnamefont {H.}~\bibnamefont {Ding}},\ }\bibfield  {title} {\emph
  {\enquote {\bibinfo {title} {Observation of Weyl nodes in TaAs},}\ }}\href
  {\doibase 10.1038/nphys3426} {\bibfield  {journal} {\bibinfo  {journal}
  {Nature Phys}\ }\textbf {\bibinfo {volume} {11}},\ \bibinfo {pages} {724}
  (\bibinfo {year} {2015}{\natexlab{a}})}\BibitemShut {NoStop}%
\bibitem [{\citenamefont {Lv}\ \emph {et~al.}(2015{\natexlab{b}})\citenamefont
  {Lv}, \citenamefont {Weng}, \citenamefont {Fu}, \citenamefont {Wang},
  \citenamefont {Miao}, \citenamefont {Ma}, \citenamefont {Richard},
  \citenamefont {Huang}, \citenamefont {Zhao}, \citenamefont {Chen},
  \citenamefont {Fang}, \citenamefont {Dai}, \citenamefont {Qian},\ and\
  \citenamefont {Ding}}]{Lv2015b}%
  \BibitemOpen
  \bibfield  {author} {\bibinfo {author} {\bibfnamefont {B.~Q.}\ \bibnamefont
  {Lv}}, \bibinfo {author} {\bibfnamefont {H.~M.}\ \bibnamefont {Weng}},
  \bibinfo {author} {\bibfnamefont {B.~B.}\ \bibnamefont {Fu}}, \bibinfo
  {author} {\bibfnamefont {X.~P.}\ \bibnamefont {Wang}}, \bibinfo {author}
  {\bibfnamefont {H.}~\bibnamefont {Miao}}, \bibinfo {author} {\bibfnamefont
  {J.}~\bibnamefont {Ma}}, \bibinfo {author} {\bibfnamefont {P.}~\bibnamefont
  {Richard}}, \bibinfo {author} {\bibfnamefont {X.~C.}\ \bibnamefont {Huang}},
  \bibinfo {author} {\bibfnamefont {L.~X.}\ \bibnamefont {Zhao}}, \bibinfo
  {author} {\bibfnamefont {G.~F.}\ \bibnamefont {Chen}}, \bibinfo {author}
  {\bibfnamefont {Z.}~\bibnamefont {Fang}}, \bibinfo {author} {\bibfnamefont
  {X.}~\bibnamefont {Dai}}, \bibinfo {author} {\bibfnamefont {T.}~\bibnamefont
  {Qian}}, \ and\ \bibinfo {author} {\bibfnamefont {H.}~\bibnamefont {Ding}},\
  }\bibfield  {title} {\emph {\enquote {\bibinfo {title} {Experimental
  Discovery of Weyl Semimetal TaAs},}\ }}\href {\doibase
  10.1103/PhysRevX.5.031013} {\bibfield  {journal} {\bibinfo  {journal} {Phys.
  Rev. X}\ }\textbf {\bibinfo {volume} {5}},\ \bibinfo {pages} {031013}
  (\bibinfo {year} {2015}{\natexlab{b}})}\BibitemShut {NoStop}%
\bibitem [{\citenamefont {Lv}\ \emph {et~al.}(2015{\natexlab{c}})\citenamefont
  {Lv}, \citenamefont {Muff}, \citenamefont {Qian}, \citenamefont {Song},
  \citenamefont {Nie}, \citenamefont {Xu}, \citenamefont {Richard},
  \citenamefont {Matt}, \citenamefont {Plumb}, \citenamefont {Zhao},
  \citenamefont {Chen}, \citenamefont {Fang}, \citenamefont {Dai},
  \citenamefont {Dil}, \citenamefont {Mesot}, \citenamefont {Shi},
  \citenamefont {Weng},\ and\ \citenamefont {Ding}}]{Lv2015c}%
  \BibitemOpen
  \bibfield  {author} {\bibinfo {author} {\bibfnamefont {B.~Q.}\ \bibnamefont
  {Lv}}, \bibinfo {author} {\bibfnamefont {S.}~\bibnamefont {Muff}}, \bibinfo
  {author} {\bibfnamefont {T.}~\bibnamefont {Qian}}, \bibinfo {author}
  {\bibfnamefont {Z.~D.}\ \bibnamefont {Song}}, \bibinfo {author}
  {\bibfnamefont {S.~M.}\ \bibnamefont {Nie}}, \bibinfo {author} {\bibfnamefont
  {N.}~\bibnamefont {Xu}}, \bibinfo {author} {\bibfnamefont {P.}~\bibnamefont
  {Richard}}, \bibinfo {author} {\bibfnamefont {C.~E.}\ \bibnamefont {Matt}},
  \bibinfo {author} {\bibfnamefont {N.~C.}\ \bibnamefont {Plumb}}, \bibinfo
  {author} {\bibfnamefont {L.~X.}\ \bibnamefont {Zhao}}, \bibinfo {author}
  {\bibfnamefont {G.~F.}\ \bibnamefont {Chen}}, \bibinfo {author}
  {\bibfnamefont {Z.}~\bibnamefont {Fang}}, \bibinfo {author} {\bibfnamefont
  {X.}~\bibnamefont {Dai}}, \bibinfo {author} {\bibfnamefont {J.~H.}\
  \bibnamefont {Dil}}, \bibinfo {author} {\bibfnamefont {J.}~\bibnamefont
  {Mesot}}, \bibinfo {author} {\bibfnamefont {M.}~\bibnamefont {Shi}}, \bibinfo
  {author} {\bibfnamefont {H.~M.}\ \bibnamefont {Weng}}, \ and\ \bibinfo
  {author} {\bibfnamefont {H.}~\bibnamefont {Ding}},\ }\bibfield  {title}
  {\emph {\enquote {\bibinfo {title} {Observation of Fermi-Arc Spin Texture in
  TaAs},}\ }}\href {\doibase 10.1103/PhysRevLett.115.217601} {\bibfield
  {journal} {\bibinfo  {journal} {Phys. Rev. Lett.}\ }\textbf {\bibinfo
  {volume} {115}},\ \bibinfo {pages} {217601} (\bibinfo {year}
  {2015}{\natexlab{c}})}\BibitemShut {NoStop}%
\bibitem [{\citenamefont {Xu}\ \emph {et~al.}(2015{\natexlab{a}})\citenamefont
  {Xu}, \citenamefont {Belopolski}, \citenamefont {Sanchez}, \citenamefont
  {Zhang}, \citenamefont {Chang}, \citenamefont {Guo}, \citenamefont {Bian},
  \citenamefont {Yuan}, \citenamefont {Lu}, \citenamefont {Chang},
  \citenamefont {Shibayev}, \citenamefont {Prokopovych}, \citenamefont
  {Alidoust}, \citenamefont {Zheng}, \citenamefont {Lee}, \citenamefont
  {Huang}, \citenamefont {Sankar}, \citenamefont {Chou}, \citenamefont {Hsu},
  \citenamefont {Jeng}, \citenamefont {Bansil}, \citenamefont {Neupert},
  \citenamefont {Strocov}, \citenamefont {Lin}, \citenamefont {Jia},\ and\
  \citenamefont {Hasan}}]{Xu2015a}%
  \BibitemOpen
  \bibfield  {author} {\bibinfo {author} {\bibfnamefont {S.-Y.}\ \bibnamefont
  {Xu}}, \bibinfo {author} {\bibfnamefont {I.}~\bibnamefont {Belopolski}},
  \bibinfo {author} {\bibfnamefont {D.~S.}\ \bibnamefont {Sanchez}}, \bibinfo
  {author} {\bibfnamefont {C.}~\bibnamefont {Zhang}}, \bibinfo {author}
  {\bibfnamefont {G.}~\bibnamefont {Chang}}, \bibinfo {author} {\bibfnamefont
  {C.}~\bibnamefont {Guo}}, \bibinfo {author} {\bibfnamefont {G.}~\bibnamefont
  {Bian}}, \bibinfo {author} {\bibfnamefont {Z.}~\bibnamefont {Yuan}}, \bibinfo
  {author} {\bibfnamefont {H.}~\bibnamefont {Lu}}, \bibinfo {author}
  {\bibfnamefont {T.-R.}\ \bibnamefont {Chang}}, \bibinfo {author}
  {\bibfnamefont {P.~P.}\ \bibnamefont {Shibayev}}, \bibinfo {author}
  {\bibfnamefont {M.~L.}\ \bibnamefont {Prokopovych}}, \bibinfo {author}
  {\bibfnamefont {N.}~\bibnamefont {Alidoust}}, \bibinfo {author}
  {\bibfnamefont {H.}~\bibnamefont {Zheng}}, \bibinfo {author} {\bibfnamefont
  {C.-C.}\ \bibnamefont {Lee}}, \bibinfo {author} {\bibfnamefont {S.-M.}\
  \bibnamefont {Huang}}, \bibinfo {author} {\bibfnamefont {R.}~\bibnamefont
  {Sankar}}, \bibinfo {author} {\bibfnamefont {F.}~\bibnamefont {Chou}},
  \bibinfo {author} {\bibfnamefont {C.-H.}\ \bibnamefont {Hsu}}, \bibinfo
  {author} {\bibfnamefont {H.-T.}\ \bibnamefont {Jeng}}, \bibinfo {author}
  {\bibfnamefont {A.}~\bibnamefont {Bansil}}, \bibinfo {author} {\bibfnamefont
  {T.}~\bibnamefont {Neupert}}, \bibinfo {author} {\bibfnamefont {V.~N.}\
  \bibnamefont {Strocov}}, \bibinfo {author} {\bibfnamefont {H.}~\bibnamefont
  {Lin}}, \bibinfo {author} {\bibfnamefont {S.}~\bibnamefont {Jia}}, \ and\
  \bibinfo {author} {\bibfnamefont {M.~Z.}\ \bibnamefont {Hasan}},\ }\bibfield
  {title} {\emph {\enquote {\bibinfo {title} {Experimental discovery of a
  topological Weyl semimetal state in TaP},}\ }}\href {\doibase
  10.1126/sciadv.1501092} {\bibfield  {journal} {\bibinfo  {journal} {Sci.
  Adv.}\ }\textbf {\bibinfo {volume} {1}},\ \bibinfo {pages} {e1501092}
  (\bibinfo {year} {2015}{\natexlab{a}})}\BibitemShut {NoStop}%
\bibitem [{\citenamefont {Xu}\ \emph {et~al.}(2015{\natexlab{b}})\citenamefont
  {Xu}, \citenamefont {Alidoust}, \citenamefont {Belopolski}, \citenamefont
  {Yuan}, \citenamefont {Bian}, \citenamefont {Chang}, \citenamefont {Zheng},
  \citenamefont {Strocov}, \citenamefont {Sanchez}, \citenamefont {Chang},
  \citenamefont {Zhang}, \citenamefont {Mou}, \citenamefont {Wu}, \citenamefont
  {Huang}, \citenamefont {Lee}, \citenamefont {Huang}, \citenamefont {Wang},
  \citenamefont {Bansil}, \citenamefont {Jeng}, \citenamefont {Neupert},
  \citenamefont {Kaminski}, \citenamefont {Lin}, \citenamefont {Jia},\ and\
  \citenamefont {Zahid~Hasan}}]{Xu2015b}%
  \BibitemOpen
  \bibfield  {author} {\bibinfo {author} {\bibfnamefont {S.-Y.}\ \bibnamefont
  {Xu}}, \bibinfo {author} {\bibfnamefont {N.}~\bibnamefont {Alidoust}},
  \bibinfo {author} {\bibfnamefont {I.}~\bibnamefont {Belopolski}}, \bibinfo
  {author} {\bibfnamefont {Z.}~\bibnamefont {Yuan}}, \bibinfo {author}
  {\bibfnamefont {G.}~\bibnamefont {Bian}}, \bibinfo {author} {\bibfnamefont
  {T.-R.}\ \bibnamefont {Chang}}, \bibinfo {author} {\bibfnamefont
  {H.}~\bibnamefont {Zheng}}, \bibinfo {author} {\bibfnamefont {V.~N.}\
  \bibnamefont {Strocov}}, \bibinfo {author} {\bibfnamefont {D.~S.}\
  \bibnamefont {Sanchez}}, \bibinfo {author} {\bibfnamefont {G.}~\bibnamefont
  {Chang}}, \bibinfo {author} {\bibfnamefont {C.}~\bibnamefont {Zhang}},
  \bibinfo {author} {\bibfnamefont {D.}~\bibnamefont {Mou}}, \bibinfo {author}
  {\bibfnamefont {Y.}~\bibnamefont {Wu}}, \bibinfo {author} {\bibfnamefont
  {L.}~\bibnamefont {Huang}}, \bibinfo {author} {\bibfnamefont {C.-C.}\
  \bibnamefont {Lee}}, \bibinfo {author} {\bibfnamefont {S.-M.}\ \bibnamefont
  {Huang}}, \bibinfo {author} {\bibfnamefont {B.}~\bibnamefont {Wang}},
  \bibinfo {author} {\bibfnamefont {A.}~\bibnamefont {Bansil}}, \bibinfo
  {author} {\bibfnamefont {H.-T.}\ \bibnamefont {Jeng}}, \bibinfo {author}
  {\bibfnamefont {T.}~\bibnamefont {Neupert}}, \bibinfo {author} {\bibfnamefont
  {A.}~\bibnamefont {Kaminski}}, \bibinfo {author} {\bibfnamefont
  {H.}~\bibnamefont {Lin}}, \bibinfo {author} {\bibfnamefont {S.}~\bibnamefont
  {Jia}}, \ and\ \bibinfo {author} {\bibfnamefont {M.}~\bibnamefont
  {Zahid~Hasan}},\ }\bibfield  {title} {\emph {\enquote {\bibinfo {title}
  {Discovery of a Weyl fermion state with Fermi arcs in niobium arsenide},}\
  }}\href {\doibase 10.1038/nphys3437} {\bibfield  {journal} {\bibinfo
  {journal} {Nature Phys}\ }\textbf {\bibinfo {volume} {11}},\ \bibinfo {pages}
  {748} (\bibinfo {year} {2015}{\natexlab{b}})}\BibitemShut {NoStop}%
\bibitem [{\citenamefont {Xu}\ \emph {et~al.}(2015{\natexlab{c}})\citenamefont
  {Xu}, \citenamefont {Belopolski}, \citenamefont {Alidoust}, \citenamefont
  {Neupane}, \citenamefont {Bian}, \citenamefont {Zhang}, \citenamefont
  {Sankar}, \citenamefont {Chang}, \citenamefont {Yuan}, \citenamefont {Lee},
  \citenamefont {Huang}, \citenamefont {Zheng}, \citenamefont {Ma},
  \citenamefont {Sanchez}, \citenamefont {Wang}, \citenamefont {Bansil},
  \citenamefont {Chou}, \citenamefont {Shibayev}, \citenamefont {Lin},
  \citenamefont {Jia},\ and\ \citenamefont {Hasan}}]{Xu2015c}%
  \BibitemOpen
  \bibfield  {author} {\bibinfo {author} {\bibfnamefont {S.-Y.}\ \bibnamefont
  {Xu}}, \bibinfo {author} {\bibfnamefont {I.}~\bibnamefont {Belopolski}},
  \bibinfo {author} {\bibfnamefont {N.}~\bibnamefont {Alidoust}}, \bibinfo
  {author} {\bibfnamefont {M.}~\bibnamefont {Neupane}}, \bibinfo {author}
  {\bibfnamefont {G.}~\bibnamefont {Bian}}, \bibinfo {author} {\bibfnamefont
  {C.}~\bibnamefont {Zhang}}, \bibinfo {author} {\bibfnamefont
  {R.}~\bibnamefont {Sankar}}, \bibinfo {author} {\bibfnamefont
  {G.}~\bibnamefont {Chang}}, \bibinfo {author} {\bibfnamefont
  {Z.}~\bibnamefont {Yuan}}, \bibinfo {author} {\bibfnamefont {C.-C.}\
  \bibnamefont {Lee}}, \bibinfo {author} {\bibfnamefont {S.-M.}\ \bibnamefont
  {Huang}}, \bibinfo {author} {\bibfnamefont {H.}~\bibnamefont {Zheng}},
  \bibinfo {author} {\bibfnamefont {J.}~\bibnamefont {Ma}}, \bibinfo {author}
  {\bibfnamefont {D.~S.}\ \bibnamefont {Sanchez}}, \bibinfo {author}
  {\bibfnamefont {B.}~\bibnamefont {Wang}}, \bibinfo {author} {\bibfnamefont
  {A.}~\bibnamefont {Bansil}}, \bibinfo {author} {\bibfnamefont
  {F.}~\bibnamefont {Chou}}, \bibinfo {author} {\bibfnamefont {P.~P.}\
  \bibnamefont {Shibayev}}, \bibinfo {author} {\bibfnamefont {H.}~\bibnamefont
  {Lin}}, \bibinfo {author} {\bibfnamefont {S.}~\bibnamefont {Jia}}, \ and\
  \bibinfo {author} {\bibfnamefont {M.~Z.}\ \bibnamefont {Hasan}},\ }\bibfield
  {title} {\emph {\enquote {\bibinfo {title} {Discovery of a Weyl fermion
  semimetal and topological Fermi arcs},}\ }}\href {\doibase
  10.1126/science.aaa9297} {\bibfield  {journal} {\bibinfo  {journal}
  {Science}\ }\textbf {\bibinfo {volume} {349}},\ \bibinfo {pages} {613}
  (\bibinfo {year} {2015}{\natexlab{c}})}\BibitemShut {NoStop}%
\bibitem [{\citenamefont {Lu}\ \emph {et~al.}(2015)\citenamefont {Lu},
  \citenamefont {Wang}, \citenamefont {Ye}, \citenamefont {Ran}, \citenamefont
  {Fu}, \citenamefont {Joannopoulos},\ and\ \citenamefont
  {Soljačić}}]{Lu2015}%
  \BibitemOpen
  \bibfield  {author} {\bibinfo {author} {\bibfnamefont {L.}~\bibnamefont
  {Lu}}, \bibinfo {author} {\bibfnamefont {Z.}~\bibnamefont {Wang}}, \bibinfo
  {author} {\bibfnamefont {D.}~\bibnamefont {Ye}}, \bibinfo {author}
  {\bibfnamefont {L.}~\bibnamefont {Ran}}, \bibinfo {author} {\bibfnamefont
  {L.}~\bibnamefont {Fu}}, \bibinfo {author} {\bibfnamefont {J.~D.}\
  \bibnamefont {Joannopoulos}}, \ and\ \bibinfo {author} {\bibfnamefont
  {M.}~\bibnamefont {Soljačić}},\ }\bibfield  {title} {\emph {\enquote
  {\bibinfo {title} {Experimental observation of Weyl points},}\ }}\href
  {\doibase 10.1126/science.aaa9273} {\bibfield  {journal} {\bibinfo  {journal}
  {Science}\ }\textbf {\bibinfo {volume} {349}},\ \bibinfo {pages} {622}
  (\bibinfo {year} {2015})}\BibitemShut {NoStop}%
\bibitem [{\citenamefont {Sun}\ \emph {et~al.}(2015)\citenamefont {Sun},
  \citenamefont {Wu},\ and\ \citenamefont {Yan}}]{Sun2015}%
  \BibitemOpen
  \bibfield  {author} {\bibinfo {author} {\bibfnamefont {Y.}~\bibnamefont
  {Sun}}, \bibinfo {author} {\bibfnamefont {S.-C.}\ \bibnamefont {Wu}}, \ and\
  \bibinfo {author} {\bibfnamefont {B.}~\bibnamefont {Yan}},\ }\bibfield
  {title} {\emph {\enquote {\bibinfo {title} {Topological surface states and
  Fermi arcs of the noncentrosymmetric Weyl semimetals TaAs, TaP, NbAs, and
  NbP},}\ }}\href {\doibase 10.1103/PhysRevB.92.115428} {\bibfield  {journal}
  {\bibinfo  {journal} {Phys. Rev. B}\ }\textbf {\bibinfo {volume} {92}},\
  \bibinfo {pages} {115428} (\bibinfo {year} {2015})}\BibitemShut {NoStop}%
\bibitem [{\citenamefont {Soluyanov}\ \emph {et~al.}(2015)\citenamefont
  {Soluyanov}, \citenamefont {Gresch}, \citenamefont {Wang}, \citenamefont
  {Wu}, \citenamefont {Troyer}, \citenamefont {Dai},\ and\ \citenamefont
  {Bernevig}}]{Soluyanov2015}%
  \BibitemOpen
  \bibfield  {author} {\bibinfo {author} {\bibfnamefont {A.~A.}\ \bibnamefont
  {Soluyanov}}, \bibinfo {author} {\bibfnamefont {D.}~\bibnamefont {Gresch}},
  \bibinfo {author} {\bibfnamefont {Z.}~\bibnamefont {Wang}}, \bibinfo {author}
  {\bibfnamefont {Q.}~\bibnamefont {Wu}}, \bibinfo {author} {\bibfnamefont
  {M.}~\bibnamefont {Troyer}}, \bibinfo {author} {\bibfnamefont
  {X.}~\bibnamefont {Dai}}, \ and\ \bibinfo {author} {\bibfnamefont {B.~A.}\
  \bibnamefont {Bernevig}},\ }\bibfield  {title} {\emph {\enquote {\bibinfo
  {title} {Type-II Weyl semimetals},}\ }}\href {\doibase 10.1038/nature15768}
  {\bibfield  {journal} {\bibinfo  {journal} {Nature}\ }\textbf {\bibinfo
  {volume} {527}},\ \bibinfo {pages} {495} (\bibinfo {year}
  {2015})}\BibitemShut {NoStop}%
\bibitem [{\citenamefont {Wang}\ \emph
  {et~al.}(2016{\natexlab{a}})\citenamefont {Wang}, \citenamefont {Gresch},
  \citenamefont {Soluyanov}, \citenamefont {Xie}, \citenamefont {Kushwaha},
  \citenamefont {Dai}, \citenamefont {Troyer}, \citenamefont {Cava},\ and\
  \citenamefont {Bernevig}}]{Wang2016a}%
  \BibitemOpen
  \bibfield  {author} {\bibinfo {author} {\bibfnamefont {Z.}~\bibnamefont
  {Wang}}, \bibinfo {author} {\bibfnamefont {D.}~\bibnamefont {Gresch}},
  \bibinfo {author} {\bibfnamefont {A.~A.}\ \bibnamefont {Soluyanov}}, \bibinfo
  {author} {\bibfnamefont {W.}~\bibnamefont {Xie}}, \bibinfo {author}
  {\bibfnamefont {S.}~\bibnamefont {Kushwaha}}, \bibinfo {author}
  {\bibfnamefont {X.}~\bibnamefont {Dai}}, \bibinfo {author} {\bibfnamefont
  {M.}~\bibnamefont {Troyer}}, \bibinfo {author} {\bibfnamefont {R.~J.}\
  \bibnamefont {Cava}}, \ and\ \bibinfo {author} {\bibfnamefont {B.~A.}\
  \bibnamefont {Bernevig}},\ }\bibfield  {title} {\emph {\enquote {\bibinfo
  {title} {${\mathrm{MoTe}}_{2}$: A Type-II Weyl Topological Metal},}\ }}\href
  {\doibase 10.1103/PhysRevLett.117.056805} {\bibfield  {journal} {\bibinfo
  {journal} {Phys. Rev. Lett.}\ }\textbf {\bibinfo {volume} {117}},\ \bibinfo
  {pages} {056805} (\bibinfo {year} {2016}{\natexlab{a}})}\BibitemShut
  {NoStop}%
\bibitem [{\citenamefont {Wang}\ \emph
  {et~al.}(2016{\natexlab{b}})\citenamefont {Wang}, \citenamefont {Li},
  \citenamefont {Yu},\ and\ \citenamefont {Liao}}]{Wang2016b}%
  \BibitemOpen
  \bibfield  {author} {\bibinfo {author} {\bibfnamefont {L.-X.}\ \bibnamefont
  {Wang}}, \bibinfo {author} {\bibfnamefont {C.-Z.}\ \bibnamefont {Li}},
  \bibinfo {author} {\bibfnamefont {D.-P.}\ \bibnamefont {Yu}}, \ and\ \bibinfo
  {author} {\bibfnamefont {Z.-M.}\ \bibnamefont {Liao}},\ }\bibfield  {title}
  {\emph {\enquote {\bibinfo {title} {Aharonov--Bohm oscillations in Dirac
  semimetal Cd3As2 nanowires},}\ }}\href {\doibase 10.1038/ncomms10769}
  {\bibfield  {journal} {\bibinfo  {journal} {Nature Commun}\ }\textbf
  {\bibinfo {volume} {7}},\ \bibinfo {pages} {10769} (\bibinfo {year}
  {2016}{\natexlab{b}})}\BibitemShut {NoStop}%
\bibitem [{\citenamefont {Moll}\ \emph {et~al.}(2016)\citenamefont {Moll},
  \citenamefont {Nair}, \citenamefont {Helm}, \citenamefont {Potter},
  \citenamefont {Kimchi}, \citenamefont {Vishwanath},\ and\ \citenamefont
  {Analytis}}]{Moll2016}%
  \BibitemOpen
  \bibfield  {author} {\bibinfo {author} {\bibfnamefont {P.~J.~W.}\
  \bibnamefont {Moll}}, \bibinfo {author} {\bibfnamefont {N.~L.}\ \bibnamefont
  {Nair}}, \bibinfo {author} {\bibfnamefont {T.}~\bibnamefont {Helm}}, \bibinfo
  {author} {\bibfnamefont {A.~C.}\ \bibnamefont {Potter}}, \bibinfo {author}
  {\bibfnamefont {I.}~\bibnamefont {Kimchi}}, \bibinfo {author} {\bibfnamefont
  {A.}~\bibnamefont {Vishwanath}}, \ and\ \bibinfo {author} {\bibfnamefont
  {J.~G.}\ \bibnamefont {Analytis}},\ }\bibfield  {title} {\emph {\enquote
  {\bibinfo {title} {Transport evidence for Fermi-arc-mediated chirality
  transfer in the Dirac semimetal Cd3As2},}\ }}\href {\doibase
  10.1038/nature18276} {\bibfield  {journal} {\bibinfo  {journal} {Nature}\
  }\textbf {\bibinfo {volume} {535}},\ \bibinfo {pages} {266} (\bibinfo {year}
  {2016})}\BibitemShut {NoStop}%
\bibitem [{\citenamefont {Xu}\ \emph {et~al.}(2016)\citenamefont {Xu},
  \citenamefont {Belopolski}, \citenamefont {Sanchez}, \citenamefont {Neupane},
  \citenamefont {Chang}, \citenamefont {Yaji}, \citenamefont {Yuan},
  \citenamefont {Zhang}, \citenamefont {Kuroda}, \citenamefont {Bian},
  \citenamefont {Guo}, \citenamefont {Lu}, \citenamefont {Chang}, \citenamefont
  {Alidoust}, \citenamefont {Zheng}, \citenamefont {Lee}, \citenamefont
  {Huang}, \citenamefont {Hsu}, \citenamefont {Jeng}, \citenamefont {Bansil},
  \citenamefont {Neupert}, \citenamefont {Komori}, \citenamefont {Kondo},
  \citenamefont {Shin}, \citenamefont {Lin}, \citenamefont {Jia},\ and\
  \citenamefont {Hasan}}]{Xu2016b}%
  \BibitemOpen
  \bibfield  {author} {\bibinfo {author} {\bibfnamefont {S.-Y.}\ \bibnamefont
  {Xu}}, \bibinfo {author} {\bibfnamefont {I.}~\bibnamefont {Belopolski}},
  \bibinfo {author} {\bibfnamefont {D.~S.}\ \bibnamefont {Sanchez}}, \bibinfo
  {author} {\bibfnamefont {M.}~\bibnamefont {Neupane}}, \bibinfo {author}
  {\bibfnamefont {G.}~\bibnamefont {Chang}}, \bibinfo {author} {\bibfnamefont
  {K.}~\bibnamefont {Yaji}}, \bibinfo {author} {\bibfnamefont {Z.}~\bibnamefont
  {Yuan}}, \bibinfo {author} {\bibfnamefont {C.}~\bibnamefont {Zhang}},
  \bibinfo {author} {\bibfnamefont {K.}~\bibnamefont {Kuroda}}, \bibinfo
  {author} {\bibfnamefont {G.}~\bibnamefont {Bian}}, \bibinfo {author}
  {\bibfnamefont {C.}~\bibnamefont {Guo}}, \bibinfo {author} {\bibfnamefont
  {H.}~\bibnamefont {Lu}}, \bibinfo {author} {\bibfnamefont {T.-R.}\
  \bibnamefont {Chang}}, \bibinfo {author} {\bibfnamefont {N.}~\bibnamefont
  {Alidoust}}, \bibinfo {author} {\bibfnamefont {H.}~\bibnamefont {Zheng}},
  \bibinfo {author} {\bibfnamefont {C.-C.}\ \bibnamefont {Lee}}, \bibinfo
  {author} {\bibfnamefont {S.-M.}\ \bibnamefont {Huang}}, \bibinfo {author}
  {\bibfnamefont {C.-H.}\ \bibnamefont {Hsu}}, \bibinfo {author} {\bibfnamefont
  {H.-T.}\ \bibnamefont {Jeng}}, \bibinfo {author} {\bibfnamefont
  {A.}~\bibnamefont {Bansil}}, \bibinfo {author} {\bibfnamefont
  {T.}~\bibnamefont {Neupert}}, \bibinfo {author} {\bibfnamefont
  {F.}~\bibnamefont {Komori}}, \bibinfo {author} {\bibfnamefont
  {T.}~\bibnamefont {Kondo}}, \bibinfo {author} {\bibfnamefont
  {S.}~\bibnamefont {Shin}}, \bibinfo {author} {\bibfnamefont {H.}~\bibnamefont
  {Lin}}, \bibinfo {author} {\bibfnamefont {S.}~\bibnamefont {Jia}}, \ and\
  \bibinfo {author} {\bibfnamefont {M.~Z.}\ \bibnamefont {Hasan}},\ }\bibfield
  {title} {\emph {\enquote {\bibinfo {title} {Spin Polarization and Texture of
  the Fermi Arcs in the Weyl Fermion Semimetal TaAs},}\ }}\href {\doibase
  10.1103/PhysRevLett.116.096801} {\bibfield  {journal} {\bibinfo  {journal}
  {Phys. Rev. Lett.}\ }\textbf {\bibinfo {volume} {116}},\ \bibinfo {pages}
  {096801} (\bibinfo {year} {2016})}\BibitemShut {NoStop}%
\bibitem [{\citenamefont {Bian}\ \emph
  {et~al.}(2016{\natexlab{b}})\citenamefont {Bian}, \citenamefont {Chang},
  \citenamefont {Zheng}, \citenamefont {Velury}, \citenamefont {Xu},
  \citenamefont {Neupert}, \citenamefont {Chiu}, \citenamefont {Huang},
  \citenamefont {Sanchez}, \citenamefont {Belopolski}, \citenamefont
  {Alidoust}, \citenamefont {Chen}, \citenamefont {Chang}, \citenamefont
  {Bansil}, \citenamefont {Jeng}, \citenamefont {Lin},\ and\ \citenamefont
  {Hasan}}]{Bian2016_drumhead}%
  \BibitemOpen
  \bibfield  {author} {\bibinfo {author} {\bibfnamefont {G.}~\bibnamefont
  {Bian}}, \bibinfo {author} {\bibfnamefont {T.-R.}\ \bibnamefont {Chang}},
  \bibinfo {author} {\bibfnamefont {H.}~\bibnamefont {Zheng}}, \bibinfo
  {author} {\bibfnamefont {S.}~\bibnamefont {Velury}}, \bibinfo {author}
  {\bibfnamefont {S.-Y.}\ \bibnamefont {Xu}}, \bibinfo {author} {\bibfnamefont
  {T.}~\bibnamefont {Neupert}}, \bibinfo {author} {\bibfnamefont {C.-K.}\
  \bibnamefont {Chiu}}, \bibinfo {author} {\bibfnamefont {S.-M.}\ \bibnamefont
  {Huang}}, \bibinfo {author} {\bibfnamefont {D.~S.}\ \bibnamefont {Sanchez}},
  \bibinfo {author} {\bibfnamefont {I.}~\bibnamefont {Belopolski}}, \bibinfo
  {author} {\bibfnamefont {N.}~\bibnamefont {Alidoust}}, \bibinfo {author}
  {\bibfnamefont {P.-J.}\ \bibnamefont {Chen}}, \bibinfo {author}
  {\bibfnamefont {G.}~\bibnamefont {Chang}}, \bibinfo {author} {\bibfnamefont
  {A.}~\bibnamefont {Bansil}}, \bibinfo {author} {\bibfnamefont {H.-T.}\
  \bibnamefont {Jeng}}, \bibinfo {author} {\bibfnamefont {H.}~\bibnamefont
  {Lin}}, \ and\ \bibinfo {author} {\bibfnamefont {M.~Z.}\ \bibnamefont
  {Hasan}},\ }\bibfield  {title} {\emph {\enquote {\bibinfo {title} {Drumhead
  surface states and topological nodal-line fermions in
  ${\mathrm{TlTaSe}}_{2}$},}\ }}\href {\doibase 10.1103/PhysRevB.93.121113}
  {\bibfield  {journal} {\bibinfo  {journal} {Phys. Rev. B}\ }\textbf {\bibinfo
  {volume} {93}},\ \bibinfo {pages} {121113} (\bibinfo {year}
  {2016}{\natexlab{b}})}\BibitemShut {NoStop}%
\bibitem [{\citenamefont {Weng}\ \emph {et~al.}(2015)\citenamefont {Weng},
  \citenamefont {Liang}, \citenamefont {Xu}, \citenamefont {Yu}, \citenamefont
  {Fang}, \citenamefont {Dai},\ and\ \citenamefont
  {Kawazoe}}]{Weng2015_drumhead}%
  \BibitemOpen
  \bibfield  {author} {\bibinfo {author} {\bibfnamefont {H.}~\bibnamefont
  {Weng}}, \bibinfo {author} {\bibfnamefont {Y.}~\bibnamefont {Liang}},
  \bibinfo {author} {\bibfnamefont {Q.}~\bibnamefont {Xu}}, \bibinfo {author}
  {\bibfnamefont {R.}~\bibnamefont {Yu}}, \bibinfo {author} {\bibfnamefont
  {Z.}~\bibnamefont {Fang}}, \bibinfo {author} {\bibfnamefont {X.}~\bibnamefont
  {Dai}}, \ and\ \bibinfo {author} {\bibfnamefont {Y.}~\bibnamefont
  {Kawazoe}},\ }\bibfield  {title} {\emph {\enquote {\bibinfo {title}
  {Topological node-line semimetal in three-dimensional graphene networks},}\
  }}\href {\doibase 10.1103/PhysRevB.92.045108} {\bibfield  {journal} {\bibinfo
   {journal} {Phys. Rev. B}\ }\textbf {\bibinfo {volume} {92}},\ \bibinfo
  {pages} {045108} (\bibinfo {year} {2015})}\BibitemShut {NoStop}%
\bibitem [{\citenamefont {Abdulla}\ \emph
  {et~al.}(2023{\natexlab{a}})\citenamefont {Abdulla}, \citenamefont {Murthy},\
  and\ \citenamefont {Das}}]{Abdulla2023_NLSM1}%
  \BibitemOpen
  \bibfield  {author} {\bibinfo {author} {\bibfnamefont {F.}~\bibnamefont
  {Abdulla}}, \bibinfo {author} {\bibfnamefont {G.}~\bibnamefont {Murthy}}, \
  and\ \bibinfo {author} {\bibfnamefont {A.}~\bibnamefont {Das}},\ }\href
  {https://arxiv.org/abs/2311.18667} {\enquote {\bibinfo {title} {Topological
  nodal line semimetals with chiral symmetry},}\ } (\bibinfo {year}
  {2023}{\natexlab{a}}),\ \Eprint
  {http://arxiv.org/abs/2311.18667}{arXiv:2311.18667}\BibitemShut {NoStop}%
\bibitem [{\citenamefont {Abdulla}\ \emph
  {et~al.}(2023{\natexlab{b}})\citenamefont {Abdulla}, \citenamefont {Murthy},\
  and\ \citenamefont {Das}}]{Abdulla2023_NLSM2}%
  \BibitemOpen
  \bibfield  {author} {\bibinfo {author} {\bibfnamefont {F.}~\bibnamefont
  {Abdulla}}, \bibinfo {author} {\bibfnamefont {G.}~\bibnamefont {Murthy}}, \
  and\ \bibinfo {author} {\bibfnamefont {A.}~\bibnamefont {Das}},\ }\href
  {https://arxiv.org/abs/2401.02966} {\enquote {\bibinfo {title} {Stable nodal
  line semimetals in the chiral classes in three dimensions},}\ } (\bibinfo
  {year} {2023}{\natexlab{b}}),\ \Eprint
  {http://arxiv.org/abs/2401.02966}{arXiv:2401.02966}\BibitemShut {NoStop}%
\bibitem [{\citenamefont {Benalcazar}\ \emph
  {et~al.}(2017{\natexlab{a}})\citenamefont {Benalcazar}, \citenamefont
  {Bernevig},\ and\ \citenamefont {Hughes}}]{benalcazar2017}%
  \BibitemOpen
  \bibfield  {author} {\bibinfo {author} {\bibfnamefont {W.~A.}\ \bibnamefont
  {Benalcazar}}, \bibinfo {author} {\bibfnamefont {B.~A.}\ \bibnamefont
  {Bernevig}}, \ and\ \bibinfo {author} {\bibfnamefont {T.~L.}\ \bibnamefont
  {Hughes}},\ }\bibfield  {title} {\emph {\enquote {\bibinfo {title}
  {{Quantized electric multipole insulators}},}\ }}\href {\doibase
  https://doi.org/10.1126/science.aah6442} {\bibfield  {journal} {\bibinfo
  {journal} {Science}\ }\textbf {\bibinfo {volume} {357}},\ \bibinfo {pages}
  {61} (\bibinfo {year} {2017}{\natexlab{a}})}\BibitemShut {NoStop}%
\bibitem [{\citenamefont {Benalcazar}\ \emph
  {et~al.}(2017{\natexlab{b}})\citenamefont {Benalcazar}, \citenamefont
  {Bernevig},\ and\ \citenamefont {Hughes}}]{benalcazarprb2017}%
  \BibitemOpen
  \bibfield  {author} {\bibinfo {author} {\bibfnamefont {W.~A.}\ \bibnamefont
  {Benalcazar}}, \bibinfo {author} {\bibfnamefont {B.~A.}\ \bibnamefont
  {Bernevig}}, \ and\ \bibinfo {author} {\bibfnamefont {T.~L.}\ \bibnamefont
  {Hughes}},\ }\bibfield  {title} {\emph {\enquote {\bibinfo {title} {{Electric
  multipole moments, topological multipole moment pumping, and chiral hinge
  states in crystalline insulators}},}\ }}\href {\doibase
  10.1103/PhysRevB.96.245115} {\bibfield  {journal} {\bibinfo  {journal} {Phys.
  Rev. B}\ }\textbf {\bibinfo {volume} {96}},\ \bibinfo {pages} {245115}
  (\bibinfo {year} {2017}{\natexlab{b}})}\BibitemShut {NoStop}%
\bibitem [{\citenamefont {Song}\ \emph {et~al.}(2017)\citenamefont {Song},
  \citenamefont {Fang},\ and\ \citenamefont {Fang}}]{Song2017}%
  \BibitemOpen
  \bibfield  {author} {\bibinfo {author} {\bibfnamefont {Z.}~\bibnamefont
  {Song}}, \bibinfo {author} {\bibfnamefont {Z.}~\bibnamefont {Fang}}, \ and\
  \bibinfo {author} {\bibfnamefont {C.}~\bibnamefont {Fang}},\ }\bibfield
  {title} {\emph {\enquote {\bibinfo {title} {{$(d\ensuremath{-}2)$-Dimensional
  Edge States of Rotation Symmetry Protected Topological States}},}\ }}\href
  {\doibase 10.1103/PhysRevLett.119.246402} {\bibfield  {journal} {\bibinfo
  {journal} {Phys. Rev. Lett.}\ }\textbf {\bibinfo {volume} {119}},\ \bibinfo
  {pages} {246402} (\bibinfo {year} {2017})}\BibitemShut {NoStop}%
\bibitem [{\citenamefont {Schindler}\ \emph {et~al.}(2018)\citenamefont
  {Schindler}, \citenamefont {Cook}, \citenamefont {Vergniory}, \citenamefont
  {Wang}, \citenamefont {Parkin}, \citenamefont {Bernevig},\ and\ \citenamefont
  {Neupert}}]{schindler2018}%
  \BibitemOpen
  \bibfield  {author} {\bibinfo {author} {\bibfnamefont {F.}~\bibnamefont
  {Schindler}}, \bibinfo {author} {\bibfnamefont {A.~M.}\ \bibnamefont {Cook}},
  \bibinfo {author} {\bibfnamefont {M.~G.}\ \bibnamefont {Vergniory}}, \bibinfo
  {author} {\bibfnamefont {Z.}~\bibnamefont {Wang}}, \bibinfo {author}
  {\bibfnamefont {S.~S.}\ \bibnamefont {Parkin}}, \bibinfo {author}
  {\bibfnamefont {B.~A.}\ \bibnamefont {Bernevig}}, \ and\ \bibinfo {author}
  {\bibfnamefont {T.}~\bibnamefont {Neupert}},\ }\bibfield  {title} {\emph
  {\enquote {\bibinfo {title} {{Higher-order topological insulators}},}\
  }}\href {\doibase https://doi.org/10.1126/sciadv.aat0346} {\bibfield
  {journal} {\bibinfo  {journal} {Science adv.}\ }\textbf {\bibinfo {volume}
  {4}},\ \bibinfo {pages} {eaat0346} (\bibinfo {year} {2018})}\BibitemShut
  {NoStop}%
\bibitem [{\citenamefont {Manna}\ \emph {et~al.}(2022)\citenamefont {Manna},
  \citenamefont {Nandy},\ and\ \citenamefont {Roy}}]{MannaPRBL2022}%
  \BibitemOpen
  \bibfield  {author} {\bibinfo {author} {\bibfnamefont {S.}~\bibnamefont
  {Manna}}, \bibinfo {author} {\bibfnamefont {S.}~\bibnamefont {Nandy}}, \ and\
  \bibinfo {author} {\bibfnamefont {B.}~\bibnamefont {Roy}},\ }\bibfield
  {title} {\emph {\enquote {\bibinfo {title} {Higher-order topological phases
  on fractal lattices},}\ }}\href {\doibase 10.1103/PhysRevB.105.L201301}
  {\bibfield  {journal} {\bibinfo  {journal} {Phys. Rev. B}\ }\textbf {\bibinfo
  {volume} {105}},\ \bibinfo {pages} {L201301} (\bibinfo {year}
  {2022})}\BibitemShut {NoStop}%
\bibitem [{\citenamefont {Ghosh}\ \emph {et~al.}(2023)\citenamefont {Ghosh},
  \citenamefont {Nag},\ and\ \citenamefont {Saha}}]{Ghosh_2024}%
  \BibitemOpen
  \bibfield  {author} {\bibinfo {author} {\bibfnamefont {A.~K.}\ \bibnamefont
  {Ghosh}}, \bibinfo {author} {\bibfnamefont {T.}~\bibnamefont {Nag}}, \ and\
  \bibinfo {author} {\bibfnamefont {A.}~\bibnamefont {Saha}},\ }\bibfield
  {title} {\emph {\enquote {\bibinfo {title} {Generation of higher-order
  topological insulators using periodic driving},}\ }}\href {\doibase
  10.1088/1361-648X/ad0e2d} {\bibfield  {journal} {\bibinfo  {journal} {J.
  Phys. Condens. Matter}\ }\textbf {\bibinfo {volume} {36}},\ \bibinfo {pages}
  {093001} (\bibinfo {year} {2023})}\BibitemShut {NoStop}%
\bibitem [{\citenamefont {Kane}\ and\ \citenamefont
  {Mele}(2005)}]{kane2005quantum}%
  \BibitemOpen
  \bibfield  {author} {\bibinfo {author} {\bibfnamefont {C.~L.}\ \bibnamefont
  {Kane}}\ and\ \bibinfo {author} {\bibfnamefont {E.~J.}\ \bibnamefont
  {Mele}},\ }\bibfield  {title} {\emph {\enquote {\bibinfo {title} {{Quantum
  Spin Hall Effect in Graphene}},}\ }}\href {\doibase
  10.1103/PhysRevLett.95.226801} {\bibfield  {journal} {\bibinfo  {journal}
  {Phys. Rev. Lett.}\ }\textbf {\bibinfo {volume} {95}},\ \bibinfo {pages}
  {226801} (\bibinfo {year} {2005})}\BibitemShut {NoStop}%
\bibitem [{\citenamefont {Bernevig}\ and\ \citenamefont
  {Zhang}(2006)}]{BHZPRL2006}%
  \BibitemOpen
  \bibfield  {author} {\bibinfo {author} {\bibfnamefont {B.~A.}\ \bibnamefont
  {Bernevig}}\ and\ \bibinfo {author} {\bibfnamefont {S.-C.}\ \bibnamefont
  {Zhang}},\ }\bibfield  {title} {\emph {\enquote {\bibinfo {title} {{Quantum
  Spin Hall Effect}},}\ }}\href {\doibase 10.1103/PhysRevLett.96.106802}
  {\bibfield  {journal} {\bibinfo  {journal} {Phys. Rev. Lett.}\ }\textbf
  {\bibinfo {volume} {96}},\ \bibinfo {pages} {106802} (\bibinfo {year}
  {2006})}\BibitemShut {NoStop}%
\bibitem [{\citenamefont {Bernevig}\ \emph {et~al.}(2006)\citenamefont
  {Bernevig}, \citenamefont {Hughes},\ and\ \citenamefont
  {Zhang}}]{bernevig2006quantum}%
  \BibitemOpen
  \bibfield  {author} {\bibinfo {author} {\bibfnamefont {B.~A.}\ \bibnamefont
  {Bernevig}}, \bibinfo {author} {\bibfnamefont {T.~L.}\ \bibnamefont
  {Hughes}}, \ and\ \bibinfo {author} {\bibfnamefont {S.-C.}\ \bibnamefont
  {Zhang}},\ }\bibfield  {title} {\emph {\enquote {\bibinfo {title} {{Quantum
  spin Hall effect and topological phase transition in ${\rm HgTe}$ quantum
  wells}},}\ }}\href {\doibase https://doi.org/10.1126/science.1133734}
  {\bibfield  {journal} {\bibinfo  {journal} {Science}\ }\textbf {\bibinfo
  {volume} {314}},\ \bibinfo {pages} {1757} (\bibinfo {year}
  {2006})}\BibitemShut {NoStop}%
\bibitem [{\citenamefont {Lin}\ and\ \citenamefont {Hughes}(2018)}]{Lin2018}%
  \BibitemOpen
  \bibfield  {author} {\bibinfo {author} {\bibfnamefont {M.}~\bibnamefont
  {Lin}}\ and\ \bibinfo {author} {\bibfnamefont {T.~L.}\ \bibnamefont
  {Hughes}},\ }\bibfield  {title} {\emph {\enquote {\bibinfo {title}
  {Topological quadrupolar semimetals},}\ }}\href {\doibase
  10.1103/PhysRevB.98.241103} {\bibfield  {journal} {\bibinfo  {journal} {Phys.
  Rev. B}\ }\textbf {\bibinfo {volume} {98}},\ \bibinfo {pages} {241103}
  (\bibinfo {year} {2018})}\BibitemShut {NoStop}%
\bibitem [{\citenamefont {Wieder}\ \emph {et~al.}(2020)\citenamefont {Wieder},
  \citenamefont {Wang}, \citenamefont {Cano}, \citenamefont {Dai},
  \citenamefont {Schoop}, \citenamefont {Bradlyn},\ and\ \citenamefont
  {Bernevig}}]{Bernevig2020NatComm}%
  \BibitemOpen
  \bibfield  {author} {\bibinfo {author} {\bibfnamefont {B.~J.}\ \bibnamefont
  {Wieder}}, \bibinfo {author} {\bibfnamefont {Z.}~\bibnamefont {Wang}},
  \bibinfo {author} {\bibfnamefont {J.}~\bibnamefont {Cano}}, \bibinfo {author}
  {\bibfnamefont {X.}~\bibnamefont {Dai}}, \bibinfo {author} {\bibfnamefont
  {L.~M.}\ \bibnamefont {Schoop}}, \bibinfo {author} {\bibfnamefont
  {B.}~\bibnamefont {Bradlyn}}, \ and\ \bibinfo {author} {\bibfnamefont
  {B.~A.}\ \bibnamefont {Bernevig}},\ }\bibfield  {title} {\emph {\enquote
  {\bibinfo {title} {Strong and fragile topological Dirac semimetals with
  higher-order Fermi arcs},}\ }}\href {\doibase 10.1038/s41467-020-14443-5}
  {\bibfield  {journal} {\bibinfo  {journal} {Nature Commun}\ }\textbf
  {\bibinfo {volume} {11}},\ \bibinfo {pages} {627} (\bibinfo {year}
  {2020})}\BibitemShut {NoStop}%
\bibitem [{\citenamefont {Wu}\ \emph {et~al.}(2020)\citenamefont {Wu},
  \citenamefont {Yu}, \citenamefont {Zhou}, \citenamefont {Zhao},\ and\
  \citenamefont {Yang}}]{Wu2020}%
  \BibitemOpen
  \bibfield  {author} {\bibinfo {author} {\bibfnamefont {W.}~\bibnamefont
  {Wu}}, \bibinfo {author} {\bibfnamefont {Z.-M.}\ \bibnamefont {Yu}}, \bibinfo
  {author} {\bibfnamefont {X.}~\bibnamefont {Zhou}}, \bibinfo {author}
  {\bibfnamefont {Y.~X.}\ \bibnamefont {Zhao}}, \ and\ \bibinfo {author}
  {\bibfnamefont {S.~A.}\ \bibnamefont {Yang}},\ }\bibfield  {title} {\emph
  {\enquote {\bibinfo {title} {Higher-order Dirac fermions in three
  dimensions},}\ }}\href {\doibase 10.1103/PhysRevB.101.205134} {\bibfield
  {journal} {\bibinfo  {journal} {Phys. Rev. B}\ }\textbf {\bibinfo {volume}
  {101}},\ \bibinfo {pages} {205134} (\bibinfo {year} {2020})}\BibitemShut
  {NoStop}%
\bibitem [{\citenamefont {Qiu}\ \emph {et~al.}(2021)\citenamefont {Qiu},
  \citenamefont {Xiao}, \citenamefont {Zhang},\ and\ \citenamefont
  {Qiu}}]{Qui2021}%
  \BibitemOpen
  \bibfield  {author} {\bibinfo {author} {\bibfnamefont {H.}~\bibnamefont
  {Qiu}}, \bibinfo {author} {\bibfnamefont {M.}~\bibnamefont {Xiao}}, \bibinfo
  {author} {\bibfnamefont {F.}~\bibnamefont {Zhang}}, \ and\ \bibinfo {author}
  {\bibfnamefont {C.}~\bibnamefont {Qiu}},\ }\bibfield  {title} {\emph
  {\enquote {\bibinfo {title} {Higher-Order Dirac Sonic Crystals},}\ }}\href
  {\doibase 10.1103/PhysRevLett.127.146601} {\bibfield  {journal} {\bibinfo
  {journal} {Phys. Rev. Lett.}\ }\textbf {\bibinfo {volume} {127}},\ \bibinfo
  {pages} {146601} (\bibinfo {year} {2021})}\BibitemShut {NoStop}%
\bibitem [{\citenamefont {Wang}\ \emph {et~al.}(2022)\citenamefont {Wang},
  \citenamefont {Liu}, \citenamefont {Teo}, \citenamefont {Wang}, \citenamefont
  {Xue},\ and\ \citenamefont {Zhang}}]{Wang2022}%
  \BibitemOpen
  \bibfield  {author} {\bibinfo {author} {\bibfnamefont {Z.}~\bibnamefont
  {Wang}}, \bibinfo {author} {\bibfnamefont {D.}~\bibnamefont {Liu}}, \bibinfo
  {author} {\bibfnamefont {H.~T.}\ \bibnamefont {Teo}}, \bibinfo {author}
  {\bibfnamefont {Q.}~\bibnamefont {Wang}}, \bibinfo {author} {\bibfnamefont
  {H.}~\bibnamefont {Xue}}, \ and\ \bibinfo {author} {\bibfnamefont
  {B.}~\bibnamefont {Zhang}},\ }\bibfield  {title} {\emph {\enquote {\bibinfo
  {title} {Higher-order Dirac semimetal in a photonic crystal},}\ }}\href
  {\doibase 10.1103/PhysRevB.105.L060101} {\bibfield  {journal} {\bibinfo
  {journal} {Phys. Rev. B}\ }\textbf {\bibinfo {volume} {105}},\ \bibinfo
  {pages} {L060101} (\bibinfo {year} {2022})}\BibitemShut {NoStop}%
\bibitem [{\citenamefont {Chen}\ \emph {et~al.}(2023)\citenamefont {Chen},
  \citenamefont {Zhou},\ and\ \citenamefont {Xu}}]{ChenQCHODSM2023}%
  \BibitemOpen
  \bibfield  {author} {\bibinfo {author} {\bibfnamefont {R.}~\bibnamefont
  {Chen}}, \bibinfo {author} {\bibfnamefont {B.}~\bibnamefont {Zhou}}, \ and\
  \bibinfo {author} {\bibfnamefont {D.-H.}\ \bibnamefont {Xu}},\ }\bibfield
  {title} {\emph {\enquote {\bibinfo {title} {Quasicrystalline second-order
  topological semimetals},}\ }}\href {\doibase 10.1103/PhysRevB.108.195306}
  {\bibfield  {journal} {\bibinfo  {journal} {Phys. Rev. B}\ }\textbf {\bibinfo
  {volume} {108}},\ \bibinfo {pages} {195306} (\bibinfo {year}
  {2023})}\BibitemShut {NoStop}%
\bibitem [{\citenamefont {Rafi-Ul-Islam}\ \emph {et~al.}(2024)\citenamefont
  {Rafi-Ul-Islam}, \citenamefont {Siu}, \citenamefont {Sahin},\ and\
  \citenamefont {Jalil}}]{rafiulislam2024engineering}%
  \BibitemOpen
  \bibfield  {author} {\bibinfo {author} {\bibfnamefont {S.~M.}\ \bibnamefont
  {Rafi-Ul-Islam}}, \bibinfo {author} {\bibfnamefont {Z.~B.}\ \bibnamefont
  {Siu}}, \bibinfo {author} {\bibfnamefont {H.}~\bibnamefont {Sahin}}, \ and\
  \bibinfo {author} {\bibfnamefont {M.~B.~A.}\ \bibnamefont {Jalil}},\
  }\bibfield  {title} {\emph {\enquote {\bibinfo {title} {Chiral surface and
  hinge states in higher-order Weyl semimetallic circuits},}\ }}\href {\doibase
  10.1103/PhysRevB.109.085430} {\bibfield  {journal} {\bibinfo  {journal}
  {Phys. Rev. B}\ }\textbf {\bibinfo {volume} {109}},\ \bibinfo {pages}
  {085430} (\bibinfo {year} {2024})}\BibitemShut {NoStop}%
\bibitem [{\citenamefont {Ghorashi}\ \emph {et~al.}(2020)\citenamefont
  {Ghorashi}, \citenamefont {Li},\ and\ \citenamefont
  {Hughes}}]{GhorashiPRL2020}%
  \BibitemOpen
  \bibfield  {author} {\bibinfo {author} {\bibfnamefont {S.~A.~A.}\
  \bibnamefont {Ghorashi}}, \bibinfo {author} {\bibfnamefont {T.}~\bibnamefont
  {Li}}, \ and\ \bibinfo {author} {\bibfnamefont {T.~L.}\ \bibnamefont
  {Hughes}},\ }\bibfield  {title} {\emph {\enquote {\bibinfo {title}
  {Higher-Order Weyl Semimetals},}\ }}\href {\doibase
  10.1103/PhysRevLett.125.266804} {\bibfield  {journal} {\bibinfo  {journal}
  {Phys. Rev. Lett.}\ }\textbf {\bibinfo {volume} {125}},\ \bibinfo {pages}
  {266804} (\bibinfo {year} {2020})}\BibitemShut {NoStop}%
\bibitem [{\citenamefont {Wang}\ \emph
  {et~al.}(2020{\natexlab{a}})\citenamefont {Wang}, \citenamefont {Lin},
  \citenamefont {Jiang}, \citenamefont {Guo},\ and\ \citenamefont
  {Jiang}}]{WangPRL2020}%
  \BibitemOpen
  \bibfield  {author} {\bibinfo {author} {\bibfnamefont {H.-X.}\ \bibnamefont
  {Wang}}, \bibinfo {author} {\bibfnamefont {Z.-K.}\ \bibnamefont {Lin}},
  \bibinfo {author} {\bibfnamefont {B.}~\bibnamefont {Jiang}}, \bibinfo
  {author} {\bibfnamefont {G.-Y.}\ \bibnamefont {Guo}}, \ and\ \bibinfo
  {author} {\bibfnamefont {J.-H.}\ \bibnamefont {Jiang}},\ }\bibfield  {title}
  {\emph {\enquote {\bibinfo {title} {Higher-Order Weyl Semimetals},}\ }}\href
  {\doibase 10.1103/PhysRevLett.125.146401} {\bibfield  {journal} {\bibinfo
  {journal} {Phys. Rev. Lett.}\ }\textbf {\bibinfo {volume} {125}},\ \bibinfo
  {pages} {146401} (\bibinfo {year} {2020}{\natexlab{a}})}\BibitemShut
  {NoStop}%
\bibitem [{\citenamefont {Rui}\ \emph {et~al.}(2022)\citenamefont {Rui},
  \citenamefont {Zheng}, \citenamefont {Hirschmann}, \citenamefont {Zhang},
  \citenamefont {Wang},\ and\ \citenamefont {Wang}}]{Rui2022}%
  \BibitemOpen
  \bibfield  {author} {\bibinfo {author} {\bibfnamefont {W.~B.}\ \bibnamefont
  {Rui}}, \bibinfo {author} {\bibfnamefont {Z.}~\bibnamefont {Zheng}}, \bibinfo
  {author} {\bibfnamefont {M.~M.}\ \bibnamefont {Hirschmann}}, \bibinfo
  {author} {\bibfnamefont {S.-B.}\ \bibnamefont {Zhang}}, \bibinfo {author}
  {\bibfnamefont {C.}~\bibnamefont {Wang}}, \ and\ \bibinfo {author}
  {\bibfnamefont {Z.~D.}\ \bibnamefont {Wang}},\ }\bibfield  {title} {\emph
  {\enquote {\bibinfo {title} {Intertwined Weyl phases emergent from
  higher-order topology and unconventional Weyl fermions via crystalline
  symmetry},}\ }}\href {\doibase 10.1038/s41535-022-00422-0} {\bibfield
  {journal} {\bibinfo  {journal} {npj Quantum Mater.}\ }\textbf {\bibinfo
  {volume} {7}},\ \bibinfo {pages} {15} (\bibinfo {year} {2022})}\BibitemShut
  {NoStop}%
\bibitem [{\citenamefont {Tanaka}\ \emph {et~al.}(2022)\citenamefont {Tanaka},
  \citenamefont {Takahashi}, \citenamefont {Okugawa},\ and\ \citenamefont
  {Murakami}}]{Tanaka2022}%
  \BibitemOpen
  \bibfield  {author} {\bibinfo {author} {\bibfnamefont {Y.}~\bibnamefont
  {Tanaka}}, \bibinfo {author} {\bibfnamefont {R.}~\bibnamefont {Takahashi}},
  \bibinfo {author} {\bibfnamefont {R.}~\bibnamefont {Okugawa}}, \ and\
  \bibinfo {author} {\bibfnamefont {S.}~\bibnamefont {Murakami}},\ }\bibfield
  {title} {\emph {\enquote {\bibinfo {title} {Rotoinversion-symmetric
  bulk-hinge correspondence and its applications to higher-order Weyl
  semimetals},}\ }}\href {\doibase 10.1103/PhysRevB.105.115119} {\bibfield
  {journal} {\bibinfo  {journal} {Phys. Rev. B}\ }\textbf {\bibinfo {volume}
  {105}},\ \bibinfo {pages} {115119} (\bibinfo {year} {2022})}\BibitemShut
  {NoStop}%
\bibitem [{\citenamefont {Wang}\ \emph {et~al.}(2019)\citenamefont {Wang},
  \citenamefont {Wieder}, \citenamefont {Li}, \citenamefont {Yan},\ and\
  \citenamefont {Bernevig}}]{Wang2019}%
  \BibitemOpen
  \bibfield  {author} {\bibinfo {author} {\bibfnamefont {Z.}~\bibnamefont
  {Wang}}, \bibinfo {author} {\bibfnamefont {B.~J.}\ \bibnamefont {Wieder}},
  \bibinfo {author} {\bibfnamefont {J.}~\bibnamefont {Li}}, \bibinfo {author}
  {\bibfnamefont {B.}~\bibnamefont {Yan}}, \ and\ \bibinfo {author}
  {\bibfnamefont {B.~A.}\ \bibnamefont {Bernevig}},\ }\bibfield  {title} {\emph
  {\enquote {\bibinfo {title} {Higher-Order Topology, Monopole Nodal Lines, and
  the Origin of Large Fermi Arcs in Transition Metal Dichalcogenides
  $X{\mathrm{Te}}_{2}$ ($X=\mathrm{Mo},\mathrm{W}$)},}\ }}\href {\doibase
  10.1103/PhysRevLett.123.186401} {\bibfield  {journal} {\bibinfo  {journal}
  {Phys. Rev. Lett.}\ }\textbf {\bibinfo {volume} {123}},\ \bibinfo {pages}
  {186401} (\bibinfo {year} {2019})}\BibitemShut {NoStop}%
\bibitem [{\citenamefont {Wang}\ \emph
  {et~al.}(2020{\natexlab{b}})\citenamefont {Wang}, \citenamefont {Dai},
  \citenamefont {Shao}, \citenamefont {Yang},\ and\ \citenamefont
  {Zhao}}]{Wang2020}%
  \BibitemOpen
  \bibfield  {author} {\bibinfo {author} {\bibfnamefont {K.}~\bibnamefont
  {Wang}}, \bibinfo {author} {\bibfnamefont {J.-X.}\ \bibnamefont {Dai}},
  \bibinfo {author} {\bibfnamefont {L.~B.}\ \bibnamefont {Shao}}, \bibinfo
  {author} {\bibfnamefont {S.~A.}\ \bibnamefont {Yang}}, \ and\ \bibinfo
  {author} {\bibfnamefont {Y.~X.}\ \bibnamefont {Zhao}},\ }\bibfield  {title}
  {\emph {\enquote {\bibinfo {title} {Boundary Criticality of
  $\mathcal{PT}$-Invariant Topology and Second-Order Nodal-Line Semimetals},}\
  }}\href {\doibase 10.1103/PhysRevLett.125.126403} {\bibfield  {journal}
  {\bibinfo  {journal} {Phys. Rev. Lett.}\ }\textbf {\bibinfo {volume} {125}},\
  \bibinfo {pages} {126403} (\bibinfo {year} {2020}{\natexlab{b}})}\BibitemShut
  {NoStop}%
\bibitem [{\citenamefont {Chen}\ \emph {et~al.}(2022)\citenamefont {Chen},
  \citenamefont {Zeng}, \citenamefont {Chen}, \citenamefont {Zhao},
  \citenamefont {Sheng},\ and\ \citenamefont {Yang}}]{Chen2022}%
  \BibitemOpen
  \bibfield  {author} {\bibinfo {author} {\bibfnamefont {C.}~\bibnamefont
  {Chen}}, \bibinfo {author} {\bibfnamefont {X.-T.}\ \bibnamefont {Zeng}},
  \bibinfo {author} {\bibfnamefont {Z.}~\bibnamefont {Chen}}, \bibinfo {author}
  {\bibfnamefont {Y.~X.}\ \bibnamefont {Zhao}}, \bibinfo {author}
  {\bibfnamefont {X.-L.}\ \bibnamefont {Sheng}}, \ and\ \bibinfo {author}
  {\bibfnamefont {S.~A.}\ \bibnamefont {Yang}},\ }\bibfield  {title} {\emph
  {\enquote {\bibinfo {title} {Second-Order Real Nodal-Line Semimetal in
  Three-Dimensional Graphdiyne},}\ }}\href {\doibase
  10.1103/PhysRevLett.128.026405} {\bibfield  {journal} {\bibinfo  {journal}
  {Phys. Rev. Lett.}\ }\textbf {\bibinfo {volume} {128}},\ \bibinfo {pages}
  {026405} (\bibinfo {year} {2022})}\BibitemShut {NoStop}%
\bibitem [{\citenamefont {Qiu}\ \emph {et~al.}(2024)\citenamefont {Qiu},
  \citenamefont {Li}, \citenamefont {Zhang},\ and\ \citenamefont
  {Qiu}}]{Qiu2024_HONLSM}%
  \BibitemOpen
  \bibfield  {author} {\bibinfo {author} {\bibfnamefont {H.}~\bibnamefont
  {Qiu}}, \bibinfo {author} {\bibfnamefont {Y.}~\bibnamefont {Li}}, \bibinfo
  {author} {\bibfnamefont {Q.}~\bibnamefont {Zhang}}, \ and\ \bibinfo {author}
  {\bibfnamefont {C.}~\bibnamefont {Qiu}},\ }\bibfield  {title} {\emph
  {\enquote {\bibinfo {title} {Discovery of Higher-Order Nodal Surface
  Semimetals},}\ }}\href {\doibase 10.1103/PhysRevLett.132.186601} {\bibfield
  {journal} {\bibinfo  {journal} {Phys. Rev. Lett.}\ }\textbf {\bibinfo
  {volume} {132}},\ \bibinfo {pages} {186601} (\bibinfo {year}
  {2024})}\BibitemShut {NoStop}%
\bibitem [{\citenamefont {Ma}\ \emph {et~al.}(2024)\citenamefont {Ma},
  \citenamefont {Pu}, \citenamefont {Ye}, \citenamefont {Lu}, \citenamefont
  {Huang}, \citenamefont {Ke}, \citenamefont {He}, \citenamefont {Deng},\ and\
  \citenamefont {Liu}}]{Ma2024}%
  \BibitemOpen
  \bibfield  {author} {\bibinfo {author} {\bibfnamefont {Q.}~\bibnamefont
  {Ma}}, \bibinfo {author} {\bibfnamefont {Z.}~\bibnamefont {Pu}}, \bibinfo
  {author} {\bibfnamefont {L.}~\bibnamefont {Ye}}, \bibinfo {author}
  {\bibfnamefont {J.}~\bibnamefont {Lu}}, \bibinfo {author} {\bibfnamefont
  {X.}~\bibnamefont {Huang}}, \bibinfo {author} {\bibfnamefont
  {M.}~\bibnamefont {Ke}}, \bibinfo {author} {\bibfnamefont {H.}~\bibnamefont
  {He}}, \bibinfo {author} {\bibfnamefont {W.}~\bibnamefont {Deng}}, \ and\
  \bibinfo {author} {\bibfnamefont {Z.}~\bibnamefont {Liu}},\ }\bibfield
  {title} {\emph {\enquote {\bibinfo {title} {Observation of Higher-Order
  Nodal-Line Semimetal in Phononic Crystals},}\ }}\href {\doibase
  10.1103/PhysRevLett.132.066601} {\bibfield  {journal} {\bibinfo  {journal}
  {Phys. Rev. Lett.}\ }\textbf {\bibinfo {volume} {132}},\ \bibinfo {pages}
  {066601} (\bibinfo {year} {2024})}\BibitemShut {NoStop}%
\bibitem [{\citenamefont {Wang}\ \emph
  {et~al.}(2023{\natexlab{a}})\citenamefont {Wang}, \citenamefont {Wang},
  \citenamefont {Sun}, \citenamefont {Chen},\ and\ \citenamefont
  {Xu}}]{WangPRBL2023}%
  \BibitemOpen
  \bibfield  {author} {\bibinfo {author} {\bibfnamefont {Z.-M.}\ \bibnamefont
  {Wang}}, \bibinfo {author} {\bibfnamefont {R.}~\bibnamefont {Wang}}, \bibinfo
  {author} {\bibfnamefont {J.-H.}\ \bibnamefont {Sun}}, \bibinfo {author}
  {\bibfnamefont {T.-Y.}\ \bibnamefont {Chen}}, \ and\ \bibinfo {author}
  {\bibfnamefont {D.-H.}\ \bibnamefont {Xu}},\ }\bibfield  {title} {\emph
  {\enquote {\bibinfo {title} {Floquet Weyl semimetal phases in
  light-irradiated higher-order topological Dirac semimetals},}\ }}\href
  {\doibase 10.1103/PhysRevB.107.L121407} {\bibfield  {journal} {\bibinfo
  {journal} {Phys. Rev. B}\ }\textbf {\bibinfo {volume} {107}},\ \bibinfo
  {pages} {L121407} (\bibinfo {year} {2023}{\natexlab{a}})}\BibitemShut
  {NoStop}%
\bibitem [{\citenamefont {Benalcazar}\ and\ \citenamefont
  {Cerjan}(2022)}]{Benalcazar2022Nxy}%
  \BibitemOpen
  \bibfield  {author} {\bibinfo {author} {\bibfnamefont {W.~A.}\ \bibnamefont
  {Benalcazar}}\ and\ \bibinfo {author} {\bibfnamefont {A.}~\bibnamefont
  {Cerjan}},\ }\bibfield  {title} {\emph {\enquote {\bibinfo {title}
  {Chiral-Symmetric Higher-Order Topological Phases of Matter},}\ }}\href
  {\doibase 10.1103/PhysRevLett.128.127601} {\bibfield  {journal} {\bibinfo
  {journal} {Phys. Rev. Lett.}\ }\textbf {\bibinfo {volume} {128}},\ \bibinfo
  {pages} {127601} (\bibinfo {year} {2022})}\BibitemShut {NoStop}%
\bibitem [{\citenamefont {Wang}\ \emph
  {et~al.}(2023{\natexlab{b}})\citenamefont {Wang}, \citenamefont {Deng},
  \citenamefont {Ji}, \citenamefont {Oudich}, \citenamefont {Benalcazar},
  \citenamefont {Ma},\ and\ \citenamefont {Jing}}]{WangPRLMCN2023}%
  \BibitemOpen
  \bibfield  {author} {\bibinfo {author} {\bibfnamefont {D.}~\bibnamefont
  {Wang}}, \bibinfo {author} {\bibfnamefont {Y.}~\bibnamefont {Deng}}, \bibinfo
  {author} {\bibfnamefont {J.}~\bibnamefont {Ji}}, \bibinfo {author}
  {\bibfnamefont {M.}~\bibnamefont {Oudich}}, \bibinfo {author} {\bibfnamefont
  {W.~A.}\ \bibnamefont {Benalcazar}}, \bibinfo {author} {\bibfnamefont
  {G.}~\bibnamefont {Ma}}, \ and\ \bibinfo {author} {\bibfnamefont
  {Y.}~\bibnamefont {Jing}},\ }\bibfield  {title} {\emph {\enquote {\bibinfo
  {title} {Realization of a $\mathbb{Z}$-Classified Chiral-Symmetric
  Higher-Order Topological Insulator in a Coupling-Inverted Acoustic
  Crystal},}\ }}\href {\doibase 10.1103/PhysRevLett.131.157201} {\bibfield
  {journal} {\bibinfo  {journal} {Phys. Rev. Lett.}\ }\textbf {\bibinfo
  {volume} {131}},\ \bibinfo {pages} {157201} (\bibinfo {year}
  {2023}{\natexlab{b}})}\BibitemShut {NoStop}%
\bibitem [{\citenamefont {Liu}\ \emph {et~al.}(2023)\citenamefont {Liu},
  \citenamefont {Huang}, \citenamefont {Yan}, \citenamefont {Lu}, \citenamefont
  {Deng},\ and\ \citenamefont {Liu}}]{LiuPRApp2023}%
  \BibitemOpen
  \bibfield  {author} {\bibinfo {author} {\bibfnamefont {H.}~\bibnamefont
  {Liu}}, \bibinfo {author} {\bibfnamefont {X.}~\bibnamefont {Huang}}, \bibinfo
  {author} {\bibfnamefont {M.}~\bibnamefont {Yan}}, \bibinfo {author}
  {\bibfnamefont {J.}~\bibnamefont {Lu}}, \bibinfo {author} {\bibfnamefont
  {W.}~\bibnamefont {Deng}}, \ and\ \bibinfo {author} {\bibfnamefont
  {Z.}~\bibnamefont {Liu}},\ }\bibfield  {title} {\emph {\enquote {\bibinfo
  {title} {Acoustic Topological Metamaterials of Large Winding Number},}\
  }}\href {\doibase 10.1103/PhysRevApplied.19.054028} {\bibfield  {journal}
  {\bibinfo  {journal} {Phys. Rev. Appl.}\ }\textbf {\bibinfo {volume} {19}},\
  \bibinfo {pages} {054028} (\bibinfo {year} {2023})}\BibitemShut {NoStop}%
\bibitem [{\citenamefont {Li}\ \emph {et~al.}(2023{\natexlab{a}})\citenamefont
  {Li}, \citenamefont {Zhang}, \citenamefont {Mei}, \citenamefont {Xie},
  \citenamefont {Lu}, \citenamefont {Ma}, \citenamefont {Xiao},\ and\
  \citenamefont {Jia}}]{LiPRApp2023}%
  \BibitemOpen
  \bibfield  {author} {\bibinfo {author} {\bibfnamefont {Y.}~\bibnamefont
  {Li}}, \bibinfo {author} {\bibfnamefont {J.-H.}\ \bibnamefont {Zhang}},
  \bibinfo {author} {\bibfnamefont {F.}~\bibnamefont {Mei}}, \bibinfo {author}
  {\bibfnamefont {B.}~\bibnamefont {Xie}}, \bibinfo {author} {\bibfnamefont
  {M.-H.}\ \bibnamefont {Lu}}, \bibinfo {author} {\bibfnamefont
  {J.}~\bibnamefont {Ma}}, \bibinfo {author} {\bibfnamefont {L.}~\bibnamefont
  {Xiao}}, \ and\ \bibinfo {author} {\bibfnamefont {S.}~\bibnamefont {Jia}},\
  }\bibfield  {title} {\emph {\enquote {\bibinfo {title} {Large-chiral-number
  corner modes in $\mathbb{Z}$-class higher-order topolectrical circuits},}\
  }}\href {\doibase 10.1103/PhysRevApplied.20.064042} {\bibfield  {journal}
  {\bibinfo  {journal} {Phys. Rev. Appl.}\ }\textbf {\bibinfo {volume} {20}},\
  \bibinfo {pages} {064042} (\bibinfo {year} {2023}{\natexlab{a}})}\BibitemShut
  {NoStop}%
\bibitem [{\citenamefont {Li}\ \emph {et~al.}(2023{\natexlab{b}})\citenamefont
  {Li}, \citenamefont {Qiu}, \citenamefont {Zhang},\ and\ \citenamefont
  {Qiu}}]{LiPRBMCN2023}%
  \BibitemOpen
  \bibfield  {author} {\bibinfo {author} {\bibfnamefont {Y.}~\bibnamefont
  {Li}}, \bibinfo {author} {\bibfnamefont {H.}~\bibnamefont {Qiu}}, \bibinfo
  {author} {\bibfnamefont {Q.}~\bibnamefont {Zhang}}, \ and\ \bibinfo {author}
  {\bibfnamefont {C.}~\bibnamefont {Qiu}},\ }\bibfield  {title} {\emph
  {\enquote {\bibinfo {title} {Acoustic higher-order topological insulators
  protected by multipole chiral numbers},}\ }}\href {\doibase
  10.1103/PhysRevB.108.205135} {\bibfield  {journal} {\bibinfo  {journal}
  {Phys. Rev. B}\ }\textbf {\bibinfo {volume} {108}},\ \bibinfo {pages}
  {205135} (\bibinfo {year} {2023}{\natexlab{b}})}\BibitemShut {NoStop}%
\bibitem [{\citenamefont {Qi}\ \emph {et~al.}(2024)\citenamefont {Qi},
  \citenamefont {He}, \citenamefont {Deng}, \citenamefont {Li},\ and\
  \citenamefont {Wang}}]{Qi2024HOmDSM}%
  \BibitemOpen
  \bibfield  {author} {\bibinfo {author} {\bibfnamefont {Y.}~\bibnamefont
  {Qi}}, \bibinfo {author} {\bibfnamefont {Z.}~\bibnamefont {He}}, \bibinfo
  {author} {\bibfnamefont {K.}~\bibnamefont {Deng}}, \bibinfo {author}
  {\bibfnamefont {J.}~\bibnamefont {Li}}, \ and\ \bibinfo {author}
  {\bibfnamefont {Y.}~\bibnamefont {Wang}},\ }\bibfield  {title} {\emph
  {\enquote {\bibinfo {title} {Multipole higher-order topological
  semimetals},}\ }}\href {\doibase 10.1103/PhysRevB.109.L060101} {\bibfield
  {journal} {\bibinfo  {journal} {Phys. Rev. B}\ }\textbf {\bibinfo {volume}
  {109}},\ \bibinfo {pages} {L060101} (\bibinfo {year} {2024})}\BibitemShut
  {NoStop}%
\bibitem [{\citenamefont {Pal}\ \emph {et~al.}(2024)\citenamefont {Pal},
  \citenamefont {Dutta},\ and\ \citenamefont {Saha}}]{Pal2024}%
  \BibitemOpen
  \bibfield  {author} {\bibinfo {author} {\bibfnamefont {A.}~\bibnamefont
  {Pal}}, \bibinfo {author} {\bibfnamefont {P.}~\bibnamefont {Dutta}}, \ and\
  \bibinfo {author} {\bibfnamefont {A.}~\bibnamefont {Saha}},\ }\bibfield
  {title} {\emph {\enquote {\bibinfo {title} {Fermi arc mediated transport in
  an inversion symmetry broken Weyl semimetal nanowire and its hybrid
  junctions},}\ }}\href {\doibase 10.1103/PhysRevB.109.235419} {\bibfield
  {journal} {\bibinfo  {journal} {Phys. Rev. B}\ }\textbf {\bibinfo {volume}
  {109}},\ \bibinfo {pages} {235419} (\bibinfo {year} {2024})}\BibitemShut
  {NoStop}%
\bibitem [{\citenamefont {C\ifmmode \u{a}\else \u{a}\fi{}lug\ifmmode~\u{a}\else
  \u{a}\fi{}ru}\ \emph {et~al.}(2019)\citenamefont {C\ifmmode \u{a}\else
  \u{a}\fi{}lug\ifmmode~\u{a}\else \u{a}\fi{}ru}, \citenamefont {Juri\ifmmode
  \check{c}\else \v{c}\fi{}i\ifmmode~\acute{c}\else \'{c}\fi{}},\ and\
  \citenamefont {Roy}}]{Roy2019}%
  \BibitemOpen
  \bibfield  {author} {\bibinfo {author} {\bibfnamefont {D.}~\bibnamefont
  {C\ifmmode \u{a}\else \u{a}\fi{}lug\ifmmode~\u{a}\else \u{a}\fi{}ru}},
  \bibinfo {author} {\bibfnamefont {V.}~\bibnamefont {Juri\ifmmode
  \check{c}\else \v{c}\fi{}i\ifmmode~\acute{c}\else \'{c}\fi{}}}, \ and\
  \bibinfo {author} {\bibfnamefont {B.}~\bibnamefont {Roy}},\ }\bibfield
  {title} {\emph {\enquote {\bibinfo {title} {{Higher-order topological phases:
  A general principle of construction}},}\ }}\href {\doibase
  10.1103/PhysRevB.99.041301} {\bibfield  {journal} {\bibinfo  {journal} {Phys.
  Rev. B}\ }\textbf {\bibinfo {volume} {99}},\ \bibinfo {pages} {041301}
  (\bibinfo {year} {2019})}\BibitemShut {NoStop}%
\bibitem [{\citenamefont {Ghosh}\ \emph {et~al.}(2024)\citenamefont {Ghosh},
  \citenamefont {Saha},\ and\ \citenamefont {Nag}}]{ghosh2024NHlongrange}%
  \BibitemOpen
  \bibfield  {author} {\bibinfo {author} {\bibfnamefont {A.~K.}\ \bibnamefont
  {Ghosh}}, \bibinfo {author} {\bibfnamefont {A.}~\bibnamefont {Saha}}, \ and\
  \bibinfo {author} {\bibfnamefont {T.}~\bibnamefont {Nag}},\ }\bibfield
  {title} {\emph {\enquote {\bibinfo {title} {Corner modes in non-Hermitian
  next-nearest-neighbor hopping model},}\ }}\href {\doibase
  10.1103/PhysRevB.110.115403} {\bibfield  {journal} {\bibinfo  {journal}
  {Phys. Rev. B}\ }\textbf {\bibinfo {volume} {110}},\ \bibinfo {pages}
  {115403} (\bibinfo {year} {2024})}\BibitemShut {NoStop}%
\bibitem [{\citenamefont {Lin}\ \emph {et~al.}(2021)\citenamefont {Lin},
  \citenamefont {Ke},\ and\ \citenamefont {Lee}}]{LinChiralWinding2021}%
  \BibitemOpen
  \bibfield  {author} {\bibinfo {author} {\bibfnamefont {L.}~\bibnamefont
  {Lin}}, \bibinfo {author} {\bibfnamefont {Y.}~\bibnamefont {Ke}}, \ and\
  \bibinfo {author} {\bibfnamefont {C.}~\bibnamefont {Lee}},\ }\bibfield
  {title} {\emph {\enquote {\bibinfo {title} {Real-space representation of the
  winding number for a one-dimensional chiral-symmetric topological
  insulator},}\ }}\href {\doibase 10.1103/PhysRevB.103.224208} {\bibfield
  {journal} {\bibinfo  {journal} {Phys. Rev. B}\ }\textbf {\bibinfo {volume}
  {103}},\ \bibinfo {pages} {224208} (\bibinfo {year} {2021})}\BibitemShut
  {NoStop}%
\bibitem [{\citenamefont {Mondal}\ \emph {et~al.}(2023)\citenamefont {Mondal},
  \citenamefont {Ghosh}, \citenamefont {Nag},\ and\ \citenamefont
  {Saha}}]{MondalPRB2023}%
  \BibitemOpen
  \bibfield  {author} {\bibinfo {author} {\bibfnamefont {D.}~\bibnamefont
  {Mondal}}, \bibinfo {author} {\bibfnamefont {A.~K.}\ \bibnamefont {Ghosh}},
  \bibinfo {author} {\bibfnamefont {T.}~\bibnamefont {Nag}}, \ and\ \bibinfo
  {author} {\bibfnamefont {A.}~\bibnamefont {Saha}},\ }\bibfield  {title}
  {\emph {\enquote {\bibinfo {title} {Topological characterization and
  stability of Floquet Majorana modes in Rashba nanowires},}\ }}\href {\doibase
  10.1103/PhysRevB.107.035427} {\bibfield  {journal} {\bibinfo  {journal}
  {Phys. Rev. B}\ }\textbf {\bibinfo {volume} {107}},\ \bibinfo {pages}
  {035427} (\bibinfo {year} {2023})}\BibitemShut {NoStop}%
\bibitem [{\citenamefont {Kaladzhyan}\ and\ \citenamefont
  {Bardarson}(2019)}]{Kaladzhyan2019}%
  \BibitemOpen
  \bibfield  {author} {\bibinfo {author} {\bibfnamefont {V.}~\bibnamefont
  {Kaladzhyan}}\ and\ \bibinfo {author} {\bibfnamefont {J.~H.}\ \bibnamefont
  {Bardarson}},\ }\bibfield  {title} {\emph {\enquote {\bibinfo {title}
  {Quantized Fermi arc mediated transport in Weyl semimetal nanowires},}\
  }}\href {\doibase 10.1103/PhysRevB.100.085424} {\bibfield  {journal}
  {\bibinfo  {journal} {Phys. Rev. B}\ }\textbf {\bibinfo {volume} {100}},\
  \bibinfo {pages} {085424} (\bibinfo {year} {2019})}\BibitemShut {NoStop}%
\bibitem [{\citenamefont {Igarashi}\ and\ \citenamefont
  {Koshino}(2017)}]{Igarashi2017}%
  \BibitemOpen
  \bibfield  {author} {\bibinfo {author} {\bibfnamefont {A.}~\bibnamefont
  {Igarashi}}\ and\ \bibinfo {author} {\bibfnamefont {M.}~\bibnamefont
  {Koshino}},\ }\bibfield  {title} {\emph {\enquote {\bibinfo {title}
  {Magnetotransport in Weyl semimetal nanowires},}\ }}\href {\doibase
  10.1103/PhysRevB.95.195306} {\bibfield  {journal} {\bibinfo  {journal} {Phys.
  Rev. B}\ }\textbf {\bibinfo {volume} {95}},\ \bibinfo {pages} {195306}
  (\bibinfo {year} {2017})}\BibitemShut {NoStop}%
\bibitem [{\citenamefont {Lin}\ \emph {et~al.}(2017)\citenamefont {Lin},
  \citenamefont {Wang}, \citenamefont {Wang}, \citenamefont {Li}, \citenamefont
  {Li}, \citenamefont {Yu},\ and\ \citenamefont {Liao}}]{Lin2017}%
  \BibitemOpen
  \bibfield  {author} {\bibinfo {author} {\bibfnamefont {B.-C.}\ \bibnamefont
  {Lin}}, \bibinfo {author} {\bibfnamefont {S.}~\bibnamefont {Wang}}, \bibinfo
  {author} {\bibfnamefont {L.-X.}\ \bibnamefont {Wang}}, \bibinfo {author}
  {\bibfnamefont {C.-Z.}\ \bibnamefont {Li}}, \bibinfo {author} {\bibfnamefont
  {J.-G.}\ \bibnamefont {Li}}, \bibinfo {author} {\bibfnamefont
  {D.}~\bibnamefont {Yu}}, \ and\ \bibinfo {author} {\bibfnamefont {Z.-M.}\
  \bibnamefont {Liao}},\ }\bibfield  {title} {\emph {\enquote {\bibinfo {title}
  {Gate-tuned Aharonov-Bohm interference of surface states in a quasiballistic
  Dirac semimetal nanowire},}\ }}\href {\doibase 10.1103/PhysRevB.95.235436}
  {\bibfield  {journal} {\bibinfo  {journal} {Phys. Rev. B}\ }\textbf {\bibinfo
  {volume} {95}},\ \bibinfo {pages} {235436} (\bibinfo {year}
  {2017})}\BibitemShut {NoStop}%
\bibitem [{\citenamefont {Wang}\ \emph {et~al.}(2018)\citenamefont {Wang},
  \citenamefont {Lin}, \citenamefont {Zheng}, \citenamefont {Yu},\ and\
  \citenamefont {Liao}}]{Wang2018a}%
  \BibitemOpen
  \bibfield  {author} {\bibinfo {author} {\bibfnamefont {S.}~\bibnamefont
  {Wang}}, \bibinfo {author} {\bibfnamefont {B.-C.}\ \bibnamefont {Lin}},
  \bibinfo {author} {\bibfnamefont {W.-Z.}\ \bibnamefont {Zheng}}, \bibinfo
  {author} {\bibfnamefont {D.}~\bibnamefont {Yu}}, \ and\ \bibinfo {author}
  {\bibfnamefont {Z.-M.}\ \bibnamefont {Liao}},\ }\bibfield  {title} {\emph
  {\enquote {\bibinfo {title} {Fano Interference between Bulk and Surface
  States of a Dirac Semimetal ${\mathrm{Cd}}_{3}{\mathrm{As}}_{2}$ Nanowire},}\
  }}\href {\doibase 10.1103/PhysRevLett.120.257701} {\bibfield  {journal}
  {\bibinfo  {journal} {Phys. Rev. Lett.}\ }\textbf {\bibinfo {volume} {120}},\
  \bibinfo {pages} {257701} (\bibinfo {year} {2018})}\BibitemShut {NoStop}%
\bibitem [{\citenamefont {Gorbar}\ \emph {et~al.}(2016)\citenamefont {Gorbar},
  \citenamefont {Miransky}, \citenamefont {Shovkovy},\ and\ \citenamefont
  {Sukhachov}}]{Gorbar2016}%
  \BibitemOpen
  \bibfield  {author} {\bibinfo {author} {\bibfnamefont {E.~V.}\ \bibnamefont
  {Gorbar}}, \bibinfo {author} {\bibfnamefont {V.~A.}\ \bibnamefont
  {Miransky}}, \bibinfo {author} {\bibfnamefont {I.~A.}\ \bibnamefont
  {Shovkovy}}, \ and\ \bibinfo {author} {\bibfnamefont {P.~O.}\ \bibnamefont
  {Sukhachov}},\ }\bibfield  {title} {\emph {\enquote {\bibinfo {title} {Origin
  of dissipative Fermi arc transport in Weyl semimetals},}\ }}\href {\doibase
  10.1103/PhysRevB.93.235127} {\bibfield  {journal} {\bibinfo  {journal} {Phys.
  Rev. B}\ }\textbf {\bibinfo {volume} {93}},\ \bibinfo {pages} {235127}
  (\bibinfo {year} {2016})}\BibitemShut {NoStop}%
\bibitem [{\citenamefont {Chen}\ \emph {et~al.}(2020)\citenamefont {Chen},
  \citenamefont {Chen},\ and\ \citenamefont {Zilberberg}}]{Chen2020}%
  \BibitemOpen
  \bibfield  {author} {\bibinfo {author} {\bibfnamefont {G.}~\bibnamefont
  {Chen}}, \bibinfo {author} {\bibfnamefont {W.}~\bibnamefont {Chen}}, \ and\
  \bibinfo {author} {\bibfnamefont {O.}~\bibnamefont {Zilberberg}},\ }\bibfield
   {title} {\emph {\enquote {\bibinfo {title} {{Field-effect transistor based
  on surface negative refraction in Weyl nanowire}},}\ }}\href {\doibase
  10.1063/1.5126033} {\bibfield  {journal} {\bibinfo  {journal} {APL Mater}\
  }\textbf {\bibinfo {volume} {8}},\ \bibinfo {pages} {011102} (\bibinfo {year}
  {2020})}\BibitemShut {NoStop}%
\bibitem [{\citenamefont {Datta}(1997)}]{datta1997electronic}%
  \BibitemOpen
  \bibfield  {author} {\bibinfo {author} {\bibfnamefont {S.}~\bibnamefont
  {Datta}},\ }\href@noop {} {\emph {\bibinfo {title} {Electronic transport in
  mesoscopic systems}}}\ (\bibinfo  {publisher} {Cambridge university press},\
  \bibinfo {year} {1997})\BibitemShut {NoStop}%
\bibitem [{\citenamefont {Groth}\ \emph {et~al.}(2014)\citenamefont {Groth},
  \citenamefont {Wimmer}, \citenamefont {Akhmerov},\ and\ \citenamefont
  {Waintal}}]{Groth_2014}%
  \BibitemOpen
  \bibfield  {author} {\bibinfo {author} {\bibfnamefont {C.~W.}\ \bibnamefont
  {Groth}}, \bibinfo {author} {\bibfnamefont {M.}~\bibnamefont {Wimmer}},
  \bibinfo {author} {\bibfnamefont {A.~R.}\ \bibnamefont {Akhmerov}}, \ and\
  \bibinfo {author} {\bibfnamefont {X.}~\bibnamefont {Waintal}},\ }\bibfield
  {title} {\emph {\enquote {\bibinfo {title} {Kwant: a software package for
  quantum transport},}\ }}\href {\doibase 10.1088/1367-2630/16/6/063065}
  {\bibfield  {journal} {\bibinfo  {journal} {New J. Phys.}\ }\textbf {\bibinfo
  {volume} {16}},\ \bibinfo {pages} {063065} (\bibinfo {year}
  {2014})}\BibitemShut {NoStop}%
\bibitem [{\citenamefont {Mikami}\ \emph {et~al.}(2016)\citenamefont {Mikami},
  \citenamefont {Kitamura}, \citenamefont {Yasuda}, \citenamefont {Tsuji},
  \citenamefont {Oka},\ and\ \citenamefont {Aoki}}]{MikamiBW2016}%
  \BibitemOpen
  \bibfield  {author} {\bibinfo {author} {\bibfnamefont {T.}~\bibnamefont
  {Mikami}}, \bibinfo {author} {\bibfnamefont {S.}~\bibnamefont {Kitamura}},
  \bibinfo {author} {\bibfnamefont {K.}~\bibnamefont {Yasuda}}, \bibinfo
  {author} {\bibfnamefont {N.}~\bibnamefont {Tsuji}}, \bibinfo {author}
  {\bibfnamefont {T.}~\bibnamefont {Oka}}, \ and\ \bibinfo {author}
  {\bibfnamefont {H.}~\bibnamefont {Aoki}},\ }\bibfield  {title} {\emph
  {\enquote {\bibinfo {title} {Brillouin-Wigner theory for high-frequency
  expansion in periodically driven systems: Application to Floquet topological
  insulators},}\ }}\href {\doibase 10.1103/PhysRevB.93.144307} {\bibfield
  {journal} {\bibinfo  {journal} {Phys. Rev. B}\ }\textbf {\bibinfo {volume}
  {93}},\ \bibinfo {pages} {144307} (\bibinfo {year} {2016})}\BibitemShut
  {NoStop}%
\bibitem [{\citenamefont {Pal}\ and\ \citenamefont
  {Ghosh}(2025)}]{pal2024multi}%
  \BibitemOpen
  \bibfield  {author} {\bibinfo {author} {\bibfnamefont {A.}~\bibnamefont
  {Pal}}\ and\ \bibinfo {author} {\bibfnamefont {A.~K.}\ \bibnamefont
  {Ghosh}},\ }\href {\doibase 10.5281/zenodo.14995911} {\enquote {\bibinfo
  {title} {Multi higher-order Dirac and nodal line semimetals},}\ } (\bibinfo
  {year} {2025})\BibitemShut {NoStop}%
\end{thebibliography}%

\end{document}